\documentclass[letterpaper,11pt]{article}
\pdfoutput=1

\usepackage[T1]{fontenc}
\usepackage[utf8]{inputenc}

\usepackage{jheppub}	
\usepackage{epstopdf}
\usepackage{multirow}
\usepackage{lmodern}
\usepackage{dsfont}

\usepackage{microtype}	
\usepackage{amsthm}
\usepackage{braket}
\usepackage{mathrsfs}
\usepackage{xspace}
\usepackage{xcolor}
\usepackage{float}
\usepackage[countmax]{subfloat}
\usepackage{accents}
\usepackage{slashed}
\usepackage[capitalize]{cleveref}    
  \crefname{section}{Sec.}{Secs.}
  \Crefname{section}{Section}{Sections}
  \crefname{appendix}{App.}{Apps.}
  \Crefname{appendix}{Appendix}{Appendices}

\usepackage[font=small]{caption}    
\usepackage{enumitem}
\usepackage{multicol}  
\usepackage{mathtools} 
\usepackage{physics}   
\usepackage{subdepth}  
\usepackage{adjustbox} 

\usepackage[most]{tcolorbox}
\tcbset{colback=white, colframe=black,
        highlight math style= {enhanced, 
            colframe=red,colback=red!10!white,boxsep=0pt}
        }

\newlength{\dhatheight}

\providecommand{\href}[2]{#2}
\usepackage{comment}

\definecolor{darkred}{rgb}{0.5,0.0,0.0}
\definecolor{darkblue}{rgb}{0.0,0.0,0.9}
\definecolor{darkerblue}{rgb}{0.0,0.0,0.5}
\definecolor{darkgreen}{rgb}{0.0,0.5,0.0}
\definecolor{black}{rgb}{0.0,0.0,0.0}
\definecolor{brown}{rgb}{0.6,0.4,0.2}

\DeclareRobustCommand{\Ref}[1]{Ref.~\cite{#1}}
\DeclareRobustCommand{\Refs}[1]{Refs.~\cite{#1}}

\providecommand{\ie}{\emph{i.e.}\xspace}
\providecommand{\eg}{\emph{e.g.}\xspace}
\providecommand{\oper}{\mathcal{O}}
\providecommand{\ord}{O}

\newcommand{\s}{\hspace{0.8pt}}

\providecommand{\MSbar}{\ensuremath{\overline{\text{MS}}}\xspace}
\providecommand{\aS}{\ensuremath{\alpha_s}\xspace}
\providecommand{\eps}{\epsilon}

\providecommand{\Lag}{\ensuremath{\mathcal{L}}}
\providecommand{\LQCD}{\ensuremath{\Lambda_\text{QCD}}}

\providecommand{\RQCD}{\ensuremath{R_Q}\xspace}
\providecommand{\RHQET}{\ensuremath{R_h}\xspace}
\providecommand{\MSbardiv}{\frac{1}{\eps} - \gamma_\text{E} + \ln 4\s \pi}

\newcommand{\D}{\text{d}}

\newcommand{\LHQETnl}{\mathcal{L}_\text{HQET}^\text{non-local}}
\newcommand{\SHQETnl}{S_\text{HQET}^\text{non-local}}
\newcommand{\ddx}[1]{\dd[d]{#1}}
\newcommand{\ddp}[1]{\dfrac{\ddx{#1}}{(2\s\pi)^d}}
\newcommand{\ddl}{\frac{\dd^d \ell}{(2\s\pi)^d}}
\providecommand{\fdd}[1]{\mathcal{D}#1}

\newcommand{\dfx}[1]{\dd[4]{#1}}
\newcommand{\dfp}[1]{\dfrac{\dfx{#1}}{(2\s\pi)^4}}

\DeclareMathOperator{\Sdet}{Sdet}

\newcommand{\RPI}{\,\,\xrightarrow[\hspace{5pt}\text{RPI}\hspace{5pt}]{}\,\,}

\usepackage[toc,page]{appendix}

\usepackage{etoolbox}
\appto\appendix{\addtocontents{toc}{\protect\setcounter{tocdepth}{1}}}

\definecolor{colorTC}{rgb}{.2,.7,.2}

\title{Functional Methods for Heavy Quark Effective Theory}
\author[1]{Timothy Cohen,}
\author[2]{Marat Freytsis,}
\author[1]{and Xiaochuan Lu\s}

\affiliation[1]{\footnotesize Institute for Fundamental Science, Department of Physics, University of Oregon, Eugene, OR 97403, USA}
\affiliation[2]{\footnotesize NHETC, Department of Physics and Astronomy, Rutgers University, Piscataway, NJ 08854, USA}

\emailAdd{tcohen@uoregon.edu}
\emailAdd{marat.freytsis@rutgers.edu}
\emailAdd{xlu@uoregon.edu}

\abstract{
We use functional methods to compute one-loop effects in Heavy Quark Effective Theory. The covariant derivative expansion technique facilitates the efficient extraction of matching coefficients and renormalization group evolution equations. This paper provides the first demonstration that such calculations can be performed through the algebraic evaluation of the path integral for the class of effective field theories that are (i) constructed using a non-trivial one-to-many mode decomposition of the UV theory, and (ii) valid for non-relativistic kinematics. We discuss the interplay between operators that appear at intermediate steps and the constraints imposed by the residual Lorentz symmetry that is encoded as reparameterization invariance within the effective description. The tools presented here provide a systematic approach for computing corrections to higher order in the heavy mass expansion; precision applications include predictions for experimental data and connections to theoretical tests via lattice QCD. A set of pedagogical appendices comprehensively reviews modern approaches to performing functional calculations algebraically, and derives contributions from a term with open covariant derivatives for the first time.
}

\begin{document}
\maketitle
\flushbottom

\section{Introduction}
\label{sec:intro}
The modern perspective on Effective Field Theories (EFTs) takes them to be well-defined quantum field theories in their own right, with all the attendant structure that implies. In particular, one can define an EFT at the quantum level using a path integral. However, the standard approach to extracting observables is to sidestep this intimidating object by organizing perturbation theory through the use of Feynman diagrams. The recent reintroduction~\cite{Henning:2014wua} of the \emph{covariant derivative expansion} (CDE)~\cite{Gaillard:1985uh,Chan:1986jq,Cheyette:1987qz} in a form tailored to tackling a broader set of EFTs has sparked a renaissance in the extraction of physics directly from the path integral using functional techniques. A noteworthy example of this progress has been a general matching formula at one-loop~\cite{Henning:2016lyp, Fuentes-Martin:2016uol} that captures the far off-shell fluctuations of heavy fields. This result can be used for the extraction of the effective operators and their Wilson coefficients that characterize the effects of the heavy physics, encoded in the so-called Universal One-Loop Effective Action (UOLEA)~\cite{Henning:2014wua, Drozd:2015rsp, Ellis:2016enq, Zhang:2016pja, Ellis:2017jns}; one of its features is that it elegantly packages together what would be multiple independent Feynman diagram calculations. This has seen ready application to beyond the Standard Model (SM) scenarios by facilitating one-loop matching onto the so-called SMEFT. However, there exist a large number of EFTs whose relationship to UV physics cannot be captured by simply integrating out an entire heavy field. Here, we present the first application of the modern functional approach to such EFTs. Specifically, we work in a particular low-energy limit of the SM~\cite{Nussinov:1986hw,Shifman:1987rj,Isgur:1989vq,Isgur:1989ed} expanded around a non-trivial background, the Heavy Quark Effective Theory (HQET)~\cite{Isgur:1989vq, Eichten:1989zv, Georgi:1990um, Grinstein:1990mj}, see also the reviews~\cite{Georgi:1991mr, Neubert:1993mb, Shifman:1995dn, Wise:1997sg, Manohar:2000dt}.

There are many novel features of HQET when compared to the SMEFT. Perhaps the most striking is that --- as a model of the long distance fluctuations of a heavy particle --- HQET is a non-relativistic EFT. Obviously, the path integral is a valid description of a non-relativistic theory (after all, it was invented as an alternative description of quantum mechanics~\cite{Feynman:1942us, Feynman:1948ur}), but the concrete demonstration that the functional approach is useful for precision non-relativistic field theory computations has been lacking until now.\footnote{To our knowledge, the only time the path integral has been discussed in the context of HQET was in~\cite{Mannel:1991mc} where the tree-level formulation of the theory was first derived, and in \cite{Chen:1993np, Kilian:1994mg, Sundrum:1997ut} where RPI was briefly discussed in the context of the path integral.} Another intriguing aspect of HQET is that it invokes the concept of a mode expansion. A single heavy quark field is decomposed into a pair of fields, which model short and long distance fluctuations. Then, \emph{assuming a particular kinematic configuration}, only the long distance modes can be accessed as external states. The short distance fluctuations can therefore be integrated out, which generates a tower of local EFT operators. It is this description that is called HQET. This type of one-to-many correspondence plays a role in many modern formulations of EFTs, and for the first time here we put such models on even firmer theoretical footing by computing observables directly from the path integral. Furthermore, a theoretically appealing aspect of this work is that the covariant derivative expansion of the functional integral manifests the symmetries of the theory in a transparent way. This is obviously true for gauge invariance, but we will also be able to approach the residual Lorentz symmetry known as Reparameterization Invariance (RPI) from a novel vantage point. In particular, we will identify an intermediate stage in our calculations where RPI becomes manifest. This provides a nice contrast to the Feynman diagram approach, where this invariance only holds once one sums the full set of relevant diagrams.

Through our application of the functional approach to HQET, we will expose some important features of the general formalism. Matching a UV theory onto an EFT, first done for HQET in \Ref{Falk:1990yz}, can be performed diagrammatically by equating matrix elements computed with the two descriptions of the theory at a common kinematic point. This requires knowing the EFT operator expansion and identifying the operators that are relevant to the EFT matrix element calculation. By contract, using functional methods the generation of operators and matching of Wilson coefficients occurs in a single step. There is no need to specify the structure of the EFT before performing a matching calculation. Working through the example of HQET will show that the problem for how to marry the mode decomposition with functional methods has a nearly universal solution. Specifically, given an implementation of the mode decomposition using operator-valued projectors, one can derive a functional equation of motion for the short distance modes. This can be used to integrate out these modes, yielding a non-local EFT description that encodes the complete dynamics of the full theory in the relevant kinematic limit. This justifies the construction of the EFT as a full path integral over a well defined field.

Deriving concrete predictions using conventional techniques typically involves several steps and many subtleties, but a functional approach makes the procedure more algorithmic. The complicated multi-mode matching calculation has a simple form whose structure is elucidated by analyzing the resulting integrals using the method of regions~\cite{Beneke:1997zp, Smirnov:2002pj}. As is well known, matching in HQET only receives support from diagrams that have loops with both a short distance mode and a mode that propagates in the EFT. Understanding how to access this exact class of diagrams using functional methods has been the subject of some confusion. Showing that we can derive matching and running in HQET with these methods is a conclusive demonstration that functional methods provide a complete framework at one-loop. These results will be summarized below as a simple \emph{master formula} that encodes the matching of QCD onto HQET at one-loop and to any order in the heavy mass expansion, see \cref{eqn:SHQET1loop}.

There are important practical implications of this work. One of the primary purposes of HQET is to provide efficient methods for precision calculations \emph{within} the SM. There are a number of EFTs widely used to facilitation precise predictions including Soft Collinear Effective Theory~\cite{Bauer:2000yr,Bauer:2001ct,Bauer:2001yt}, theories of non-relativistic bound states such as  nrQCD~\cite{Caswell:1985ui, Bodwin:1994jh, Grinstein:1997gv}, and others. Improving on the precision of a calculation often requires the application of novel theoretical approaches, with a goal to provide computational benefits over a naive perturbative expansion evaluated using Feynman diagrams. Specifically, functional technology has seen little use in this context, despite the fact that some of the simplifications it provides are arguably most relevant to the questions these kinematic EFTs are designed to answer. By showing we can reproduce non-trivial matching and running results for HQET here, we open the door to understanding how to apply functional techniques in these other contexts. We additionally lay the foundation for performing new calculations within HQET itself. In particular, one can now perform matching calculations to higher order in the heavy mass expansion, which would be relevant to high precision measurements made at experiments such as LHCb and Belle II, and for connecting lattice gauge theory calculations performed in the heavy quark limit to their continuum limits~\cite{Heitger:2003nj, Blossier:2010jk}.

The rest of the paper is organized as follows. In the rest of \cref{sec:intro}, we provide a summary of the known results that we reproduce in a novel way throughout this paper. Next, \cref{sec:HQET} provides a condensed introduction to HQET. \Cref{sec:DiagrammaticResidue} provides a summary of the traditional method of calculating the simplest piece of the one-loop HQET matching, the residue matching. (Readers familiar with HQET could skip these two sections.) In \cref{sec:FunctionalMethods}, we review how to extract matching and running from a functional determinant. (Readers familiar with functional methods could skip this section.) In \cref{sec:MatchingHQET}, we introduce the use of functional integration to construct kinematic EFTs, and clarify how the method of regions simplifies the derivation of the resulting master matching formula. \Cref{sec:FunctionalMatching} is then dedicated to an explanation of how such matching is performed to one-loop order. This is followed by \cref{sec:Examples}, which strengthens the case for functional methods by providing additional matching calculations, along with an example that shows how operator running can be derived in the formalism as well. \Cref{sec:Conc} then concludes.

An extensive set of pedagogical appendices provide an introduction to many of the relevant technical details. The ``covariant derivative expansion''  technique used here was originally invented in 1980s \cite{Gaillard:1985uh,Chan:1986jq,Cheyette:1987qz}, we refer to this as ``original CDE'' in \cref{appsec:CDE}. This was reintroduced in the context of modern EFT calculations in~\Ref{Henning:2014wua}, and has been applied by~\Refs{Drozd:2015rsp, Ellis:2016enq, Zhang:2016pja, Ellis:2017jns, Summ:2018oko} to develop one-loop universal effective actions. A closely related variant of the original CDE, which we call ``simplified CDE'' in \cref{appsec:CDE}, was proposed in \Ref{Henning:2016lyp}. This more rudimentary version of the CDE turns out to be significantly more convenient for extracting operators that do not involve a gauge field strength. Most of the results in the main text of this paper were derived using this simplified CDE. In \cref{appsec:CDE}, we clarify the relations between these two versions of the CDE. For completeness, in \cref{appsec:CDE} we also provide some simple universal results (tabulated in \cref{appsubsubsec:functionaltraces}) for functional traces derived using the CDE. These include the famous elliptic operator, see \cref{eqn:UniversalDU}, which is the central object of study in the development of the UOLEA. Additionally, we provide the first computation of the contributions to the UOLEA from a term with an open covariant derivative (truncated at dimension four). This result is provided in \cref{eqn:UniversalDUJ}, and was previously unknown as emphasized by Refs.~\cite{Ellis:2017jns, Brivio:2017vri, Kramer:2019fwz}. A variety of RGE example calculations are given in \cref{appsec:ExamplesRGE}, and a generalization of the Heavy-Heavy matching calculation is provided in \cref{sec:HHMatch2}.

\subsection{Summary of Results}
\label{sec:Summary}
In what follows, we will present our methods by way of a few canonical matching and running calculations. First, we derive the high scale HQET Wilson coefficients by matching QCD onto the EFT for the \emph{heavy--light} currents $\bar{q}\, \gamma^\mu (\gamma^5) \,Q$ and the \emph{heavy--heavy} currents $\bar{Q}_1\, \gamma^\mu (\gamma^5)\, Q_2$ using purely functional methods. We also present a functional derivation of running effects, using the first subleading operators in the HQET Lagrangian as a concrete example.

In order to fix our notation and make comparison to standard results straightforward, we provide a brief compendium of the results we will reproduce in this paper. The conventional derivation is presented in much more detail in standard references, \eg, \Refs{Neubert:1993mb, Manohar:2000dt}.\footnote{\textbf{Notation:} We have mostly chosen to follow the notation of \Ref{Manohar:2000dt}, but have made a number of minor changes, which is partially why we provide this summary here. For dimensional regularization (dim.\ reg.), our convention is to work in $d=4 - 2\eps$ dimensions. We have also chosen to use a different standard notation of the EFT heavy quark fields. Additionally, we have reserved the superscript numeral in parenthesis notation to denote loop order, $R^{(0)}$ is tree-level, $R^{(1)}$ is the one-loop correction, and so on.  This is different from the standard notation in the HQET community, \eg in \Ref{Manohar:2000dt} the one-loop correction to the residue is denoted as $R_1$.  Additionally, we note that we will only denote the loop order of the renormalized terms, \ie, counterterms are implicit.} This list provides a useful context for our goals here.

\subsubsection*{Propagator Residue}
When performing a matching calculation in any off-shell scheme, such as \MSbar, it is critical to track the difference in propagator residues (necessary for obtaining the desired $S$-matrix elements using the LSZ reduction procedure) when moving from the full theory to the EFT. Taking the functional point of view, the resulting effect shows up as a rescaling of the kinetic terms for the EFT fields. This can then be moved to its canonical position in the Wilson coefficients by a field redefinition. Thus, the first step performed in what follows is to derive the one-loop corrections to the kinetic terms from QCD:
\begin{equation}
  \Lag_\text{HQET} \supset  \Big(1- \Delta R^{(1)}\s \aS\Big) \, \bar{h}_v\, (i\s v\cdot D)\, h_v \,,
\end{equation}
where $h_v$ models the long distance fluctuations of the heavy quark field, $v_\mu$ is the reference vector defined in \cref{eq:pmuHQET} below, $D_\mu$ is the covariant derivative, $\alpha_s$ is the strong fine structure constant, and $\Delta R^{(1)}$ is the matching correction for the propagator residue.  In a traditional matching calculation, the correction comes from the difference of the residues between the full and EFT descriptions:
\begin{equation}
 \Delta R^{(1)} \s \aS \equiv \bigg(R_Q^{(1)} - R_h^{(1)}\bigg)\, \aS = - \frac{1}{3} \frac{\aS}{\pi} \pqty{ 3\ln\frac{\mu^2}{m_Q^2} + 4} \,.
 \label{eqn:DeltaRdef}
\end{equation}
While this result is sensitive to the choice of matching scale $\mu$ due to presence of a UV divergence; the cancellation of terms that depend on the IR regulator serves as a check that the IR behavior of the two theories is identical.

\subsubsection*{Heavy--Light Current}
At leading order in the heavy-mass expansion, two HQET operators have the same quantum numbers as the heavy--light vector current of QCD, implying that they can appear in the matching:
\begin{equation}
  \bar{q}\, \gamma^\mu \, Q = C_{V,1}\pqty{\frac{m_Q}{\mu}, \aS(\mu)}\, \bar{q}\, \gamma^\mu\, h_v + C_{V,2}\pqty{\frac{m_Q}{\mu}, \aS(\mu)}\, \bar{q}\, v^\mu\,  h_v \,,
\label{eq:C1V}
\end{equation}
where $C_{V,i}$ is the to-be-calculated matching coefficient for the vector current, which is a function of the heavy quark mass $m_Q$ and the strong coupling; there is an analogous expression for the axial current derived by replacing $C_{V,i} \rightarrow C_{A,i}$, $\gamma^\mu \rightarrow \gamma^\mu \s\gamma^5$, and  $v^\mu \rightarrow v^\mu \s\gamma^5$ in \cref{eq:C1V}. The answer up to one-loop (see Eq.~(3.48) in \Ref{Manohar:2000dt}) can be written as
\begin{subequations}
\label{eqn:HLresult}
\begin{align}
\setlength{\jot}{7pt}
  C_{V,1} &= 1 + \Bigg[ \frac{1}{2} \Delta R^{(1)} + V_{\text{HL},1}^{(1)} - V_\text{eff}^{(1)} \Bigg]\, \aS + \dotsb \,, \\
  C_{V,2} &= V_{\text{HL},2}^{(1)}\, \aS + \dotsb \,,
\end{align}
\end{subequations}
where $\Delta R^{(1)}$ is defined in \cref{eqn:DeltaRdef}, and the other terms (see Eqs.~(3.66) and (3.73) in \Ref{Manohar:2000dt}, respectively) are
\begin{subequations}
\label{eqn:HLcomponents}
\begin{align}
 \setlength{\jot}{5pt}
V_{\text{HL},1}^{(1)}\, \aS &= - \frac{1}{3} \frac{\aS}{\pi} \pqty{\MSbardiv + 2} \,, \\
V_{\text{HL},2}^{(1)}\, \aS &= + \frac{2}{3} \frac{\aS}{\pi} \,, \\
V_\text{eff}^{(1)}\, \aS &= - \frac{1}{3} \frac{\aS}{\pi} \pqty{\MSbardiv} \,,
\end{align}
\end{subequations}
where $\MSbardiv$ is the standard factor that is subtracted when using the $\overline{\text{MS}}$ scheme, with $\gamma_\text{E}$ denoting the Euler-Mascheroni constant. Then the axial matching coefficients are given by $C_{A,1} = C_{V,1}$ and $C_{A,2} = - C_{V,2}$. The one-loop matching coefficients are thus (compare with Eq.~(3.74) in \Ref{Manohar:2000dt})
\begin{subequations}
\begin{align}
  C_{V,1} &= 1 + \frac{\aS}{\pi} \pqty{\ln\frac{m_Q}{\mu} - \frac{4}{3}} + \ord\Big(\aS^2\Big) \,, \\
  C_{V,2} &= \frac{2}{3} \frac{\aS}{\pi}+ \ord\Big(\aS^2\Big) \,.
\end{align}
\end{subequations}

\subsubsection*{Heavy--Heavy Current}
In case of the heavy-heavy currents, for simplicity we will take the special kinematic choice of zero recoil, corresponding to $v_1 = v_2$ in the EFT (the generalization to $v_1 \neq v_2$, first done in \Ref{Neubert:1992tg}, is discussed in \cref{sec:HHMatch2}). In this limit, all possible HQET operators at leading order in the mass expansion are equal by the equations of motion, and the matching between QCD and HQET is simply\footnote{The choice of notation for the HQET fields here is made for of ease of legibility in later sections, but deserves comment here so as to not mislead. The labels $v_{1,2}$ differ only to keep track of finite-mass corrections. They do not indicate different velocities, and the relation between the QCD and HQET operators is not valid except in the zero-recoil limit. (Compare with Eq.~(3.89) in \Ref{Manohar:2000dt}.)}
\begin{equation}
  \bar{Q}_1\, \gamma^\mu\, Q_2 = \eta_{V}\, \bar{h}_{v_1} \, \gamma^\mu\, h_{v_2} \,,
\end{equation}
with an analogous expression for the axial current given by the replacement $\eta_V \rightarrow \eta_A$ and $\gamma^\mu \rightarrow \gamma^\mu \s \gamma^5$. The results can be parameterized as (compare with Eqs.~(3.98) and~(3.101) in \Ref{Manohar:2000dt})
\begin{subequations}\label{eqn:hhresult}
\begin{align}
  \eta_V &= 1 + \frac{1}{2}\, \pqty{ \Delta R^{(1)}_1 + \Delta R^{(1)}_2}\,\aS + \Delta V_\text{HH}^{(1)}\, \aS \,, \\
  \eta_A &= \eta_V - \frac{2}{3}\frac{\aS}{\pi} \,,
\end{align}
\end{subequations}
where $\Delta R_1^{(1)}$ and $\Delta R_2^{{(1)}}$ are \cref{eqn:DeltaRdef} for the heavy quarks $Q_1$ and $Q_2$ respectively, and
\begin{equation}
\label{eqn:hhcomponents}
  \Delta V_\text{HH}^{(1)} = - \frac{2}{3} \frac{\aS}{\pi} \,
  \bqty{ 1 + \frac{3}{m_1-m_{2}} \pqty{m_1\ln\frac{m_{2}}{\mu} - m_{2}\ln\frac{m_1}{\mu}} } \,,
\end{equation}
where $m_{1,2}$ corresponds to the mass of $Q_{1,2}$. The matching coefficients then take the form
\begin{subequations}
\begin{align}
  \eta_V &= 1 + \frac{\aS}{\pi}\, \pqty{ -2 + \frac{m_1+m_{2}}{m_1-m_{2}} \,\ln\frac{m_1}{m_{2}}} \,, \\[2pt]
  \eta_A &= 1 + \frac{\aS}{\pi}\, \pqty{ -\frac{8}{3} + \frac{m_1+m_{2}}{m_1-m_{2}}\, \ln\frac{m_1}{m_{2}}} \,,
\end{align}
\end{subequations}
in agreement with Eqs.~(3.97), (3.99), and (3.101) of \Ref{Manohar:2000dt}.

\subsubsection*{\boldmath $\beta$-functions}
Finally, we reproduce the expressions for the running of the HQET matching coefficients at one-loop and at $\ord(1/m_Q)$ (compare with Eq.~(4.8) in \Ref{Manohar:2000dt}):
\begin{equation}
  \Lag_1 = -\, c_\text{kin}(\mu)\, \bar{h}_v\, \frac{D_\perp^2}{2\s m_Q}\, h_v
         - c_\text{mag}(\mu)\, g_s\, \bar{h}_v\, \frac{\sigma_{\mu\nu}\s G^{\mu\nu}}{4\s m_Q}\, h_v \,.
\end{equation}
The Renormalization Group Equations (RGEs) are~\cite{Luke:1992cs,Eichten:1990vp,Falk:1990pz}
\begin{subequations}
\label{eq:RGESummary}
\begin{align}
\setlength{\jot}{7pt}
  \mu\s\dv{\mu}\s c_\text{kin} &= 0 \,,\\
  \mu\s\dv{\mu}\s c_\text{mag} &= \frac{\aS}{4\s\pi}\, 2\s C_A\, c_\text{mag} \,,
\end{align}
\end{subequations}
where $C_A$ denotes the Casimir factor for the adjoint representation

In what follows, we will show how to reproduce all of these results using functional methods equipped with the CDE technique.

\section{Heavy Quark Effective Theory}
\label{sec:HQET}
One of our main goals here is to initiate the study of functional methods for higher-order calculations in EFTs that are derived by performing multi-modal decomposition of full theory fields. Such EFTs are quite common; they occur when the kinematics of the process being studied selects a preferred reference frame due to, \eg a conservation law preventing the decay of a heavy particle, or a measurement function that forces the external states into a particular region of phase space. These restrictions imply that there are full theory modes which cannot be put on-shell within the EFT regime of validity, and so it is sensible to integrate them out. This procedure breaks the full theory space-time symmetries to some subgroup, while potentially also introducing new internal ones. A theory of this type is the Heavy Quark Effective Theory (HQET), which describes the fluctuations of a heavy quark ($m_Q \gg \LQCD$) in the presence of light QCD charged degrees of freedom. The simplifications and universal behavior of QCD in the heavy-mass limit were first appreciated by \Refs{Nussinov:1986hw,Shifman:1987rj} and especially \Refs{Isgur:1989vq,Isgur:1989ed}. A non-covariant EFT making this behavior manifest was later developed and shown to be well-behaved in perturbation theory~\cite{Eichten:1989zv,Grinstein:1990mj}, and finally given a covariant formulation~\cite{Georgi:1990um}. We provide a review of HQET here, with a particular emphasis on the equation of motion due to the critical role it plays in what follows. The reader familiar with HQET can skip ahead to \cref{sec:FunctionalMethods}, while more details can be found in, \eg Sec.~4.1 of \Ref{Manohar:2000dt}.

The full theory (QCD) Lagrangian for a heavy quark includes
\begin{equation}
\label{eq:QLag}
  \Lag_\text{QCD} \supset \bar{Q}\,\big (i\s\slashed{D} -m_Q\big)\, Q \,,
\end{equation}
where $Q$ is our heavy quark, and the covariant derivative only includes the QCD interactions, $D_\mu = \partial_\mu - i\s g_s \,G_\mu^a\s T^a$. In the following, additional interactions of the heavy quark will be modeled through the introduction of current operators as needed.

Naively, it does not seem that the Lagrangian of \cref{eq:QLag} has a good expansion about the $m_Q \to \infty$ limit. The key insight that allows one to circumventing this issue lies in an appropriately chosen phase redefinition of the field, whose purpose is to cancel the mass term for certain components. The full quark field can then be separated into so-called short distance and long distance fields, where the latter become approximately mass-independent. Physically, this is motivated by the realization that the heavy quark cannot be pushed very far off-shell by degrees of freedom for which $|q| \lesssim \LQCD$. To make this manifest, we decompose a heavy quark's momentum as
\begin{equation}
\label{eq:pmuHQET}
  p^\mu = m_Q\s v^\mu + k^\mu \,,
\end{equation}
where $v^\mu$ is a unit time-like vector, and $k^\mu$ is the residual heavy quark momentum which models small fluctuations about its mass shell. For kinematic configurations such that all Lorentz invariants depending on $k^\mu$ are small compared to $m_Q$, a truncation at finite order in the $|k|/m_Q$ expansion is justified.\footnote{When we expand assuming $|k| \ll m_Q$, we take each element of the $k^\mu$ vector to be much smaller than $m_Q$.} Since the theory is expanded around the $m_Q \to \infty$ limit, the structure of HQET does not know about the mass of the heavy quark except through various non-dynamical quantities. In particular, no sensitivity to $m_Q$ appears in any of the calculations beyond what is encoded in the structure of the matching coefficients.

As a brief aside, we note that the decomposition in \cref{eq:pmuHQET} is not unique. In particular, a simultaneous transformation of $k^\mu$ and $v^\mu$ by a fixed vector:
\begin{equation}
\label{eq:RPIdef}
  k^\mu \RPI k^\mu + \delta k^\mu \qquad \text{and} \qquad
  v^\mu \RPI v^\mu - \frac{\delta k^\mu}{m_Q} \,,
\end{equation}
leaves $p^\mu$ unchanged. Enforcing that $v^\mu$ remains a unit vector implies that $\delta k^\mu$ must satisfy $v \cdot \delta k = \delta k^2/(2\s m_Q)$. Reparameterization invariance (RPI) is then the statement that physical observables cannot depend on $\delta k^\mu$, thereby enforcing the residual Lorentz invariance of the underlying full theory. We will explore the interplay between RPI and the functional approach in \cref{sec:RPI} below.

In order to make use of \cref{eq:pmuHQET}, we decompose the heavy quark field into two fields by first extracting the rapidly-varying phase that remains unchanged by low-energy interactions, and then using $v^\mu$-dependent projectors to split the remaining field into short distance and long distance components:
\begin{subequations}\label{eq:QProjections}
\begin{align}
\setlength{\jot}{7pt}
  h_v(x) &= e^{i\s m_Q\s v\cdot x}\, \frac{1 + \slashed{v}}{2} Q(x) \,, \\
  H_v(x) &= e^{i\s m_Q\s v\cdot x}\, \frac{1 - \slashed{v}}{2}\, Q(x)\,,
\end{align}
\end{subequations}
or equivalently
\begin{align}
  Q(x) = e^{-i\s m_Q\s v\cdot x}\, \Big[h_v(x) + H_v(x)\Big]\,.
  \label{eq:QSplit}
\end{align}
Plugging \cref{eq:QSplit} into \cref{eq:QLag} yields
\begin{equation}
  \Lag \supset \bar{h}_v \,\big(i\s v \cdot D\big)\, h_v - \bar{H}_v\,  \big(i\s v \cdot D + 2\s m_Q\big)\, H_v
               + \bar{H}_v\, i\s \slashed{D}_\perp\, h_v + \bar{h}_v\, i\s\slashed{D}_\perp\, H_v\,,
 \label{eqn:LQCD}
\end{equation}
since the projectors enforce $\slashed{v}\, h_v = h_v$ and $\slashed{v}\, H_v = -H_v$ (and hence $\bar{H}_v\, \slashed{v}\, h_v = \bar{h}_v\, \slashed{v}\, H_v = 0$), and we have defined
\begin{align}
  D_\perp^\mu \equiv D^\mu - v^\mu\, (v\cdot D)\,.
\label{eq:Dperp}
\end{align}
The Lagrangian in \cref{eqn:LQCD} makes the interpretation of $H_v$ as the heavy mode manifest --- this field has an effective mass of $2\s m_Q$, permitting a description at lower energies in terms of $h_v$ alone.

To formally integrate out $H_v$ at tree-level, we solve for its equation of motion
\begin{equation}
 H_v = \frac{1}{i\s v \cdot D + 2\s m_Q}\, i\s \slashed{D}_\perp\, h_v\,,
\end{equation}
and plug it into \cref{eqn:LQCD}, which yields
\begin{equation}
  \LHQETnl \supset \bar{h}_v\, \Bigg( i\s v \cdot D + i\s \slashed{D}_\perp\, \frac{1}{i\s v \cdot D + 2\s m_Q}\, i\s \slashed{D}_\perp \Bigg)\, h_v\,.
\label{eqn:LHQETnonlocal}
\end{equation}
Provided that the momentum of the field $h_v$ satisfies $|k| \ll m_Q$, \cref{eqn:LHQETnonlocal} can be expressed into a convergent series of local terms
\begin{equation}
  \frac{1}{i\s v \cdot D + 2\s m_Q}
    = \frac{1}{2\s m_Q} - \frac{1}{\big(2\s m_Q\big)^2} (i\s v \cdot D) + \frac{1}{\big(2\s m_Q\big)^3}(i\s v \cdot D)^2 + \dotsb\;.
\label{eq:HQETExpansion}
\end{equation}
Since the leading term is then insensitive to the Dirac structure of the spinor components of the field, the theory has gained an approximate $SU(2)$ heavy-spin symmetry, which can be expanded to $SU(2\s n_h)$ in the presence of $n_h$ heavy flavors. This is an example of an aforementioned emergent symmetry of the EFT.

To understand how this procedure for deriving the EFT Lagrangian can be interpreted at the path integral level, we note that the projection operators in \cref{eq:QProjections} imply that $h_v$ and $H_v$ can be treated as two orthogonal projections of $Q$. As such, the path-integral measure factorizes:
\begin{equation}
  \int \fdd{Q} = \int\fdd{h_v}\, \int\fdd{H_v}\,.
\end{equation}
Since the resulting Lagrangian is quadratic in $H_v$, the Gaussian integral over the short-distance field immediately yields \cref{eqn:LHQETnonlocal}. Therefore, the procedure described here allows us to literally integrate out the short distance modes of our original quark. This elegant derivation of the tree-level HQET Lagrangian by way of the path-integral measure was first presented in \Ref{Mannel:1991mc}. Furthermore, it also justifies the use of the functional approach to matching and running developed in \Ref{Henning:2016lyp}. Note that there would be no significant obstructions even if our projectors were quantum operators instead of simply being functions of the kinematics as above, since all objects in the path integral are treated as operators acting on fields. As long as the mode decomposition that is used to define an EFT can be written in terms of operator-valued projectors, we expected the methods developed here to be universally applicable.

\subsection{Decoupling the Heavy Quark}
\label{sec:Decoupling}
The decomposition of the full QCD Lagrangian given in \cref{eqn:LQCD} makes it clear that $H_v$ can be identified as a short distance mode which one can integrate out in the limit that $|k| \ll m_Q$ for all fields in a given process. This procedure yields the non-local Lagrangian given in \cref{eqn:LHQETnonlocal}, which makes predictions that are equivalent to QCD for processes involving only $h_v$ modes in the external states. Expanding the non-local Lagrangian using \cref{eq:HQETExpansion} and truncating the series at some order in $1/m_Q$ yields an EFT which is valid for momenta $|k| \ll m_Q$. In this subsection, we will briefly review the connection between this procedure and the fact that the heavy quark should not contribute to the running of the QCD gauge coupling at scales below $m_Q$. This both has a conceptual benefit, and will also be of practical importance since similar arguments will be used below when we derive our master formula for one-loop matching given in \cref{eqn:SHQET1loop}.

For this argument, we will work directly with the $h_v$ and $H_v$ fields, and we will use the terms that are diagonal in these fields to derive propagators, while the mixed terms will be treated as interactions. Two features are of critical importance, the fact that the kinetic terms are linear in the momentum of the state, and the relative minus sign between the $h_v$ and the $H_v$ kinetic terms. The later fact implies that the $i\s \eps$ factors in the propagators will take the opposite sign such that the same Wick rotation can be used to Euclideanize diagrams involving both $h_v$ and $H_v$ propagators:

The QCD Lagrangian written as \cref{eqn:LQCD} has three types of couplings between the heavy quark modes and the gluon: diagonal couplings $h_v h_v G_\mu$, $H_v H_v G_\mu$, and an off-diagonal coupling $H_v h_v G_\mu$. The integral that results when computing the contribution to the vacuum polarization for the gluon from the two diagonal loops schematically take the form
\begin{align}
  I_\text{diag} \sim \int \ddp{q} \left(\frac{1}{v\cdot q + m \pm i\s\eps}\right)
                       \left( \frac{1}{v\cdot (q+p) + m \pm i\s\eps}\right) \,,
    \label{eqn:Idiagonal}
\end{align}
where $m$ is an IR regulator for the $h_v$ loop and is equal to $2\s m_Q$ for the $H_v$ loop. Due to the linear nature of the kinetic terms and the sign on the factor of $i\s\eps$, when integrating over $q^0$, the poles reside on only one side of the real axis, and one can deform the contour away from all them yielding zero contribution. However, the situation is different for the mixed loop, where the integral takes the form
\begin{align}
  I_\text{off-diag} \sim \int \ddp{q} \left(\frac{1}{v\cdot q + m + i\s\eps}\right)
                           \left( \frac{1}{v\cdot (q+p) + 2\s m_Q - i\s\eps}\right)\, .
  \label{eqn:Ioffdiagonal}
\end{align}
Now we see that the opposite sign on the $i\s \eps$ terms implies that there is a pole in both the positive and negative $\Im q^0$ half-planes. The contour will enclose a pole for any possible deformation, yielding a non-zero contribution to the $\beta$-function.

The argument above makes it clear why in HQET the heavy quark does not contribute to the RGEs of the gauge coupling. Once we construct HQET by integrating out $H_v$ and expanding, the heavy field $H_v$ is non-propagating. Therefore, there are no diagrams that yield contributions of the type in \cref{eqn:Ioffdiagonal}. The mode $h_v$ only yields potentially relevant integrals of the type in \cref{eqn:Idiagonal}, which vanish as we have argued. Similar reasoning will be critical to the derivation of the master formula for matching in \cref{eqn:SHQET1loop} below.

\section{Residues Using Feynman Diagrams}
\label{sec:DiagrammaticResidue}
In this section, we will review the diagrammatic approach to calculating matching coefficients, using the simple example of the residue of the on-shell propagator for concreteness. Along the way, we will encounter an order of limits issue, which is a manifestation of the IR divergence structure of QCD. We will then revisit the calculation by relying on the so-called method of regions~\cite{Beneke:1997zp, Smirnov:2002pj} that will avoid the need to deal with this subtlety directly. This has the additional benefit of providing a familiar setting to review the method of regions, which will be a critical tool in the derivation of our master formula for HQET matching coefficients below.

When performing matching calculations, one typically equates matrix elements as calculated in the full theory and the EFT. Particular care must be taken to include the appropriate residue factors for the external states to ensure that the LSZ reduction is correctly implemented. One way to extract the residue for the heavy quark $Q$ is to take the derivative of the 1PI corrections to the propagator $-i\s \Sigma(\slashed{p})$, and evaluate it on the mass shell. We will compute this factor for a quark in QCD $\RQCD =1 + \RQCD^{(1)} \aS + \dotsb$, and for a quark in HQET $\RHQET = 1 + \RHQET^{(1)} \aS + \dotsb$, from which we get the quantity that appears in matching calculations $\Delta R^{(1)} = \RQCD^{(1)} - \RHQET^{(1)}$, see \cref{eqn:DeltaRdef}.

\subsection{Residue in QCD}
The one-loop QCD residue $\RQCD^{(1)}$ can be obtained by computing the two-point 1PI function
\begin{align}
 -i\s\Sigma _\text{QCD}(\slashed{p})
   &= -\frac{4}{3}\, g_s^2\, \mu^{2\eps}
       \int \ddp{q} \frac{(2-d)(\slashed{q} + \slashed{p}) + d\,m_Q}{q^2\bqty{(q+p)^2 - m_Q^2}}
       - \text{c.t.} \notag\\
   &= -i\bqty{ (A - A_\text{ct}) \,m_Q + (B - B_\text{ct})\,\slashed{p} }\, .
\label{eqn:SigmaQCDIntegral}
\end{align}
Here as well as throughout this paper, ``c.t.'' denotes the counter term contributions, and we take the Feynman gauge $\xi=1$ for gauge boson propagators. Performing standard manipulations, we derive\footnote{See Eqs.~(3.53), (3.55), and~(3.57) in \Ref{Manohar:2000dt}, noting again that we use $d = 4-2\s\eps$, while \Ref{Manohar:2000dt} uses $d = 4-\eps$.}
\begin{subequations}
\begin{align}
  A\big(\s p^2\s \big) &= \frac{\aS}{3\s\pi} \Big(4\s\pi\,\mu^2\Big)^\eps\, \Gamma(\eps)\,
    4\left(1 - \frac{\eps}{2}\right) \int_0^1 \D x \left[m_Q^2\, x - p^2\, x(1-x)\right]^{-\eps}\, , \\
  B\big(\s p^2\s \big) &= -\frac{\aS}{3\s\pi} \Big(4\s\pi\,\mu^2\Big)^\eps\, \Gamma(\eps)\,
    2\,(1 - \eps) \int_0^1 \D x \,(1-x) \left[ m_Q^2\, x - p^2\, x\s(1-x) \right]^{-\eps}\, , \\
  A_\text{ct} &= \frac{\aS}{3\s\pi}\, 4\left(\MSbardiv\right)\, , \\
  B_\text{ct} &= -\frac{\aS}{3\s\pi} \left(\MSbardiv\right) \,,
\end{align}
\end{subequations}
where the \MSbar counter terms $A_\text{ct}$ and $B_\text{ct}$ are derived by taking the $\eps$ expansion of $A\big(\s p^2\s\big)$ and $B\big(\s p^2\s\big)$. This yields
\begin{subequations}
  \begin{align}
    \lim_{\eps \to 0} (A - A_\text{ct}) &= \frac{4}{3}\,\frac{\aS}{\pi}\, \pqty{ \frac{3}{2} - \frac{m_Q^2}{p^2} \ln\frac{m_Q^2}{\mu^2} + \frac{m_Q^2-p^2}{p^2} \ln\frac{m_Q^2 - p^2}{\mu^2} }\,, \\[3pt]
    \lim_{\eps \to 0} (B - B_\text{ct}) &= -\frac{\aS}{3\s\pi}\, \pqty{ 1 - \frac{m_Q^4}{p^4} \ln\frac{m_Q^2}{\mu^2} + \frac{m_Q^4-p^4}{p^4} \ln\frac{m_Q^2 - p^2}{\mu^2} + \frac{m_Q^2}{p^2} }\,.
\end{align}
\end{subequations}
These results are finite but non-analytic at $p^2=m_Q^2$. In particular, their derivatives with respect to $p^2$ are divergent when evaluated at $p^2=m_Q^2$. These are a manifestation of IR divergences that appear when taking on-shell kinematics.

One way to side step this issue, thereby allowing us to extract the residue, is to keep $\eps \neq 0$ until after taking the derivative. It is then straightforward to derive
\begin{align}
  \RQCD^{(1)}\aS &= \eval{\dv{\Sigma_\text{QCD}(\slashed{p})}{\slashed{p}}}_{\slashed{p}=m_Q}\notag\\[3pt]
           &= 2\,m_Q^2 \eval{\bqty{ \dv{(A - A_\text{ct})}{p^2} + \dv{(B - B_\text{ct})}{p^2} }}_{p^2 = m_Q^2}
             + \eval{(B - B_\text{ct})}_{p^2 = m_Q^2} \notag\\
    &= -\frac{\aS}{3\s\pi} \bqty{ {2\,\pqty{\MSbardiv} + 3\,\ln\frac{\mu^2}{m_Q^2} + 4} }\,,
\label{eqn:RQCDresult}
\end{align}
where $\epsilon$ is specifically regulating the IR divergence. Next, we will derive the residue in HQET, where we will see that the same IR divergent terms appear. Therefore, the object of interest $\Delta R^{(1)}$ is IR finite. This is to be expected, since one of the standard tests that one has correctly implemented the matching procedure, namely that one is working with the correct low energy description, is to check that the IR of the full theory and EFT have the same divergence structure. We will provide a procedure for directly extracting $\Delta R^{(1)}$ that avoids this IR subtlety utilizing the method of regions, see \cref{subsec:ResidueRegions} below.

\subsection{Residue in HQET}
\label{sec:ResInHQET}
Next, we perform the two-point function calculation in HQET. Diagrammatically, the structure is identical to the QCD calculation where the relativistic quark propagator is replaced by the HQET propagator, yielding
\begin{align}
  -i\s\Sigma_\text{HQET}(v \cdot k) &= -\frac{4}{3}\, g_s^2\, \mu^{2\eps}
    \int \ddp{q} \frac{1}{q^2\s \big(v \cdot (q + k)\big)} - \text{c.t.} \notag\\
   &= - i\s\Big[ C(v \cdot k) - C_\text{ct}(v \cdot k)\Big]\,,
 \label{eqn:SigmaHQETIntegral}
\end{align}
which evaluates to\footnote{See Eqs.~(3.67) and~(3.69) in \Ref{Manohar:2000dt}.}
\begin{subequations}
\begin{align}
  C(v \cdot k) &= -\frac{2}{3}\frac{\aS}{\pi}\s \pqty{4\s\pi\,\mu^2}^\eps\, \Gamma(\eps)(-v \cdot k)^{1 - 2\eps}\,\frac{\Gamma(1 - \eps)\, \Gamma\Big(\frac{1}{2} + \eps\Big)}{(1 - 2\s\eps)\,\Gamma\Big(\frac{1}{2}\Big)}\,, \\
  C_\text{ct}(v \cdot k) &= \frac{2}{3}\frac{\aS}{\pi}\s v \cdot k\,\pqty{\MSbardiv}\,,\label{eq:Cct}
\end{align}
\end{subequations}
where the counter term is again determined in the $\overline{\text{MS}}$ scheme.\footnote{Note that when computing the counter term in dim.\ reg., one must be careful to isolate the UV divergence.  This is done here by keeping $v\cdot k \neq 0$ as an IR regulator at intermediate steps.  This is why we must take a derivative of \cref{eq:Cct} before sending $v\cdot k \to 0$ to derive \cref{eq:dCct}.  If instead we took $v\cdot k \to 0$ first, we would effectively be using dim.\ reg.\ to regulate the IR divergence as well.} Noting that to zeroth order in $1/m_Q$ the on-shell condition for $h_v$ is $v\cdot k =0$, we evaluate
\begin{subequations}
\begin{align}
  \eval{ \dv{C(v \cdot k)}{(v\cdot k)} }_{v \cdot k = 0} &=
    \eval{ \frac{2}{3}\frac{\aS}{\pi}\s \pqty{4\s\pi\,\mu^2}^\eps\, \Gamma(\eps)\, (-v \cdot k)^{-2\eps}\, \frac{\Gamma(1-\eps)\Gamma\Big( \frac{1}{2} + \eps\Big)}{\Gamma(\frac{1}{2})} }_{v \cdot k = 0} = \,0\,, \\[5pt]
  \eval{ \dv{C_\text{ct}(v \cdot k)}{(v\cdot k)} }_{v \cdot k = 0} &= \frac{2}{3}\frac{\aS}{\pi}\s\pqty{\MSbardiv}\,,
  \label{eq:dCct}
\end{align}
\end{subequations}
which yields
\begin{equation}
  \RHQET^{(1)} \aS =\frac{\dd C(v \cdot k) - \dd C_\text{ct}(v \cdot k)}{\dd (v \cdot k)}\bigg|_{v \cdot k = 0} =  -\frac{2}{3}\frac{\aS}{\pi}\s \pqty{\MSbardiv}\,.
\label{eqn:RHQETresult}
\end{equation}
Similar to above, the $\lim_{\eps \to 0} \Sigma_\text{HQET}$ is not analytic at $v \cdot k=0$ due to an IR divergence, and so we had to defer taking the $\eps \to 0$ limit until after taking the derivative with respect to $v\cdot k$.

\subsection{Residue Difference from the Method of Regions}
\label{subsec:ResidueRegions}
The IR divergences in \cref{eqn:RQCDresult,eqn:RHQETresult} are the same, as they must be if the EFT correctly captures the dynamics of the full theory below a certain scale. This implies that the residue difference is IR finite:
\begin{equation}
  \Delta R^{(1)} \aS = \pqty{\RQCD^{(1)} - \RHQET^{(1)}}\, \aS =  -\frac{\aS}{3\s\pi} \pqty{ 3\s\ln\frac{\mu^2}{m_Q^2} + 4}\,.
\label{eqn:DeltaRresult}
\end{equation}
Operationally, we were forced to maintain $\epsilon \neq 0$ until we evaluated this difference. Therefore, it would be convenient to have an approach that would allow us to compute $\Delta R^{(1)}$ directly. This can be accomplished by exploiting a technique known as the method of regions~\cite{Beneke:1997zp,Smirnov:2002pj}.

To begin, we rewrite some expressions in order to make the comparison between $\RQCD^{(1)}$ and $\RHQET^{(1)}$ more obvious. Within QCD, we set $p^\mu = m_Q\, v^\mu + k^\mu$ and define
\begin{equation}
  \Xi_\text{QCD}(k) \equiv \Sigma_\text{QCD}(\slashed{p})\, \frac{1 + \slashed{v}}{2}\,.
\end{equation}
so that
\begin{align}
  \eval{ v^\mu\, \dv{\Xi_\text{QCD}(k)}{k^\mu} }_{v \cdot k = 0} &= \slashed{v}\, \eval{ \dv{\Sigma_\text{QCD}(\slashed{p})}{\slashed{p}} }_{\slashed{p} = m_Q} \frac{1 + \slashed{v}}{2} \,+ \,\ord\pqty{\frac{1}{m_Q^2}} \notag\\
    &= \RQCD^{(1)} \aS\, \frac{1 + \slashed{v}}{2} + \ord\pqty{\frac{1}{m_Q^2}}\,.
\end{align}
Note that we have changed the evaluation condition from $v \cdot k = 0$ to $\slashed{p} = m_Q$ in the first line, which is valid up to $\ord(1/m_Q)$. Similarly, we define the HQET quantity
\begin{equation}
  \Xi _\text{HQET}(k) \equiv \Sigma_\text{HQET}(v \cdot k)\, \frac{1 + \slashed{v}}{2}\,,
\end{equation}
which is related to $\RHQET^{(1)}$ as
\begin{equation}
  \eval{ v^\mu\, \dv{\Xi_\text{HQET}(k)}{k^\mu} }_{v \cdot k = 0}
    = \eval{ \dv{\Xi_\text{HQET}(v \cdot k)}{(v \cdot k)} }_{v \cdot k = 0} \frac{1 + \slashed{v}}{2}
    = \RHQET^{(1)}\, \aS \frac{1 + \slashed{v}}{2}\,.
\end{equation}
This allows us to simply express the difference as
\begin{equation}
  \Delta R^{(1)} \alpha_s \frac{1 + \slashed{v}}{2}
    = \eval{ v^\mu\, \dv{\Xi_\text{QCD}(k) - \dd\Xi_\text{HQET}(k)}{k^\mu} }_{v \cdot k = 0} \,+\, \ord\pqty{\frac{1}{m_Q^2}}\,.
    \label{eqn:DelRXiSubtract}
\end{equation}

In order to evaluate this difference, we take the integral expressions for $\Sigma_\text{QCD}$ and $\Sigma_\text{HQET}$ given in \cref{eqn:SigmaQCDIntegral,eqn:SigmaHQETIntegral} to write
\begin{subequations}
\begin{align}
  -i\s\Xi_\text{QCD}(k) &= -\frac{4}{3}\, g_s^2\, \mu^{2\eps} \int \ddp{q}
    \frac{(2 - d)(\slashed{q} + \slashed{p}) + d\,m_Q}{q^2\,\bqty{ (q + p)^2 - m_Q^2}} \frac{1 + \slashed{v}}{2}
    - \text{c.t.} \,, \label{eq:XiQCD} \\
  -i\s\Xi_\text{HQET}(k) &=  -\frac{4}{3}\, g_s^2\, \mu^{2\eps} \int \ddp{q}
    \frac{1}{q^2\,\bqty{v \cdot (q + k)}} \frac{1 + \slashed{v}}{2} - \text{c.t.} \,. \label{eq:XiHQET}
\end{align}
\end{subequations}
Note that the integrands of $\Xi_\text{QCD}$ and $\Xi_\text{HQET}$ are equal in the heavy quark limit:
\begin{equation}
  \lim_{m_Q \to \infty}
    \frac{(2 - d)(\slashed{q} + \slashed{p}) + d\,m_Q}{q^2\, \bqty{(q + p)^2 - m_Q^2}} \frac{1 + \slashed{v}}{2}
    = \frac{(2 - d) \slashed{v} + d}{2\,q^2\, \bqty{v \cdot (q + k)}} \frac{1 + \slashed{v}}{2}
    = \frac{1}{q^2\,\bqty{v \cdot (q + k)}} \frac{1 + \slashed{v}}{2}\,.
\label{eqn:ExpansionNaive}
\end{equation}
Naively, one might be tempted to conclude that $\Xi_\text{QCD} = \Xi_\text{HQET}$ at leading order in $1/m_Q$, which would imply that $\Delta R^{(1)}$ vanishes. This would be in conflict with the result derived in \cref{eqn:DeltaRresult}. This is due to an order of limits issue:  since the integral is taken over all momenta, one cannot expanding the integrand for large $m_Q$ before integrating.

The method of regions~\cite{Beneke:1997zp,Smirnov:2002pj} is a technique for consistently expanding the integrands such that non-analytic dependence on small parameters is correctly reproduced after integration. The key is to isolate regions of the integration domain that are dominated by single scale contributions. For example, the QCD integral contains two physical scales that are separated by a large hierarchy $|k| \ll m_Q$. To expand the integrand, we introduce an intermediate cutoff scale $\Lambda$ such that $|k| \ll \Lambda \ll m_Q$, which can be used to split the integral over $q^\mu$ into two pieces. The first receives support from the soft region $|q| < \Lambda$, while the second is non-zero due to the hard region $|q| > \Lambda$:
\begin{equation}
  \Xi_\text{QCD}(k) = \Xi_\text{QCD,soft}(k,\Lambda) + \Xi_\text{QCD,hard}(k,\Lambda)\,.
\end{equation}

The domain of integration is now bounded, so that one can expand the integrand according to the assumed scaling in each region, while maintaining $|k| \ll m_Q$ fixed. The soft region can be isolated by assuming $|q| < \Lambda \ll m_Q$ holds, while the hard region where $|k| \ll \Lambda < |q|$ yields a different expansion. Once the integrand has been expanded, the domain of integration can be restored to infinity when using dimensional regularization so that the explicit cutoff $\Lambda$ no longer appears --- the contribution from the extended integration limits is scaleless and therefore vanishes. This is reflected by our notation since we drop the $\Lambda$ dependence for the expressions where the domain of the integral is taken to infinity.

Applying this procedure to the two-point function integrals, we recognize that the soft expansion is identical to the naive approach in \cref{eqn:ExpansionNaive}. Therefore, we conclude that
\begin{equation}
  \Xi_\text{QCD,soft}(k) = \Xi_\text{HQET}(k) \,.
\label{eqn:Cancellation}
\end{equation}
This tells us that the residue difference is fully determined by the hard region from QCD:
\begin{equation}
  \Delta R^{(1)} \aS\s \frac{1 + \slashed{v}}{2}
    = \eval{ v^\mu\, \dv{\Xi_\text{QCD}(k) - \dd \Xi_\text{HQET}(k)}{k^\mu} }_{v \cdot k = 0}
    = \eval{ v^\mu\, \dv{\Xi_\text{QCD,hard}(k)}{k^\mu} }_{v \cdot k = 0}\,.
\label{eqn:DeltaRnonzero}
\end{equation}
Note that \cref{eqn:Cancellation} holds to all order in $1/m_Q$. In \cref{eqn:ExpansionNaive}, it was demonstrated only at the leading order, \textit{i.e.} $m_Q\to\infty$. To explicitly verify this at higher orders in $1/m_Q$, one needs to include additional diagrams that contribute to \cref{eq:XiHQET} due to higher order operators in HQET. Then the result in \cref{eqn:DeltaRnonzero} still holds, with an appropriate modification of the evaluation condition $v\cdot k=0$.

The practical implication of \cref{eqn:DeltaRnonzero} is that we no longer have to track the individual IR divergences. Instead, the residue difference $\Delta R^{(1)}$ is determined by a single loop integral $\Xi_\text{QCD,hard}$, and the only divergences that appear are from the UV, which can be subtracted using counter terms.

For completeness, we evaluate this hard region integral. Starting with \cref{eq:XiQCD}, we expand the integrand assuming $|k| \ll |q|$ and $|k| \ll m_Q$:
\begin{align}
  -i\s\Xi _\text{QCD,hard}(k)
    &= -\frac{4}{3}\, g_s^2\, \mu^{2\eps} \int \ddp{q} \frac{1}{q^2}\,
      \Bigg\{ \frac{2\,m_Q + (2 - d)(\slashed{q} + \slashed{k})}{(q + m_Q\, v)^2 - m_Q^2} \notag\\
     &\hspace{80pt} - \frac{\bqty{2\,m_Q + (2 - d)\slashed{q}} 2\,(q + m_Q v) \cdot k}{\bqty{(q + m_Q\, v)^2 - m_Q^2}^2} \Bigg\}
      \frac{1 + \slashed{v}}{2} - \text{c.t.} \,,
\end{align}
where we have truncated the expansion up to the linear order in $k$, since this is all that is required to isolate the residue. Evaluating the integral yields
\begin{align}
  \Delta R^{(1)} \aS\s \frac{1 + \slashed{v}}{2}
    &= \eval{ v^\mu\, \dv{\Xi_\text{QCD,hard}(k)}{k^\mu} }_{v \cdot k = 0} \notag\\[5pt]
    &= -\frac{\aS}{3\s\pi^2} \pqty{\frac{4\s\pi\,\mu^2}{m_Q^2}}^\eps \Gamma(\eps)\s
         \bqty{1 - \frac{4\s\eps}{(1 - 2\s\eps)(-2\s\eps)}} \frac{1 + \slashed{v}}{2} - \text{c.t.} \notag\\[5pt]
    &= -\frac{\aS}{3\s\pi^2} \pqty{3\s\ln\frac{\mu^2}{m_Q^2} + 4}\, \frac{1 + \slashed{v}}{2} \,,
\end{align}
which reproduces the result in \cref{eqn:DeltaRresult}.

\section{Matching and Running Using Functional Methods}
\label{sec:FunctionalMethods}
This section reviews how to utilize the functional approach for calculating matching coefficients and RGEs. To set the stage, we first briefly review the one-particle irreducible (1PI) effective action, which is a key object in functional methods that one could compute directly by evaluating the path integral. The general matching condition can be compactly expressed by equating the 1PI effective actions of the UV theory and the EFT, whose solution gives us a direct link between the Lagrangians of the two theories, \eg see~\Ref{Henning:2016lyp}. Finally, we also provide a brief review on how to extract RGEs from the 1PI effective action; more details for RGE calculations are provided in \cref{appsec:ExamplesRGE}.

\subsection{1PI Effective Action from a Functional Determinant}
\label{subsec:1PIGamma}
In modern functional methods, the central object of study is the so-called 1PI effective action $\Gamma[\phi]$, where $\phi$ collectively denotes all the fields in the theory. It is a generating functional for the 1PI correlation functions:
\begin{equation}
  \Big\langle \phi(x_1) \dotsm \phi(x_n)\Big\rangle_\text{1PI}
    = \, i\s \frac{\var[n]{\Gamma[\phi]}}{\var{\phi(x_1)} \dotsi \var{\phi(x_n)}}
  \qquad \,\, \text{for}\qquad n > 2 \,.
\end{equation}
One can in principle extract any perturbative quantum field theoretic prediction from $\Gamma[\phi]$. Concretely, the 1PI effective action up to one-loop order can be organized as a loop expansion:
\begin{equation}
  \Gamma[\phi] \supset \Gamma^{(0)}[\phi] + \Gamma^{(1)}[\phi] \,,
\end{equation}
where each piece can be extracted from the Lagrangian as
\begin{subequations}
\begin{align}
  \Gamma^{(0)}[\phi] &= S[\phi] \\
  \Gamma^{(1)}[\phi] &=  \frac{i}{2} \ln\Sdet\pqty{- \fdv[2]{S[\phi]}{\phi}} \,,
\end{align}
\end{subequations}
with $S[\phi] = \int \dd[4]{x} \Lag(\phi)$, and ``$\Sdet$'' is the so-called super-determinant, which tracks the minus sign difference between fermionic and bosonic loops. The normalization factor $\frac{i}{2}$ assumes that we are tracing over ``real'' degrees of freedom: a complex scalar field should be separated into its real and imaginary parts, and Dirac fermions should be decomposed as discussed in Sec.~4 of \Ref{Henning:2016lyp}.

\subsection{Matching from a Functional Determinant}
\label{subsec:ReviewFunctionalMatching}
In general, one performs a matching calculation by equating the EFT $\mathcal{L}_\text{EFT}(\phi)$ with a UV theory $\mathcal{L}_\text{UV}(\phi, \Phi)$ at a matching scale, where $\phi$ ($\Phi$) denotes the light (heavy) particles. The general matching condition is that all the one-light-particle irreducible (1LPI) diagrams agree at the matching scale \cite{Georgi:1991ch,Georgi:1992xg}. The natural definition of this statement when using functional methods is to enforce that the so-called 1LPI effective action $\Gamma_\text{L}[\phi]$ --- the generating functional for all the 1LPI correlation functions --- of each theory coincides at the matching scale:
\begin{equation}
\Gamma_\text{L,EFT}[\phi] = \Gamma_\text{L,UV}[\phi] \,.
\label{eqn:MatchingCondition}
\end{equation}
One can relate the 1LPI to the 1PI effective action for the UV theory and the EFT:
\begin{subequations}
\begin{align}
\Gamma_\text{L,EFT}[\phi] &= \Gamma_\text{EFT}[\phi] \,, \\
\Gamma_\text{L,UV}[\phi] &= \Gamma_\text{UV}[\phi,\Phi]\big |_{\Phi = \Phi_c[\phi]} \,.
\end{align}
\end{subequations}
The relation for the EFT is trivial. By contrast, to derive $\Gamma_\text{L, UV}$, one must integrate out the heavy field by plugging in the solution $\Phi=\Phi_c[\phi]$ to its equation of motion.

Solving \cref{eqn:MatchingCondition} in general is nontrivial. However, \Ref{Henning:2016lyp} was able to make significant progress towards this goal by deriving the following general expression for the EFT Lagrangian up to one-loop order:\footnote{Note that as before, we use superscript numbers in parenthesis to denote loop order.  We need not specify the loop order for the full theory action, since we treat the counterterm contributions implicitly. However, matching results in nontrivial loop orders in the EFT Lagrangian.}
\begin{subequations}\label{eqn:SEFT}
\begin{align}
  S^{(0)}_\text{EFT} &= \int \dd[4]{x} \sum_i C_i^{(0)}\, \oper_i(\phi)
    = S_\text{UV}[\phi,\Phi]\big |_{\Phi = \Phi_c[\phi]} \,,
  \label{eqn:SEFT0} \\
  S^{(1)}_\text{EFT} &= \int \dd[4]{x} \sum_i \pqty{ C_{i,\text{heavy}}^{(1)} + C_{i,\text{mixed}}^{(1)} }\, \oper_i(\phi) \,,
  \label{eqn:SEFT1}
\end{align}
\end{subequations}
where
\begin{subequations}\label{eqn:cLoop}
\begin{align}
  \int \dd[4]{x} \sum_i C_{i,\text{heavy}}^{(1)}\, \oper_i(\phi)
    &= \frac{i}{2} \ln\Sdet \pqty{ \eval{ -\fdv[2]{S_\text{UV}[\phi,\Phi]}{\Phi} }_{\Phi = \Phi_c[\phi]} \,} \,, \label{eqn:cheavy} \\[5pt]
  \int \dd[4]{x} \sum_i C_{i,\text{mixed}}^{(1)}\, \oper_i(\phi)
      &= \frac{i}{2} \ln\Sdet \pqty{ -\fdv[2]{S_\text{EFT}^\text{non-local}[\phi]}{\phi} } - \frac{i}{2} \ln\Sdet \pqty{ -\fdv[2]{S_\text{EFT}^{(0)}[\phi]}{\phi} } \notag\\
      &= \eval{ \frac{i}{2} \ln\Sdet \pqty{ -\fdv[2]{S_\text{EFT}^\text{non-local}[\phi]}{\phi} } }_\text{hard} \,.
  \label{eqn:cmixed}
\end{align}
\end{subequations}
A few clarifications are in order:
\begin{itemize}
  \item In this approach, we do need to know the EFT operators $\oper_i(\phi)$ in advance. We simply obtain them (with the appropriate one-loop coefficient) by evaluating the right-hand sides of \cref{eqn:SEFT,eqn:cLoop}.
  \item The one-loop contributions to the Wilson coefficients derive from two classes of loops in the UV theory. The ``heavy'' Wilson coefficients $C_{i,\text{heavy}}^{(1)}$ collect contributions from loops where only $\Phi$ appear, while the ``mixed'' Wilson coefficients $C_{i,\text{mixed}}^{(1)}$ are generated by loops with both $\Phi$ and $\phi$.
  \item The non-local Lagrangian appearing in \cref{eqn:cmixed} is given by
        \begin{equation}
          S_\text{EFT}^\text{non-local}[\phi] = S_\text{UV}[\phi,\Phi]\big |_{\Phi = \Phi_c[\phi]}\,,
          \label{eq:SEFTnonlocal}
        \end{equation}
        where the heavy field is integrated out using the tree-level solution to its equations of motion $\Phi_c[\phi]$. No expansion in the heavy masses is performed at this stage, and hence the resulting Lagrangian contains non-local terms, \eg see~\cref{eqn:LHQETnonlocal}. This is in contrast to the tree-level EFT action $S_\text{EFT}^{(0)}[\phi]$, which is derived by expanding the non-local Lagrangian in the heavy mass limit; therefore, it only contains local terms, \eg see~\cref{eq:HQETExpansion}. As we have discussed extensively in \cref{subsec:ResidueRegions}, expanding the action in the heavy particle limit yields a critical difference between the two descriptions, where the non-trivial effects result from being careful about the order of limits --- performing the heavy mass expansion does not commute with taking the functional determinant (which is essentially equivalent to performing the loop integration). Although, the two terms in the first line of \cref{eqn:cmixed} are not equal, they are intimately related in such a way that allows for the following simplification: Using method of regions,\footnote{Simplifying the functional matching calculation with method of regions was also discussed in \Ref{Fuentes-Martin:2016uol,Zhang:2016pja}, where a different (but equivalent) approach to the row reduction procedure was used to diagonalize the functional determinant matrix in the space of $(\Phi, \phi)$.} we can identify the second term as equivalent to the soft region of the first term. Their difference leaves behind the hard region alone, as shown in the second line of \cref{eqn:cmixed}.
\end{itemize}

\subsection{RGEs from the 1PI Effective Action}\label{subsec:ReviewFunctionalRunning}
In this section, we briefly review the general procedure for using the 1PI effective action to compute the RGEs~\cite{Henning:2016lyp}. Given a Lagrangian
\begin{equation}
  \Lag(\phi) \supset \oper_K(\phi) + \lambda\, \oper_\lambda(\phi) + \dotsb \,,
\end{equation}
we are interested in deriving the RGEs for a coupling $\lambda$, where $\oper_\lambda$ is the corresponding operator, $\oper_K$ denotes the kinetic terms, and the ``$\dotsb$'' allow for the possibility of additional interactions. The first step is to compute the 1PI effective action, which takes the form
\begin{equation}
  \Gamma[\phi] \supset
    \int \dd[4]{x} \Big[(1 + a_K)\, \oper_K(\phi) + (\lambda + a_\lambda)\, \oper_\lambda(\phi)\Big] \,,
\end{equation}
where $a_K$ and $a_\lambda$ encode the one-loop corrections. Next, we renormalize the kinetic terms to their canonical value by rescaling the fields $\phi$:
\begin{equation}
  \Gamma[\phi] \,\,\longrightarrow\,\,
    \int \dd[4]{x} \Big[\oper_k(\phi) + (\lambda  + b_\lambda) \oper_\lambda(\phi)\Big] \,.
\end{equation}
where $b_\lambda$ is derived by Taylor expanding the version of $\Gamma[\phi]$ with canonical kinetic terms as a series in $\lambda$. Finally, the RGE for $\lambda$ is obtained:
\begin{equation}
  \mu\dv{\mu} \big(\lambda  + b_\lambda\big) = 0 \,.
\label{eqn:RGE1PIgeneral}
\end{equation}
For reference, a number of standard RGEs are derived using this method in \cref{appsec:ExamplesRGE}; additionally, functional methods were used to compute the bosonic dimension-6  SMEFT RGEs in \Ref{Buchalla:2019wsc}. In \cref{sec:Examples}, we will apply this formalism in HQET to show that the kinetic term does not run, and will derive the RGEs for the Wilson coefficient of the HQET magnetic dipole moment operator.

\section{HQET Matching from a Functional Determinant}\label{sec:MatchingHQET}
In this section, we apply the general matching result \cref{eqn:SEFT,eqn:cLoop} to HQET. As we will see, salient features of HQET will result in further simplifications for matching calculations, which will be summarized by our master formula \cref{eqn:SHQET1loop} for HQET one-loop matching coefficients. In this case, we identify $\Phi = \big(\bar{H}_v, H_v\big)$ as the short distance modes of the heavy quark, and $\phi$ collectively as all the propagating long distance degrees of freedom modeled by HQET. There are two special features of matching QCD onto HQET that simplify the resulting master formula.

First, taking second functional derivatives of \cref{eqn:LQCD} with respect to $H_v$ and $\bar{H}_v$, we see that the matching coefficient $C_{i,\text{heavy}}^{(1)}$ vanishes at one-loop:
\begin{align}
  \int \dd[4]{x} \sum_i C_{i,\text{heavy}}^{(1)}\, \oper_i(\phi)
    &= \frac{i}{2} \ln\Sdet \pqty{ \eval{ -\fdv[2]{S_\text{QCD}[\phi,\Phi]}{\Phi} }_{\Phi = \Phi_c[\phi]} \,} \notag\\
    &\propto -i \ln \Sdet \big( i\s v\cdot D + 2\, m_Q\big) = 0 \,.
\end{align}
This functional determinant is zero due to the same contour arguments presented in \cref{sec:Decoupling} (see also~\cite{Mannel:1991mc}), where we discussed decoupling the heavy quark.\footnote{There is an even simpler (albeit gauge dependent) argument also given in \Ref{Mannel:1991mc}: if one takes the $v\cdot A = 0$ gauge, the determinant no longer depends on any field, and its evaluation simply yields a constant that is absorbed into the path integral measure.} This indicates that all the one-loop matching contributions for HQET are mixed loop contributions of the type given in \cref{eqn:cmixed}.
Therefore, the non-local HQET Lagrangian
\begin{equation}
  \SHQETnl[\phi] = S_\text{QCD}[\phi,\Phi]\big |_{\Phi = \Phi_c[\phi]}\,,
\label{eq:SHQETnonlocal}
\end{equation}
which is given explicitly in \cref{eqn:LHQETnonlocal}, is equivalent to QCD with respect to the dynamics of the light fields $\phi$.

Another useful feature of HQET is that loop corrections are scaleless~\cite{Finkemeier:1997re}, and therefore they vanish when using dim.\ reg.. This means that the second term in the first line of \cref{eqn:cmixed} is also zero:
\begin{equation}
 \frac{i}{2} \ln\Sdet \pqty{ -\fdv[2]{S_\text{HQET}^{(0)}[\phi]}{\phi} } =0 \,.
 \label{eqn:HQETLoopVanish}
\end{equation}
We emphasize that this expression is valid when (i) we are computing on-shell $S$-matrix elements, and (ii) when  dim.\ reg.\ is used to regularize both UV and IR divergences (see \cref{sec:ResInHQET} for a calculation where we regulated the IR by keeping the long distance fluctuations off shell).

Taking both simplifications into account, we arrive at our master formula for computing the Wilson coefficients of HQET\footnote{There is one caveat to keep track of when applying this formula, which is that it is valid when one sets all the light masses that appear in a loop integral to zero.  If one is interested in computing power corrections proportional to a light mass, then \cref{eqn:cmixed} should be used: one should keep the light masses non-zero, and the hard region must be isolated before integrating.}
\begin{equation}
  \tcboxmath{
  S_\text{HQET}^{(1)} = \frac{i}{2} \ln\Sdet \pqty{ -\fdv[2]{\SHQETnl[\phi]}{\phi} }\,.}
\label{eqn:SHQET1loop}
\end{equation}
In particular, note that we have restored the vanishing region, implying that when using this result we do not need to perform any method of regions style expansion of the integrals that result from taking the functional trace, and can simple evaluate the integrals that result from this procedure directly. In \cref{sec:FunctionalMatching,sec:Examples}, we will demonstrate how to apply this formula by working out a number of explicit examples. We will additionally see that this formula makes symmetry properties such as gauge invariance and RPI more manifest.

\section{Residue Difference from a Functional Determinant}
\label{sec:FunctionalMatching}
This section provides a first non-trivial example of the procedure for matching between QCD and HQET at one-loop using functional methods: the calculation of the difference between the propagator residue for QCD and HQET. Along the way, we will highlight many of the simplifying benefits of performing these calculations directly from the path integral. Then before moving on to a number of additional (more complicated) examples in the next section, we briefly discuss how RPI manifests in the functional language.

Before getting into the details, we provide a simple road map highlighting the critical steps of the calculation:
\begin{enumerate}[label=(\roman*)]
\item Beginning with the UV Lagrangian, we will integrate out the short distance mode using the tree-level equations of motion.  This will result in a non-local Lagrangian (using background fields as necessary to implement gauge fixing) that encodes a complete description of the dynamics of the light modes.  See~\cref{eq:HQETNL}.
\item Given this non-local Lagrangian, we will then derive the matrix of the second-order functional derivatives (see~\cref{eqn:ResidueVariationMatrix}), whose determinant is the central object to evaluate according to our master formula \cref{eqn:SHQET1loop}.
\item Next, we will rewrite the determinant into an efficient form for explicit evaluation via row reduction. See~\cref{eq:residueDeterminant}.
\item Finally, we will evaluate this trace using the methodology of the covariant derivative expansion as reviewed in \cref{appsec:CDE}.  From this expansion, we will isolate the operators of interest.  This yields the second line of \cref{eq:EvalResInt}, which is an expression for one-loop Wilson coefficients multiplied by the appropriate operators.  The last step will be to explicitly evaluate the loop integral.  See the final line of \cref{eq:EvalResInt}.
\end{enumerate}
Note that along the way, we will drop any terms which do not contribute to the operator of interest for simplicity.

The rest of this section is devoted to the explicit calculation of the residue difference. Our goal is to apply \cref{eqn:SHQET1loop} in order to derive the one-loop correction to the two-point function of heavy quarks. This implies that we only need to take variations of the Lagrangian with respect to the gluon and the heavy quark field. The first step is to derive the non-local HQET action. Following \cref{sec:HQET}, where we reviewed the derivation of the HQET Lagrangian, we start with QCD (including the gluon kinetic term, gauge fixing, and ghost contributions) and integrate out the short distance quark modes:
\begin{equation}
  \LHQETnl = \bar{h}_v\, \pqty{ i\s v \cdot D + i\s\slashed{D}_\perp \frac{1}{i\s v \cdot D + 2\s m_Q} i\s\slashed{D}_\perp }\, h_v - \frac{1}{4}\s G_{\mu\nu}^a \, G^{\mu\nu,a}  + \Lag_\text{gf} + \Lag_\text{gh} \,,
\label{eq:HQETNL}
\end{equation}
where the explicit gauge-fixing term $\Lag_\text{gf}$ and ghost term $\Lag_\text{gh}$ are specified in \cref{eq:gaugefix,eq:ghost}.

Next, we take functional variations of $\SHQETnl$. We will treat gauge bosons using the background field prescription described in \cref{appsubsec:BackgroundField}:
\begin{align}
  G_\mu &= G_{B,\mu} + A_\mu \,,
\end{align}
where $G_{B,\mu}$ is a background field and $A_\mu$ encodes the fluctuations. Covariant derivatives are expanded similarly:
\begin{align}
  D_\mu &= D_{B,\mu } - i\s g_s\s A_\mu \,,
\end{align}
where $D_{B,\mu}$ is the covariant derivative that includes the background gauge field. After taking functional variations with respect to $A_\mu$, we replace
\begin{subequations}
\begin{align}
  G_{B,\mu} &\,\,\longrightarrow\,\, G_\mu \,, \\
  D_{B,\mu} &\,\,\longrightarrow\,\, D_\mu \,.
\end{align}
\end{subequations}

We only need to take the second variation of the Lagrangian in \cref{eq:HQETNL} with respect to $A_\mu^a$, $h_v$, and $\bar{h}_v$ to compute the residue difference:
\begin{equation}
  \delta^2 \SHQETnl \supset
    \pmqty{ \delta A_\mu^a \,\,& \delta h_v^T\,\, & \delta \bar{h}_v}
    \pmqty{ C^{\mu\nu,ab}             & \bar{\Gamma}^{\mu,a} & -\Big(\Gamma^{\mu,a}\Big)^T \\
            -\Big(\bar{\Gamma}^{\nu,b}\Big)^T & 0                    & -B^T \\[6pt]
            \Gamma^{\nu,b}            & B                    & 0}
    \pmqty{\delta A_\nu^b  \\[6pt] \delta h_v \\[7pt] \delta \bar{h}_v^T} \,,
\label{eqn:ResidueVariationMatrix}
\end{equation}
where terms not relevant for the residue operator in $U^{\mu\nu,ab}$, $\Gamma^{\mu,a}$, and $\bar{\Gamma}^{\mu,a}$ are discarded:
\begin{subequations}\label{eqn:ResidueDetparameters}
\begin{align}
  C^{\mu\nu,ab} &= \eta^{\mu\nu} \big(D^2\big)^{ab} - 2\s U^{\mu\nu,ab} \,, \\
  U^{\mu\nu,ab} &= g_s\, f^{abc}\, G_B^{\mu\nu,c} - g_s^2\,\bar{h}_v \gamma_\perp^\mu\, \frac{1}{i\s v \cdot D + 2\s m_Q} \gamma_\perp^\nu\, T^a\s T^b\, h_v \,, \\
  B &= i\s v \cdot D + i\s \slashed D_\perp \frac{1}{i\s v \cdot D + 2\s m_Q} i\s \slashed D_\perp \,, \\
  \Gamma^{\mu,a} &= g_s\s \bqty{ v^\mu
    + i\s \slashed{D}_\perp \frac{1}{i\s v \cdot D + 2\s m_Q} \gamma_\perp^\mu}\, T^a\, h_v \,, \\
  \bar{\Gamma}^{\mu,a} &= g_s\, \bar{h}_v\, T^a\, \bqty{ v^\mu + \gamma_\perp^\mu \frac{1}{i\s v \cdot D + 2\s m_Q} i\s \slashed D_\perp } \,.
\end{align}
\end{subequations}
Here $\gamma_\perp^\mu\equiv\gamma^\mu-v^\mu\slashed{v}$ is defined in parallel with $D_\perp^\mu$. The reality of the Lagrangian implies that $B^\dagger = \gamma^0\s B\s \gamma^0$ and $\bar\Gamma^{\mu,a} = (\Gamma^{\mu,a})^\dagger \gamma^0$ must hold, and it is straightforward to verify these with the explicit expressions in \cref{eqn:ResidueDetparameters}.

Next, we evaluate the functional determinant by row reducing \cref{eqn:ResidueVariationMatrix}:
\begin{align}
  \fdv[2]{\SHQETnl}{(A_\mu^a,\bar{h}_v,h_v)}
    &= \pmqty{ C^{\mu\nu,ab}             & \bar{\Gamma}^{\mu,a} & -\Big(\Gamma^{\mu,a}\Big)^T \\[3pt]
            -\Big(\bar{\Gamma}^{\nu,b}\Big)^T & 0                    & -B^T \\[6pt]
            \Gamma^{\nu,b}            & B                    & 0} \nonumber \\[7pt]
    &= \pmqty{ C^{\mu\nu,ab} - \bar{\Gamma}^{\mu,a} \s B^{-1}\s \Gamma^{\nu,b} - \bar{\Gamma}^{\nu,b}\s B^{-1}\s \Gamma^{\mu,a} & 0 & 0 \\[3pt]
               -\Big(\bar{\Gamma}^{\nu,b}\Big)^T & 0 & -B^T \\[6pt]
               \Gamma^{\nu,b} & B & 0}  , \label{eqn:ResidueRowReduction}
\end{align}
where in the last line we have used $\big (\Gamma^{\mu,a}\big)^T \big(B^{-1}\big)^T \big(\bar{\Gamma}^{\nu,b}\big)^T = -\bar{\Gamma}^{\nu,b}\s B^{-1}\s \Gamma^{\mu,a}$. Following the prescription in \cref{subsec:ReviewFunctionalMatching}, we extract the residue operator part of the one-loop HQET Lagrangian using
\begin{align}
    S_\text{HQET}^{(1)}
     &= \frac{i}{2} \ln\Sdet\bqty{ -\fdv[2]{\SHQETnl}{(A_\mu^a, \bar{h}_v, h_v)} }  \notag \\[5pt]
     &= \frac{i}{2} \ln\det\nolimits_G \Big[-C^{\mu\nu,ab} + \bar{\Gamma}^{\mu,a}\s B^{-1}\s \Gamma^{\nu,b} + \bar{\Gamma}^{\nu,b} B^{-1} \Gamma^{\mu,a}\Big]
        - \frac{i}{2}\ln\det\nolimits_h \pmqty{ 0 & B^T \\ -B & 0} \notag \\[5pt]
     &\supset \frac{i}{2} \ln\det\nolimits_G \Big[ -\eta^{\mu\nu} \big(D^2\big)^{ab} + 2\s U^{\mu\nu,ab} + \bar{\Gamma}^{\mu,a}\s B^{- 1}\s \Gamma^{\nu,b} + \bar{\Gamma}^{\nu,b} \s B^{-1}\s \Gamma^{\mu,a} \Big] \notag \\[5pt]
     &\supset -i\Tr \bqty{ \pqty{\frac{1}{D^2}}^{ba} \eta_{\mu\nu}\s \Big(U^{\mu\nu,ab} + \bar{\Gamma}^{\mu,a}\s B^{-1}\s \Gamma^{\nu,b}\Big) }\,,
\label{eq:residueDeterminant}
\end{align}
where the subscripts denote the field that is being traced over, \ie, the state that is running in the loop, and starting with the third line, we have truncated the series and dropped terms that do not contribute to the residue operator. Note that in the last line, the trace over the gluon space is taken (\textit{i.e.} $\mu\nu$ and $ab$ indices are contracted). Next, we evaluate this expression using the explicit objects given in \cref{eqn:ResidueDetparameters}, and simplify the two traces as follows:
\begin{align}
  -i\Tr \bqty{ \pqty{\frac{1}{D^2}}^{ba} \eta_{\mu\nu}\s U^{\mu\nu,ab} }
    &= -i\s g_s^2\s C_F\, \Tr \bqty{ (d-1) \frac{1}{(i\s D)^2}\, \bar{h}_v\, \frac{1}{i\s v \cdot D + 2\s m_Q} h_v } \,,
\end{align}
where we used $T_h^a\s T_h^a = C_F = 4/3$, and
\begin{align}
     \hspace{-20pt} -&i\Tr \bqty{ \pqty{\frac{1}{D^2}}^{ba} \eta_{\mu\nu}\s \bar{\Gamma}^{\mu,a}\s B^{-1}\s \Gamma^{\nu,b} }  \\[5pt]
     &\quad\supset i\s g_s^2\, C_F\,
                \Tr \Bigg\{ \frac{1}{(i\s D)^2}\, \bar{h}_v\, \frac{1}{(i\s D)^2 + 2\s m_Q\,( i\s v \cdot D)} (i\s v \cdot D + 2\s m_Q)\, h_v \notag\\[5pt]
     &\quad\qquad + (d - 1) \frac{1}{(i\s D)^2}\, \bar{h}_v\, \frac{1}{i\s v \cdot D + 2\s m_Q}
                  \frac{1}{(i\s D)^2 + 2\s m_Q\, (i\s v \cdot D)} \bqty{ (i\s D)^2 - (i\s v \cdot D)^2}\, h_v \Bigg\}\,.\notag
\end{align}
In performing these manipulations, we set $\comm{D_\mu}{D_\nu} = 0$ since this commutator returns the field strength $G_{\mu\nu}^a$, which does not contribute to the residue difference. Putting these two results together, we obtain
\begin{align}
  S_\text{HQET}^{(1)} &=
-i\s g_s^2\, C_F\, \Tr \Bqty{ \frac{1}{(i\s D)^2}\, \bar{h}_v \,\frac{1}{(i\s D)^2 + 2\s m_Q \, (i\s v \cdot D)} \Big[ (d-2) (i\s v \cdot D) - 2\s m_Q \Big]}\, h_v\,.
\label{eqn:ResidueTrace}
\end{align}

Finally, we evaluate this functional trace using the simplified CDE technique described in~\cref{appsubsubsec:naiveCDE} to derive\footnote{
Since we treat $D_\mu$ as a commuting object here, the simplified CDE approach is sufficient.  If one were interested in extracting Wilson coefficients for operators that involve the field strength,  the more sophisticated original CDE approach described in~\cref{appsubsubsec:elegantCDE} would be more convenient.}
\begin{align}
  S_\text{HQET}^{(1)}
    &\supset -i\s g_s^2\, \mu^{2\eps}\, C_F\,\int \ddx{x} \int \ddp{q}
               \frac{1}{(i\s D + q)^2} \,\bar{h}_v\, \frac{1}{(i\s D + q)^2 + 2\s m_Q\, v \cdot (i\s D + q)}\notag\\[-2pt]
               &\hspace{210pt}\times \Big[ (d-2)\, v \cdot (i\s D + q) - 2\s m_Q\Big]\, h_v - \text{c.t.} \notag \\
    &\supset -i\s g_s^2\, \mu^{2\eps}\, C_F\, \int \ddx{x} \bqty{\bar{h}_v\, i\s D_\mu\, h_v}\notag\\[-2pt]
           &\hspace{70pt} \times  \int \ddp{q} \frac{(d-2)\, q^2\, v^\mu - 2\s (d-2)\, q^\mu\, v \cdot q + 4\s m_Q\, q^\mu + 4\s  m_Q^2\, v^\mu}{q^2\, \pqty{q^2 + 2\s m_Q\, v \cdot q}^2}   - \text{c.t.} \notag \\
    &= \int \ddx{x} \bqty{\bar{h}_v \, i\s v \cdot D\, h_v}\, \frac{\aS}{3\s\pi} \pqty{\frac{4\s\pi\,\mu^2}{m_Q^2}}^\eps\,
         \Gamma(\eps) \bqty{ 1 - \frac{4\s \eps}{(1-\eps)(-2\s\eps)} }  - \text{c.t.}\notag \\
    &= \int \ddx{x} \bqty{\bar{h}_v \,i\s v \cdot D\, h_v}\, \frac{\aS}{3\s\pi}\,
         \pqty{ 3\,\ln\frac{\mu^2}{m_Q^2} + 4 }\, .
\label{eq:EvalResInt}
\end{align}
The first line follows from \cref{eqn:ResidueTrace} by taking the steps shown in \cref{eqn:T0naiveCDE}. In the second line, we have expanded in the covariant derivative $D_\mu$ --- this is the explicit step that uses the Covariant Derivative Expansion. Note that in taking this expansion we need to invoke $\big|D\big| \ll m_Q$ to justify only keeping the residue operator. Its Wilson coefficient is given by a loop integral with a nonzero hard region but a vanishing soft region. This verifies our general statement made in \cref{sec:MatchingHQET}, in particular \cref{eqn:HQETLoopVanish}. In the third line, we have evaluated the loop integral, with the \MSbar scheme counter terms. The result agrees with \cref{eqn:DeltaRdef}, providing our first demonstration for using functional methods to compute HQET loop effects. Before moving on to additional examples in \cref{sec:Examples}, we will briefly discuss RPI in the context of the residue difference calculation.

\subsection{When Does RPI Become Manifest?}
\label{sec:RPI}
In this section, we briefly explore the interplay between RPI and the calculation detailed in the previous section.\footnote{RPI relations between coefficients are typically obscured in the course of a conventional calculation, since different combinations of Feynman diagrams will contribute to the matching of operators at different orders in the mass expansion. In practice, RPI relations are derived independently and then externally imposed to minimize the number of matching calculations to be performed~\cite{Neubert:1993za}.} Our goal here is to understand at what point RPI holds when computing with the functional approach. For simplicity, we will only explore this question to leading order in $1/m_Q$ in the RPI transformations. The full RPI transformations are significantly more complicated, since the eigenstates $h_v$ and $H_v$ rotate into each other at subleading order, see~\cite{Chen:1993np, Kilian:1994mg, Sundrum:1997ut, Heinonen:2012km, Hill:2012rh} for a discussion.

The RPI transformation shifts\footnote{To reduce notational clutter, we have opted to use $k$ as opposed to $\delta k$ here to track the change induced by RPI when comparing with \cref{eq:RPIdef}.   Given the form of the relevant expressions, there is no ambiguity.}
\begin{align}
  v \RPI v^\prime &= v - \frac{k}{{{m_Q}}} \,, \notag\\[2pt]
  {h_v} \RPI {h_{v^\prime}} &= {e^{ - i\s k \cdot x}}\left( {1 - \frac{{\slashed k}}{{2\s m_Q}}} \right){h_v} + O\left( {\frac{1}{{{m_Q^2}}}} \right) \,.
\label{eq:RPIonhnew}
\end{align}
We will now show that this is a good symmetry of HQET by noting that the non-local HQET Lagrangian is built using the operator $B$ defined in~\cref{eqn:ResidueDetparameters}. In particular, this object transforms covariantly as
\begin{align}
  B &\equiv i\s v \cdot D + i\s \slashed D_\perp \frac{1}{i\s v \cdot D + 2\s m_Q} i\s\slashed D_\perp \notag\\
  \RPI B' &= B - \frac{{i\s k \cdot D}}{{{m_Q}}} + O\left( {\frac{1}{{{m_Q^2}}}} \right) = {e^{ - i\s k \cdot x}}B{e^{i\s k \cdot x}} + O\left( {\frac{1}{{{m_Q^2}}}} \right) \,,
\label{eq:BDefAgain}
\end{align}
and hence
\begin{align}
  \LHQETnl &\supset {{\bar h}_v}\,B\,{h_v}\notag\\
  \RPI {{\bar h}_{v'}}\,B'\,{h_{v'}} &= {{\bar h}_v}\left( {1 - \frac{{\slashed k}}{{2\s m_Q}}} \right)B\left( {1 - \frac{{\slashed k}}{{2\s m_Q}}} \right){h_v} = {{\bar h}_v}\,B\,{h_v} + O\left(\frac{1}{m_Q^2}\right) \,,
\label{eq:RPILNL}
\end{align}
where we used the fact that $\slashed{v}\s h_v = h_v$,  $\slashed{v}\s \slashed{k} + \slashed{k}\s \slashed{v} = 2\s v\cdot k$, and $v \cdot k = k^2/(2\s m_Q)$.

Alternatively, one can check the interplay between the two terms in the first line of~\cref{eq:BDefAgain}. We begin by analyzing the explicit transformation of the local term. Keeping all terms to $\ord(1/m_Q)$, we find
\begin{align}
  {{\bar h}_v}\left( {i\s v \cdot D} \right){h_v} &\RPI {{\bar h}_v}\left( {1 - \frac{{\slashed k}}{{2\s m_Q}}} \right){e^{i\s k \cdot x}}\left( {i\s v \cdot D - \frac{{i\s k \cdot D}}{{{m_Q}}}} \right){e^{ - i\s k \cdot x}}\left( {1 - \frac{{\slashed k}}{{2\s m_Q}}} \right){h_v} \nonumber \\
  &\hspace{15pt}= {{\bar h}_v}\left( {i\s v \cdot D} \right){h_v} - {{\bar h}_v}\left( {\frac{{i\s k \cdot D}}{{{m_Q}}} + \frac{{{k^2}}}{{2\s m_Q}}} \right){h_v} + O\left(\frac{1}{m_Q^2}\right)\,.
\end{align}
This change is compensated by a corresponding shift in the non-local piece
\begin{align}
  &{{\bar h}_v}\s i\s {{\slashed D}_ \bot }\frac{1}{{i\s v \cdot D + 2\s m_Q}}\s i\s {{\slashed D}_ \bot }{h_v} \notag\\
  \RPI\hspace{10pt}& {{\bar h}_v}\left( {1 - \frac{{\slashed k}}{{2\s m_Q}}} \right){e^{i\s k \cdot x}}\left( {i\s{{\slashed D}_ \bot }\frac{1}{{i\s v \cdot D + 2\s m_Q}}\s i\s{{\slashed D}_ \bot }} \right){e^{ - i\s k \cdot x}}\left( {1 - \frac{{\slashed k}}{{2\s m_Q}}} \right){h_v} \nonumber \\
  &= {{\bar h}_v}\s i\s {{\slashed D}_ \bot }\frac{1}{{i\s v \cdot D + 2\s m_Q}}\s i\s{{\slashed D}_ \bot }{h_v} + {{\bar h}_v}\left( {\frac{{i\s k \cdot D}}{{{m_Q}}} + \frac{{{k^2}}}{{2\s m_Q}}} \right){h_v} + O\left(\frac{1}{m_Q^2}\right)\,.
\end{align}
where we used the identities
\begin{align}
{i\s {\slashed k}_ \bot }{{\slashed D}_ \bot } + i\s{{\slashed D}_ \bot }{{\slashed k}_ \bot } &= 2\s i\,{k_ \bot } \cdot {D_ \bot } = 2\s {k_\mu }\, i\Big( {\s {D^\mu } - {v^\mu } \s v \cdot D} \Big) + O\left( {\frac{1}{{{m_Q}}}} \right) = 2\s i\, k \cdot D + O\left( {\frac{1}{{{m_Q}}}} \right) \,, \notag \\
{{\slashed k}_\bot }{{\slashed k}_\bot } &= {\left( {\slashed k - \slashed v\s v \cdot k} \right)^2} = {k^2} + O\left( {\frac{1}{{{m_Q}}}} \right)\,.
\end{align}
Then clearly \cref{eq:RPILNL} holds.

The loop-level Lagrangian is given by the super-determinant of the second variation of \cref{eq:HQETNL}. Therefore, it is also invariant under RPI. Let us explicitly check this to the order $\ord(1/m_Q)$ for the residue difference calculation presented in the previous section. The various components defined in \cref{eqn:ResidueDetparameters} shift under RPI as
\begin{subequations}
\begin{align}
{U^{\mu \nu ,ab}} &\RPI {U^{\mu \nu ,ab}} + O\left( {\frac{1}{{m_Q^2}}} \right) \,, \\
B &\RPI {e^{ - i\s k \cdot x}}\,B\,{e^{i\s k \cdot x}} + O\left( {\frac{1}{{m_Q^2}}} \right)  \,, \\
{\Gamma ^{\mu ,a}} &\RPI  {e^{ - i\s k \cdot x}}\left[ {{\Gamma ^{\mu ,a}} - {g_s}\left( {{\gamma ^\mu } + 2\s{v^\mu }} \right)\frac{{\slashed k}}{{2\s m_Q}}\,{T^a}\s{h_v}} \right] + O\left( {\frac{1}{{m_Q^2}}} \right)  \,, \\
{{\bar \Gamma }^{\mu ,a}} &\RPI \left[ {{{\bar \Gamma }^{\mu ,a}} - {g_s}\,{{\bar h}_v}\,{T^a}\frac{{\slashed k}}{{2\s m_Q}}\Big( {{\gamma ^\mu } + 2\s{v^\mu }} \Big)} \right]{e^{i\s k \cdot x}} + O\left( {\frac{1}{{m_Q^2}}} \right) \,.
 \end{align}
\end{subequations}
From here, we can check the transformations of each term that appears in the argument of the functional trace in \cref{eq:residueDeterminant}:
\begin{subequations}
\begin{align}
  \left(\frac{1}{D^2}\right)^{ab} &\RPI \left(\frac{1}{D^2}\right)^{ab} \,, \\
  {\eta_{\mu\nu}\s U^{\mu \nu ,ab}} &\RPI \eta_{\mu\nu}\s{U^{\mu \nu ,ab}} + O\left( {\frac{1}{{m_Q^2}}} \right) \,, \\
  {\eta_{\mu\nu}\s{\bar \Gamma }^{\mu ,a}}\,{B^{ - 1}}\,{\Gamma ^{\nu ,b}} &\RPI \eta_{\mu\nu} \left[ {{{\bar \Gamma }^{\mu ,a}} - {g_s}\,{{\bar h}_v}\,{T^a}\frac{{\slashed k}}{{2\s m_Q}}\Big( {{\gamma ^\mu } + 2\s{v^\mu }} \Big)} \right]{B^{ - 1}}\notag\\
    &\hspace{70pt}\times\left[ {{\Gamma ^{\nu ,b}} - {g_s}\,\Big( {{\gamma ^\nu } + 2\s{v^\nu }} \Big)\frac{{\slashed k}}{{2\s m_Q}}\,{T^b}\,{h_v}} \right] \nonumber \\
    &\hspace{45pt}=\eta_{\mu\nu}\s{{\bar \Gamma }^{\mu ,a}}\,{B^{ - 1}}\,{\Gamma ^{\nu ,b}} + O\left( {\frac{1}{{m_Q^2}}} \right) \,,
\end{align}
\end{subequations}
where in the second derivation we used
\begin{equation}
  \eta_{\mu\nu}\s{{\bar h}_v}\,{T^a}\,{v^\mu }\,{B^{ - 1}}\Big( {{\gamma ^\nu } + 2\s{v^\nu }} \Big)\,\frac{{\slashed k}}{{2\s{m_Q}}}\,{T^b}\,{h_v} = {{\bar h}_v}\,{T^a}\,{B^{ - 1}}\,\Big( {\slashed v + 2} \Big)\frac{{{k^2}}}{{4\s m_Q^2}}\,{T^b}\,{h_v} \,.
\end{equation}
This demonstrates an elegant feature of functional methods when applied to HQET. Specifically, in the last line of \cref{eq:residueDeterminant} the terms within the square brackets are manifestly invariant under RPI \emph{before} evaluating the trace. This is in contrast with the Feynman diagram approach, where one must sum the full set of Feynman diagrams before the RPI symmetry becomes apparent.

\section{More Matching and Running Calculations}
\label{sec:Examples}
We have provided an explicit demonstration of a functional calculation in HQET for the simplest non-trivial example above. However, we have not yet made contact with an actual observable quantity. The purpose of this section is to do so by exploring more examples of matching calculations and a derivation of the RGEs for some Wilson coefficients. This will provide overwhelming evidence that these techniques capture all of relevant physics. Using the formalism developed in \Refs{Falk:1990yz,Falk:1991nq}, once the one-loop matching of operators is known, one-loop relations between exclusive quantities, such as decay constants and form factors, are straightforward to extract. Furthermore, explaining these examples will provide us with the opportunity to highlight some additional subtle aspects of applying functional methods.

\subsection{Heavy-Light Current Matching}
This section is devoted to our next example, matching the heavy-light current. The heavy-light operator in QCD is defined as
\begin{equation}
  \oper^\mu = \bar{q}\, \gamma^\mu\, Q \,,
\end{equation}
where $q$ is a light quark and $Q$ is a heavy quark that will be treated as an HQET field. Since one should use the reference vector $v^\mu$ when constructing HQET operators, two operators can be written down at leading order in $\ord(1/m_Q)$ that manifest the same symmetry properties as the heavy-light QCD operator:
\begin{subequations}
\begin{align}
  \oper_1^\mu &= \bar{q}\, \gamma^\mu\, h_v \,,\\
  \oper_2^\mu &= \bar{q}\, v^\mu\, h_v \,.
\end{align}
\end{subequations}
The matching condition requires that matrix elements derived using QCD and HQET at a convenient matching scale $\mu$ are equal,\footnote{Note that due to confinement, the matching is being done with unphysical external states. This does not cause any problems since the matching condition between the two theories is universal.}
\begin{equation}
  \Big\langle q\big(0,s'\big)\,\Big|\, \oper^\mu \,\Big|\, Q\big(p,s\big)\Big\rangle_\text{QCD}
    = \Big\langle q\big(0,s'\big) \,\Big|\, C_1\, \oper_1^\mu + C_2\, \oper_2^\mu \,\Big|\, h_v(k,s)\Big\rangle_\text{HQET}\,\,.
\end{equation}
where the Wilson coefficients for the HQET operators $C_{1,2} = C_{1,2}(m_Q/\mu, \alpha_s(\mu))$ are functions of $\mu$ and $\alpha_s$.

At one-loop order, the two matrix elements can be expressed schematically:
\begin{subequations}
\label{eq:HLoperatorsExpanded}
\begin{align}
&\hspace{-7pt}\Big\langle q\big(0,s'\big)\,\Big|\, \oper^\mu \,\Big|\, Q\big(p,s\big)\Big\rangle_\text{QCD} \! = \sqrt{ \RQCD\, R_{q} }\,  \bar{u}\big(0,s'\big) \Big[ \Big(1 + V_{\text{HL},1}^{(1)}\,\aS\Big)\, \gamma^\mu + V_{\text{HL},2}^{(1)}\,\aS\, v^\mu\Big]\, u\big(p,s\big)\,, \\[3pt]
  &\hspace{-7pt}  \Big\langle q\big(0,s'\big) \,\Big|\, C_1\, \oper_1^\mu + C_2\, \oper_2^\mu \,\Big|\, h_v(k,s)\Big\rangle_\text{HQET} = \sqrt{ \RHQET\, R_{q} }\,
    \bar{u}\big(0,s'\big)\, \Big(1+V_\text{eff}^{(1)}\,\aS\Big) \notag\\[-3pt]
     &\hspace{273pt}\times \Big(C_1\,\gamma^\mu + C_2\, v^\mu\Big)\, u\big(k,s\big)\,,
\end{align}
\end{subequations}
where the one-loop vertex corrections are $V^{(1)}_{\text{HL},i}$, the one-loop residue corrections are $\RQCD^{(1)} \aS = R_Q-1$ and $\RHQET^{(1)} \aS = R_h-1$, and $R_{q}$ denotes the residue for the light quark propagator, which is the same in QCD and HQET. Equating the two expressions in \cref{eq:HLoperatorsExpanded} lead to the simplified form given in \cref{eqn:HLresult}. In particular, $R_q$ drops out and the matching coefficient only depends on the residue difference $\Delta R^{(1)}=\RQCD^{(1)} - \RHQET^{(1)}$, computed in the previous section.

In order to extract the heavy-light matching coefficient from the master matching formula given in~\cref{eqn:SHQET1loop}, we follow the same steps outlined at the beginning of \cref{sec:FunctionalMatching}. The set of fluctuating fields that contribute are $A_\mu^a, h_v, \bar{h}_v, q,$ and $\bar{q}$. We need the equation of motion for the $H_v$ field, which can be derived from the UV Lagrangian (where sources $J_\mu^\pm$ for the heavy-light current are now included)
\begin{align}
  \Lag_\text{QCD} \supset \,&\bar{Q} \,\Big(i\s \slashed D - m_Q\Big)\, Q + \bar{q}\, i\s \slashed{D} q + \Big( \bar{q}\, \gamma^\mu\, Q\, J^-_\mu + \text{h.c.}\Big) - \frac{1}{4}\, G_{\mu\nu}^a\s G^{\mu\nu,a}    + \Lag_\text{gf} + \Lag_\text{gh} \,.
\label{eq:LHeavyLight}
\end{align}
Note that the sources $J_\mu^\pm$ are not dynamical fields, but must be included to ensure that the desired operators appear when matching the full theory and EFT actions. We split the heavy quark field as in \cref{eq:QSplit} above, and derive the equation of motion for the short distance mode:
\begin{equation}
  H_v = \frac{1}{i\s v \cdot D + 2\s m_Q} \pqty{ i\s \slashed{D}_\perp h_v + J^+_\mu\, \gamma^\mu\, e^{i\s m_Q\, v \cdot x}\,q}\,.
            \label{eq:HvHeavyLight}
\end{equation}
Since $J_\mu^+$ is merely a source, we are free to redefine it to absorb the phase, $J^+_\mu \to J^+_\mu e^{-i\s m_Q\, v \cdot x}$. Plugging the equation of motion \cref{eq:HvHeavyLight} into the UV Lagrangian \cref{eq:LHeavyLight} yields the tree-level non-local HQET Lagrangian:
\begin{align}
    \LHQETnl &= \bar{h}_v \, \pqty{ i\s v \cdot D + i\s \slashed{D}_\perp \frac{1}{i\s v \cdot D + 2\s m_Q}i\s \slashed{D}_\perp }\, h_v
           -\frac{1}{4}\, G_{\mu\nu}^a\s G^{\mu\nu,a} + \Lag_\text{gf} + \Lag_\text{gh} \notag\\
      &\hspace{20pt} + \bar{q}\, \big(i\s \slashed{D}\big)\, q
              + \pqty{ \bar{q}\, \gamma^\mu\, J^-_\mu\, h_v + \bar{q}\, \gamma^\mu\, J^-_\mu\, \frac{1}{i\s v \cdot D + 2\s m_Q} i\s\slashed{D}_\perp\, h_v + \text{h.c.} } \,.
\end{align}
We then take the second variation of the non-local action with respect to the relevant fluctuating fields, again following the prescription described in \cref{appsubsec:BackgroundField} for the gluons:
\begin{align}
 \hspace{-5pt}\delta^2 \SHQETnl \supset
    \pmqty{ \var{A_\mu^a} & \var{h_v^T} & \var{\bar{h}_v} & \var{q^T} & \var{\bar{q}}\, }
    \pmqty{ C^{\mu\nu,ab} & \bar{\Gamma}_1^{\mu,a} & -\Big(\Gamma_1^{\mu,a}\Big)^T
            & \bar{\Gamma}_2^{\mu,a} & -\Big(\Gamma_2^{\mu,a}\Big)^T \\[5pt]
            -\Big(\bar{\Gamma}_1^{\nu,b}\Big)^T & 0 & -B_1^T & 0 & -S_2^T \\[8pt]
            \Gamma _1^{\nu,b} & B_1 & 0 & S_1 & 0 \\[5pt]
            -\Big(\bar{\Gamma}_2^{\nu,b}\Big)^T & 0 & -S_1^T & 0 & -B_2^T \\[8pt]
            \Gamma_2^{\nu,b} & S_2 & 0 & B_2 & 0 }
    \pmqty{\var{A_\nu^b} \\[11pt] \var{h_v} \\[11pt] \var{\bar{h}_v^T} \\[10pt] \var{q} \\[10pt] \var{\bar{q}^T} }\,,\notag\\
\label{eq:secondVarHeavyLight}
\end{align}
where terms not relevant for the heavy-light operator are discarded in $\Gamma_i^{\mu,a}$ and $\bar{\Gamma}_i^{\mu,a}$
\begingroup
\allowdisplaybreaks
\begin{subequations}\label{eqn:HLDetParameters}
\begin{align}
  C^{\mu\nu,ab} &= \eta^{\mu\nu}\s \Big(D^2\Big)^{ab} \,, \\
  B_1 &= i\s v \cdot D + \slashed{D}_\perp\, \frac{1}{i\s v \cdot D + 2\s m_Q}\, \slashed{D}_\perp \,, \\
  B_2 &= i\s \slashed D \, , \\[5pt]
  \Gamma_1^{\mu,a} &= g_s\, T^a\, \bqty{ v^\mu + \slashed{D}_\perp \frac{1}{i\s v \cdot D + 2\s m_Q}
                                       \gamma_\perp^\mu }\, h_v  \,, \\
  \bar{\Gamma}_1^{\mu,a} &= g_s\, \bar{h}_v\, T^a\, \bqty{ v^\mu + \gamma_\perp^\mu
                                                     \frac{1}{i\s v \cdot D + 2\s m_Q} i\s \slashed{D}_\perp } \,, \\
  \Gamma_2^{\mu,a} &= g_s\, T^a\, \bqty{ \gamma^\mu\, q + \gamma^\alpha\, J^-_\alpha\, \frac{1}{i\s v \cdot D + 2\s m_Q}\gamma_\perp^\mu\, h_v }  \,, \\
  \bar{\Gamma}_2^{\mu,a} &= g_s\, \bqty{ \bar{q}\, \gamma^\mu + \bar{h}_v\, \gamma_\perp^\mu \frac{1}{i\s v \cdot D + 2\s m_Q}\,\gamma^\alpha \,J^+_\alpha } \,T^a \, , \\
  S_1 &= \bqty{ 1 + i\s \slashed{D}_\perp \frac{1}{i\s v \cdot D + 2\s m_Q} } \,\gamma^\alpha \, J^+_\alpha\, , \\
  S_2 &= \gamma^\alpha \, J^-_\alpha \, \bqty{ 1 + \frac{1}{i\s v \cdot D + 2\s m_Q}\,\slashed{D}_\perp } \,.
\end{align}
\end{subequations}
\endgroup
Again, only relevant terms are kept here. Since the Lagrangian is real, we expect
\begin{equation}
  B_i^\dag = \gamma^0\, B_i\, \gamma^0 \, \qc\quad
  \bar{\Gamma}_i^{\mu,a} = \Big(\Gamma_i^{\mu,a}\Big)^\dag\, \gamma^0\, \qc\quad
  S_1^\dag = \gamma^0\, S_2\, \gamma^0\,,
\end{equation}
which is explicitly satisfied by the expressions in \cref{eqn:HLDetParameters}.

In order to evaluate this functional determinant, we row reduce the matrix and obtain the one-loop HQET action
\begin{align}
  S_\text{HQET}^{(1)}
    \supset i\Tr\bqty{ \frac{1}{(i\s D)^2}\, \delta^{ab}\, \eta_{\mu\nu} \Big(\bar{\Gamma}_2^{\mu,a}\, B_2^{-1}\, \Gamma_2^{\nu,b}
            - \bar{\Gamma}_1^{\mu,a}\, B_1^{-1}\, S_1 \,B_2^{-1}\, \Gamma_2^{\nu,b}
            - \bar{\Gamma}_2^{\mu,a}\, B_2^{-1}\, S_2\, B_1^{-1}\, \Gamma_1^{\nu,b} \Big) }\,.\notag\\[2pt]
\end{align}
Next, we use \cref{eqn:HLDetParameters} to derive
\begin{align}
  S_\text{HQET}^{(1)} &\supset -i\s g_s^2\, C_F\, \Tr\left[ \frac{1}{(i\s D)^2}\, \bar{q}\, \gamma^\mu\, \frac{1}{i\s\slashed{D}}\, \gamma^\alpha\, J_\alpha^-\, \right.\notag\\[5pt]
&\hspace{70pt}\left.  \times  \frac{\bqty{ 2m_Q + (1 + \slashed{v})i\slashed{D} } v_\mu
          + \bqty{ i\s \slashed{D} - \big(1 + \slashed{v}\big)\, i\s v \cdot D}\, \gamma_\mu}
         {(i\s D)^2 + 2\, m_Q\, i\s v \cdot D} \, h_v
    + \text{h.c.} \right]\,.
\end{align}
This functional trace can be converted into integral expressions using the method described in \cref{appsubsubsec:naiveCDE} (since we do not need any operators involving the field strength). This procedure yields
\begin{align}
  S_\text{HQET}^{(1)} &\supset
    -i\s g_s^2\, \mu^{2\eps}\, C_F\, \int \ddx{x} \int \ddp{p}
    \left[ \frac{1}{\big(i\s D + p\big)^2}\, \bar{q} \, \gamma^\mu
              \frac{1}{i\s \slashed{D} + \slashed{p}}\, \gamma^\alpha\, J_\alpha^-\right. \notag\\[5pt]
    &\hspace{25pt}\left. \times\frac{\bqty{2\s m_Q + \big(1 + \slashed{v}\big)\big(i\s\slashed{D} + \slashed{p}\big) }\, v_\mu
                        + \bqty{ i\s \slashed{D} + \slashed{p} - \big(1 + \slashed{v}\big)\, v \cdot \big(i\s D + p\big) }\, \gamma_\mu}
                       {\big(i\s D + p\big)^2 + 2\s m_Q \, v \cdot \big(i\s D + p\big)}\, h_v + \text{h.c.} \right] \notag\\[5pt]
    &\supset \int \ddx{x} \bqty{ \bar{q}\, \pqty{-i\s g_s^2\, \mu^{2\eps}\, C_F\, \int \ddp{p}\, \gamma^\mu\, \slashed{p}\, \gamma^\alpha\,
       \frac{\slashed{p}\, \gamma_\mu + 2\s m_Q \, v_\mu}{p^4\, \Big(p^2 + 2\s m_Q\, v \cdot p\Big)} }\, J_\alpha^-\, {h_v} + \text{h.c.}} \notag\\[5pt]
    &\supset \int \ddx{x} \pqty{ \bar{q}\, I_\text{HL}^\alpha\, J_\alpha^-\, h_v + \text{h.c.} } \,,
\end{align}
where we have implicitly defined the integral $I_\text{HL}^\alpha$ in the last line. Evaluating it yields
\begin{align}
      I_\text{HL}^\alpha &= -i\s g_s^2\, \mu^{2\eps}\, C_F\, \int \ddp{p}\,\frac{-(d - 2)\, \slashed{p}\, \gamma^\alpha\, \slashed{p} + 2\s m_Q\, \slashed{v}\, \slashed{p}\, \gamma^\alpha}
                           {p^4 \Big(p^2 + 2\s m_Q v \cdot p\Big)} \notag\\[5pt]
                 &= -\frac{\aS}{3\s \pi} \pqty{\frac{4\s \pi\,\mu^2}{m_Q^2}}^\eps\, \Gamma(\eps)\, \frac{2\s \eps}{1 - 2\s \eps} \Big(\gamma^\alpha - v^\alpha\Big)
                  \,\,\longrightarrow\,\,  -\frac{2}{3}\frac{\aS}{\pi} \Big(\gamma^\alpha - v^\alpha\Big)\,,
\end{align}
and hence
\begin{equation}
  S_\text{HQET}^{(1)} \supset -\frac{2}{3}\frac{\aS}{\pi} \int \ddx{x} \bqty{ \bar{q}\, \Big(\gamma^\alpha - v^\alpha\Big)\, J^-_\alpha\, h_v + \text{h.c.}} \,.
\end{equation}
This reproduces the one-loop matching coefficient for the heavy-light current, see \cref{eqn:HLresult,eqn:HLcomponents}.

\subsection{Heavy-Heavy Current Matching}
\label{sec:HHMatch}
The last matching example we will present is the derivation of the coefficient for the heavy-heavy current. The steps are essentially identical to the previous calculation, but the details are a bit more involved here since there are two flavors of heavy quarks to track. Note that while we will allow the two heavy quark masses $m_1$ and $m_2$ to differ, we will make the simplifying kinematic assumption that $v_1 = v_2=v$. The velocity labels on the fields are thus to be taken as flavor indices to keep track of masses only, and not to be taken as arbitrary. Otherwise, various simplification relations, \eg $\bar{h}_{v_1} v^\mu h_{v_2} = \bar{h}_{v_1} \gamma^\mu h_{v_2}$ are no longer valid. In \cref{sec:HHMatch2}, we work out the generalization allowing $v_1 \neq v_2$.

We start with the UV Lagrangian:
\begin{align}
\label{eq:LagUVHH}
  \Lag_\text{QCD} =\,& \bar{Q}_1\, \big(i\s \slashed{D} - m_1\big)\, Q_1 + \bar{Q}_2\, \big(i\s \slashed{D} - m_{2}\big)\, Q_2
  + \Big[\bar{Q}_1\, \Big(J^+_\alpha\, \gamma^\alpha + J^+_{5\alpha}\,\gamma^\alpha\,\gamma^5\Big)\, Q_2 + \text{h.c.}\Big] \notag\\
 & - \frac{1}{4} G_{\mu\nu}^a G^{\mu\nu,a} + \Lag_\text{gf} + \Lag_\text{gh} \,.
\end{align}
where $J^\pm_\alpha$ and $J^\pm_{5\alpha}$ are sources for the vector and axial heavy-heavy currents respectively. To reduce the clutter and absorb the phase, we introduce the shorthand
\begin{align}
J^\pm \equiv \Big(J^\pm_\alpha\,\gamma^\alpha + J^\pm_{5\alpha}\,\gamma^\alpha\,\gamma^5\Big)\, e^{\pm i\s\Delta m\, v\cdot x}\,,
\end{align}
where $\Delta m = m_1 - m_{2}$. Note that due to the gamma matrices, their conjugation relation is $\Big(J^+\Big)^\dag=\gamma^0 J^- \gamma^0$. A new feature of the heavy-heavy current matching calculation is that now the two heavy quarks mix:
\begin{subequations}
\begin{align}
  0 &= \fdv{S}{\bar{H}_{v_1}} = -\big(i\s v \cdot D + 2\s m_1\big)\,H_{v_1} + i\s {\slashed D_{\perp}}\,h_{v_1} + J^+\, \big(h_{v_2} + H_{v_2}\big)\,, \\[6pt]
  0 &= \fdv{S}{\bar{H}_{v_2}} = -\big(i\s v \cdot D + 2\s m_{2}\big)\,H_{v_2} + i\s {\slashed D_{\perp}}\,h_{v_2} + J^-\, \big(h_{v_1} + H_{v_1}\big)\,,
\end{align}
\end{subequations}
For our purposes here, we only need to solve these equations to linear order in $J$:
\begin{align}\label{eq:HSolUVHH}
  \pmqty{ H_{v_1} \\ H_{v_2} } =\,&
    \pmqty{ \frac{1}{i\s v \cdot D + 2\s m_1}\, i\s\slashed{D}_\perp\, h_{v_1} \\[4pt]
            \frac{1}{i\s v \cdot D + 2\s m_{2}}\, i\s\slashed{D}_\perp\, h_{v_2} } + \pmqty{ \frac{1}{i\s v \cdot D + 2\s m_1}\, J^+
                \pqty{ 1 + \frac{1}{i\s v \cdot D + 2\s m_{2}}\, i\s\slashed{D}_\perp }\, h_{v_2} \\[4pt]
              \frac{1}{i\s v \cdot D + 2\s m_{2}} J^-
                \pqty{ 1 + \frac{1}{iv \cdot D + 2\s m_1} i\slashed{D}_\perp } h_{v_1} }
    + \ord\Big(J^2\Big)\,.
\end{align}
Plugging this solution into the UV Lagrangian, and taking the relevant second variations yields
\begin{small}
\begin{align}
  \delta^2 \SHQETnl =
    \pmqty{ \var{A_\mu^a} & \var{h_{v_1}^T} & \var{\bar{h}_{v_1}} & \var{h_{v_2}^{T}} & \var{\bar{h}_{v_2}} }
    \pmqty{ C^{\mu\nu,ab} & \bar{\Gamma}_1^{\mu,a} & -\Big(\Gamma_1^{\mu,a}\Big)^T
            & \bar{\Gamma}_2^{\mu,a} & -\Big(\Gamma_2^{\mu,a}\Big)^T \\[5pt]
            -\Big(\bar{\Gamma}_1^{\nu,b}\Big)^T & 0 & -B_1^T & 0 & -S_2^T \\[8pt]
            \Gamma_1^{\nu,b} & B_1 & 0 & S_1 & 0 \\[5pt]
            -\Big(\bar{\Gamma}_2^{\nu,b}\Big)^T & 0 & -S_1^T & 0 & -B_2^T \\[8pt]
            \Gamma_2^{\nu,b} & S_2 & 0 & B_2& 0}
    \pmqty{ \var{A_\nu^b} \\[10pt] \var{h_{v_1}} \\[10pt] \var{\bar{h}^T_{v_1}} \\[10pt] \var{h_{v_2}} \\[9pt] \var{\bar{h}^{T}_{v_2}} }\,,\notag\\[3pt]
\end{align}
\end{small}\noindent
with again irrelevant terms for the heavy-heavy operator are discarded in $\Gamma_i^{\mu,a}$ and $\bar{\Gamma}_i^{\mu,a}$
\begingroup
\allowdisplaybreaks
\begin{subequations}
\begin{align}
 \hspace{-7pt} C^{\mu\nu,ab}\! &= \eta^{\mu\nu} \big(D^2\big)^{ab} - 2\,\Big(U_1^{\mu\nu,ab} + 1 \leftrightarrow 2\Big)\, , \\[5pt]
 \hspace{-7pt}  U_{1,2}^{\mu\nu,ab}\! &= -g_s^2\, \bar{h}_{v_{1,2}}\, T^a\s T^b\, \frac{\gamma_\perp^\mu}{i\s v \cdot D + 2\s m_{1,2}}
                     \,J^\pm\,
                     \frac{\gamma_\perp^\nu}{i\s v \cdot D + 2\s m_{2,1}}\, h_{v_{2,1}}\,, \\[5pt]
 \hspace{-7pt}  B_{1,2} &= i\s v \cdot D + i\s \slashed{D}_\perp\, \frac{1}{i\s v \cdot D + 2\s m_{1,2}} i\s \slashed{D}_\perp\,, \\[5pt]
 \hspace{-7pt}  \Gamma_{1,2}^{\mu,a} &= g_s\, T^a\, \Bigg\{ \bqty{ v^\mu + i\s \slashed{D}_\perp
                                       \frac{\gamma_\perp^\mu}{i\s v \cdot D + 2\s m_{1,2}} }\, h_{v_{1,2}} \notag\\[2pt]
    &\hspace{50pt}+ \bqty{ 1 + i\slashed{D}_\perp \frac{1}{iv \cdot D + 2\s m_{1,2}} }
        \,J^\pm\,
        \frac{\gamma_\perp^\mu}{i\s v \cdot D + 2\s m_{2,1}}\, h_{v_{2,1}} \Bigg\} \,, \\[5pt]
 \hspace{-7pt}  \bar{\Gamma}_{1,2}^{\mu,a} &= g_s\, \Bigg\{ \bar{h}_{v_{1,2}}\, \bqty{ v^\mu
    + \frac{\gamma_\perp^\mu}{i\s v \cdot D + 2\s m_{1,2}} i\slashed{D}_\perp }\notag\\[2pt]
    &\hspace{50pt}
    + \bar{h}_{v_{2,1}} \frac{\gamma_\perp^\mu}{iv \cdot D + 2\s m_{2,1}}
                   \,J^\mp\,
                                      \bqty{ 1 + \frac{1}{i\s v \cdot D + 2\s m_{1,2}} i\s \slashed{D}_\perp } \Bigg\}\s T^a\,, \\[5pt]
 \hspace{-7pt}  S_{1,2} &= \bqty{ 1 + i\s \slashed{D}_\perp \frac{1}{i\s v \cdot D + 2\s m_{1,2}} }
           \,J^\pm\,
           \bqty{ 1 + \frac{1}{i\s v \cdot D + 2\s m_{2,1}} i\s \slashed{D}_\perp }\,.
\end{align}
\end{subequations}
\endgroup
Note that this matrix takes the same form as \cref{eq:secondVarHeavyLight}, so the row reduction is identical. The functional determinant is then
\begin{align}
   \MoveEqLeft S^{(1)}_\text{HQET}
     \supset
      i\Tr\bqty{ \frac{1}{(i\s D)^2}\, \delta^{ab}\, \eta_{\mu\nu}\, \Big(U_1^{\mu\nu,ab}
                         + \bar{\Gamma}_1^{\mu,a}\, B_1^{-1}\, \Gamma_1^{\nu,b}
                         - \bar{\Gamma}_1^{\mu,a}\, B_1^{-1}\, S_1\, B_2^{-1}\, \Gamma_2^{\nu,b}\Big) } + 1 \leftrightarrow 2 \notag\\[8pt]
     &\supset -i\s g_s^2\, C_F\, \Tr\left\{\frac{1}{(i\s D)^2}\, \bar{h}_{v_1}\, \frac{i\s \slashed{D} + 2\s m_1}{(i\s D)^2 + 2\s m_1\, i\s v \cdot D}\,J^+\, \frac{i\s \slashed{D} + 2\s m_{2}}{(i\s D)^2 + 2\s m_{2}\, i\s v \cdot D}\, h_{v_2} \right.\notag \\[5pt]
 &\hspace{30pt}- \frac{1}{(i\s D)^2}\, \bar{h}_{v_1}\, \frac{i\s \slashed{D} - 2\s i\s v \cdot D}{(i\s D)^2 + 2\s m_1\, i\s v \cdot D} \,J^+\, \frac{i\s \slashed{D} - 2\s i\s v \cdot D}{(i\s D)^2 + 2\s m_{2}\, i\s v \cdot D}\, h_{v_2} \notag\\[5pt]
&\hspace{30pt} \left.+ \frac{1}{(i\s D)^2}\, \bar{h}_{v_1}\, \gamma^\mu\, \frac{i\s \slashed{D}_\perp - i\s v \cdot D}{(i\s D)^2 + 2\s m_1\, i\s v \cdot D} \,J^+\, \frac{i\s \slashed{D}_\perp - i\s v \cdot D}{(i\s D)^2 + 2\s m_{2}\, i\s v \cdot D}\, \gamma_\mu\, h_{v_2}
\right \} \notag\\[5pt]
&\hspace{13pt}+ \left( 1 \leftrightarrow 2 ,\, J^+ \leftrightarrow J^-\right) .
  \label{eq:HHSDetForm}
\end{align}
Our notation $\left( 1 \leftrightarrow 2 ,\, J^+ \leftrightarrow J^-\right)$ here (as well as later in the paper) represents only one term, which results from exchanging both $1 \leftrightarrow 2$ and $J^+ \leftrightarrow J^-$ in the previous term. Applying the CDE prescription in~\cref{appsubsubsec:naiveCDE} to evaluate these functional traces, we get
\begin{align}
\label{eq:HHIntForm}
     &S^{(1)}_\text{HQET} \supset -i\s g_s^2\, \mu^{2\eps}\, C_F\,
      \int \ddx{x} \int \ddp{p}\notag\\[6pt]
     &\hspace{10pt} \times \left[\frac{1}{{{{\left( {i\s D + p} \right)}^2}}} \,\bar{h}_{v_1}\, \frac{i\s \slashed{D} + \slashed{p} + 2\s m_1}{(\s iD + p)^2 + 2\s m_1\, v \cdot (i\s D + p)}\, J^+\, \frac{{i\s \slashed{D} + \slashed{p} + 2\s m_2}}{{{{\left( {i\s D + p} \right)}^2} + 2\s {m_2}\, v \cdot \left( {i\s D + p} \right)}}\, h_{v_2} \right.\notag\\[6pt]
  &\hspace{20pt}   - \frac{1}{{{{\left( {iD + p} \right)}^2}}}{\bar{h}_{v_1}}\frac{{i\slashed D + \slashed p - 2\s v \cdot \left( {i\s D + p} \right)}}{{{{\left( {i\s D + p} \right)}^2} + 2\s {m_1}\,v \cdot \left( {i\s D + p} \right)}}J^+\frac{{i\s \slashed D + \slashed p - 2\s v \cdot \left( {i\s D + p} \right)}}{{{{\left( {i\s D + p} \right)}^2} + 2\s {m_2}\,v \cdot \left( {i\s D + p} \right)}}{h_{v_2}}\notag \\[6pt]
  & \hspace{20pt} \left. + \frac{1}{{{{\left( {i\s D + p} \right)}^2}}}{\bar{h}_{v_1}}{\gamma ^\mu }\frac{{i\s \slashed D_\perp + \slashed p_\perp - v \cdot \left( {i\s D + p} \right)}}{{{{\left( {i\s D + p} \right)}^2} + 2\s {m_1}\,v \cdot \left( {i\s D + p} \right)}}\,J^+\,\frac{{i\s \slashed D_\perp + \slashed p_\perp - v \cdot \left( {i\s D + p} \right)}}{{{{\left( {i\s D + p} \right)}^2} + 2\s {m_2}\,v \cdot \left( {i\s D + p} \right)}}{\gamma _\mu }{h_{v_2}}
 \right]\notag\\[6pt]
 &\hspace{10pt}  + \left( 1 \leftrightarrow 2 ,\, J^+ \leftrightarrow J^-\right) \,.
\end{align}
This can be simplified down to
 \begin{align}
 S^{(1)}_\text{HQET}&\supset \int \ddx{x} \bqty{ \bar{h}_{v_1} \, \big(J^+_\alpha\, I_\text{HH}^\alpha + J^+_{5\alpha}\, I_{\text{HH},5}^\alpha\big)\, h_{v_2} }
              + \left( 1 \leftrightarrow 2 ,\, J^+ \leftrightarrow J^-\right) \,,
\end{align}
where the loop integrals are
\begin{subequations}
\begin{align}
    I_\text{HH}^\alpha &\equiv
      -i\s g_s^2\, \mu^{2\eps}\, C_F\, \int \ddp{p}\,\frac{1}{p^2 \big(p^2 + 2\s m_1\, v \cdot p\big) \big(p^2 + 2\s m_{2}\, v \cdot p\big)}\notag\\[2pt]
        &\hspace{107pt}\times\bigg\{\big(\slashed{p} + 2\s m_1\big)\, \gamma^\alpha\, \big(\slashed{p} + 2\s m_{2}\big) - \big(\slashed{p} - 2\s v \cdot p\big) \,\gamma^\alpha\, \big(\slashed{p} - 2\s v \cdot p\big)\notag\\[-3pt]
              &\hspace{145pt} +\gamma^\mu \big( \slashed{p}_\perp - v \cdot p\big)
                \gamma^\alpha \big( \slashed{p}_\perp -  v \cdot p\big)\, \gamma_\mu
             \bigg\} \,,
 \\[15pt]
    I_{\text{HH},5}^\alpha &\equiv I^\alpha\big|_{\gamma^\alpha \rightarrow\, \gamma^\alpha\,\gamma^5}\,.
\end{align}
\end{subequations}
Evaluating $I_\text{HH}^\alpha$ gives
\begin{align}
    I_\text{HH}^\alpha &= \gamma^\alpha\, \frac{\aS}{\pi}\, \bqty{ \MSbardiv - \frac{2}{3}
           - \frac{2}{\Delta m} \pqty{ m_1 \ln\frac{m_{2}}{\mu} - m_{2} \ln\frac{m_1}{\mu} }} \notag\\[7pt]
      &\longrightarrow\,\, -\gamma^\alpha\, \frac{2}{3} \frac{\aS}{ \pi} \bqty{ 1 + \frac{3}{\Delta m} \pqty{ m_1 \ln\frac{m_{2}}{\mu} - m_{2} \ln\frac{m_1}{\mu} } }\,,
\end{align}
and $I_{\text{HH},5}^\alpha$ can be evaluated in the same way. Putting it all together yields
\begin{align}
  S^{(1)}_\text{HQET} &\supset - \frac{2}{3}\frac{\aS}{ \pi} \int \ddx{x} \Bigg\{ \bqty{ 1 + \frac{3}{\Delta m} \pqty{ m_1 \ln\frac{m_{2}}{\mu} - m_{2} \ln\frac{m_1}{\mu} } }\, \bar{h}_{v_1}\, J^+_\alpha\, \gamma^\alpha\, h_{v_2} \notag\\[5pt]
     &\hspace{25pt}+ \bqty{ 2 + \frac{3}{\Delta m} \pqty{ m_1 \ln\frac{m_{2}}{\mu} - m_{2} \ln\frac{m_1}{\mu} } }  \, \bar{h}_{v_1}\, J^+_{5\alpha} \,\gamma^\alpha\, \gamma^5\, h_{v_2} +\text{h.c.} \Bigg\} \,,\notag\\[-3pt]
\end{align}
which agree with \cref{eqn:hhresult} and \cref{eqn:hhcomponents}. We note that the case of heavy-heavy matching at finite recoil ($v_1 \ne v_2$) is provided in \cref{sec:HHMatch2}. The next section turns to calculation of the RGEs, focusing on the particular examples of the Wilson coefficients for two operators that appear at $\ord(1/m_Q)$.

\subsection{HQET RGEs}
In this section, we will derive the RGEs for the two operators that appear at subleading order in the $1/m_Q$ HQET expansion:
\begin{align}
  \Lag_\text{HQET} \supset\,& \bar{h}_v\, (i\s v \cdot D)\, h_v + \frac{c_\text{kin}}{2\s m_Q}\, \bar{h}_v\,(i\s D_\perp)^2\, h_v + \frac{c_\text{mag}}{4\s m_Q}\,g_s\, \bar{h}_v\,\sigma^{\mu\nu}\, G_{\mu\nu}^a\, T^a\, h_v \notag \\[3pt]
&  -\frac{1}{4}\, G_{\mu\nu}^a\, G^{\mu\nu,a} + \Lag_\text{gf} + \Lag_\text{gh}\,,
\end{align}
with
\begin{equation}
  D_\perp^\mu  \equiv D^\mu - v^\mu (v \cdot D) \qquad \text{and} \qquad  \sigma^{\mu\nu} \equiv \frac{i}{2}\, \Big[\gamma^\mu, \gamma^\nu\Big]\,.
\end{equation}
The Wilson coefficient $c_\text{kin}$ is related to the leading order kinetic term through RPI, and as such it will not be renormalized; we will derive this explicitly at one-loop in this section. The Wilson coefficient for the chromomagnetic moment operator $c_\text{mag}$ does run, and has phenomenological consequences such as predicting the mass splitting between the ground-state vector and pseudoscalar mesons that contain a heavy quark. A precision determination of this mass splitting requires evolving $c_\text{mag}$ between the bottom mass and the charm mass.

It is straightforward to match onto these Wilson coefficients at tree-level by expanding~\cref{eqn:LHQETnonlocal}. Note that because
\begin{equation}
{{\bar h}_v}\, \Big[ {{\sigma _{\mu \nu }}\,{v^\mu }\,( {i\s v \cdot D} )} \Big] \,{h_v} = {{\bar h}_v}\,\bigg[ {\frac{i}{2}\,\big( {\slashed v\, {\gamma _\nu } - {\gamma _\nu }\,\slashed v} \big)\,\big( {i\s v \cdot D} \big)} \bigg]\,{h_v} = 0 \,,
\end{equation}
we can replace $D_\perp^\mu$ with $D^\mu$ when it is multiplied by $\sigma_{\mu\nu}$:
\begin{align}
 {i\s{{\slashed D}_ \bot }\,i\s{{\slashed D}_ \bot }}  &= {\big( {i\s {D_ \bot }} \big)^2} - \frac{i}{2}\,{\sigma _{\mu \nu }}\Big[ {i\s D_ \bot ^\mu ,i\s D_ \bot ^\nu } \Big] \nonumber \\[5pt]
 &\longrightarrow\,\, {\big( {i\s{D_ \bot }} \big)^2} - \frac{i}{2}\,{\sigma _{\mu \nu }}\,\Big[ {i\s{D^\mu },i\s{D^\nu }} \Big] = {\big( {i\s {D_ \bot }} \big)^2} + \frac{1}{2}\,g_s\,{\sigma ^{\mu \nu }}\,G_{\mu \nu }^a\,{T^a} \,.
\end{align}
This yields the tree-level matching conditions
\begin{equation}
  c_\text{kin}^{(0)} = c_\text{mag}^{(0)} = 1\,.
\end{equation}
These results serve as the boundary conditions for integrating the RGEs whose derivation are the subject of what follows.

We follow the procedure outlined in \cref{subsec:ReviewFunctionalRunning}. The first step is to compute the 1PI effective action.  We need the second variation of the tree-level action with respect to $A_\mu$, $h_v$ and $\bar{h}_v$. This yields an equation of the same form as \cref{eqn:ResidueVariationMatrix}. As opposed to the residue difference calculation, which uses the non-local form of the Lagrangian, here we are computing the 1PI effective action for HQET directly. Therefore, we must Taylor expand in $1/m_Q$ and truncate. In particular, we only keep terms up to order $1/m_Q$, and we drop everything that does not include quark fields:
\begingroup
\allowdisplaybreaks
\begin{subequations}
  \label{eqn:HQETDetParameters}
  \begin{align}
    C^{\mu\nu,ab} &= \eta^{\mu\nu}\, \big(D^2\big)^{ab} - 2\s U^{\mu\nu,ab}\,, \\[5pt]
    U^{\mu\nu,ab} &= g_s\, f^{abc}\, G^{\mu\nu,c} - g_s^2\, \bar{h}_v\, \bqty{ \frac{c_\text{kin}}{2\s m_Q} \big(\eta^{\mu\nu} - v^\mu v^\nu\big)\, T^a\, T^b
                       + \frac{c_\text{mag}}{4\s m_Q}\, \sigma^{\mu\nu}\, f^{abc}\, T^c } h_v\,, \\[5pt]
    B &= i\s v \cdot D + \frac{c_\text{kin}}{2\s m_Q} \big(i\s D_\perp\big)^2
         + \frac{c_\text{mag}}{4\s m_Q} \, g_s\, \sigma^{\mu\nu}\, G_{\mu\nu}^a\, T^a\, , \\[5pt]
    \Gamma^{\mu,a} &= g_s\, T^a\, \bqty{ v^\mu\, h_v + \frac{c_\text{kin}}{2\s m_Q}\, i\s D_\perp^\mu\, h_v
      + \frac{c_\text{kin}}{2\s m_Q} \big(i\s D_\perp^\mu\, h_v\big)_x - \frac{c_\text{mag}}{2\s m_Q}\,\sigma^{\mu\nu}\, h_v\, D_\nu}\,, \\[5pt]
    \bar{\Gamma}^{\mu,a} &= g_s\, \bqty{ \bar{h}_v\, v^\mu + \frac{c_\text{kin}}{2m_Q}\, \bar{h}_v\, i\s D_\perp^\mu
      - \frac{c_\text{kin}}{2\s m_Q} \big(i\s D_\perp^\mu\, \bar{h}_v\big)_x
      + \frac{c_\text{mag}}{2\s m_Q}\, D_\nu\, \bar{h}_v \,\sigma^{\mu\nu} }\, T^a \,.
\end{align}
\end{subequations}
\endgroup
Note that we have used a non-standard notation in the above --- a bracket with a subscript $x$ is to indicate that the covariant derivative is \emph{closed}. This is in contrast with the other terms, where the covariant derivatives are \emph{open}. A detailed elaboration of this notation is given in \cref{appsubsubsec:Territory}, in particular \cref{eqn:defDUix}. Reusing the row reduction result \cref{eqn:ResidueRowReduction}, we follow the same intermediate steps given in \cref{eq:residueDeterminant} to obtain the one-loop effective action
\begin{align}
  \Gamma^{(1)}_\text{HQET} &= \frac{i}{2} \ln\Sdet\bqty{ -\fdv[2]{S_\text{HQET}}{(A_\mu^a, \bar{h}_v, h_v)} } \notag\\[5pt]
    &\supset  -i\Tr \Bigg\{ \eta_{\mu\nu}\, \pqty{\frac{1}{D^2}}^{ba} \Big(U^{\mu\nu,ab} + \bar{\Gamma}^{\mu,a}\, B^{-1}\, \Gamma^{\nu,b}\Big) + \pqty{\frac{1}{D^2}}^{bc} \, U_{\nu\mu}^{cd}\, \pqty{\frac{1}{D^2}}^{da}\, U^{\mu\nu,ab} \notag\\[3pt]
    &\qquad\qquad\quad + \pqty{\frac{1}{D^2}}^{bc}\, U_{\nu\mu}^{cd}\, \pqty{\frac{1}{D^2}}^{da}
       \Big(\bar{\Gamma}^{\mu,a}\, B^{-1}\, \Gamma^{\nu,b} + \bar{\Gamma}^{\nu,b}\, B^{-1}\, \Gamma^{\mu,a}\Big) \Bigg\} \,.
\end{align}
where again we emphasize that we are not including higher order $1/m_Q$ terms than above. Next, we use the expressions in \cref{eqn:HQETDetParameters} to derive
\begin{align}
  \Gamma^{(1)}_\text{HQET}
    &\supset i\Tr \Bigg\{ \frac{c_\text{kin}}{2\s m_Q}\, g_s^2\, (d-1) \,C_F\, \frac{1}{D^2}\, \bar{h}_v\, h_v
      - g_s^2\, C_F\, \frac{1}{D^2}\, \bar{h}_v\, \frac{1}{i\s v \cdot D}\, h_v \notag\\[3pt]
    &\hspace{30pt} + g_s^2\, C_F\, \frac{c_\text{kin}}{2\s m_Q}\, \frac{1}{D^2}\,\bar{h}_v\, \frac{1}{i\s v \cdot D}\big(i\s D_\perp\big)^2 \frac{1}{i\s v \cdot D} \,h_v \notag\\[3pt]
    &\hspace{30pt} + g_s^2\, \pqty{ C_F - \frac{1}{2}\, C_A } \frac{c_\text{mag}}{4\s m_Q}\, \frac{1}{D^2}\, \bar{h}_v\,\frac{1}{i\s v \cdot D}\, g_s\, \sigma^{\mu\nu}\, G_{\mu\nu}^a\, T^a\, \frac{1}{i\s v \cdot D}\, h_v \notag\\[3pt]
    &\hspace{30pt} + 2\,g_s^2\, C_A\, \frac{c_\text{mag}}{4\s m_Q}\, \frac{1}{D^2}\, g_s\, G_{\mu\nu}^a\, \frac{1}{D^2}\, \pqty{ \bar{h}_v\, \sigma^{\mu\nu} \,T^a\, h_v - i\s D_\alpha\, \bar{h}_v\, v^\nu\, \sigma^{\mu\alpha}\, \frac{1}{i\s v \cdot D} \,T^a\, h_v } \Bigg\} \notag\\[5pt]
  &\equiv \Gamma_1 + \Gamma_2 + \Gamma_3 + \Gamma_4 + \Gamma_5 + \Gamma_6\,, \label{eqn:GammaHQET1}
\end{align}
where we have used
\begin{equation}
  T_F^a\, T_F^a = C_F \,,\qquad
  f^{abc}\, f^{abd} = C_A \,\delta ^{cd}\,,\qquad\text{and} \qquad
  T_F^a\, T_F^b\, T_F^a = \pqty{C_F - \frac{1}{2}\, C_A} \,T_F^b\,,
\end{equation}
with $C_F$ and $C_A$ denote the Casimir factor for the fundamental and adjoint representations respectively. Note that we defined the six terms in \cref{eqn:GammaHQET1} as $\Gamma_{i=1,\cdots,6}$ in the order that they appear. Since we want to derive the RGEs, we need to isolate the UV divergence. To this end, we regulate the IR using a non-zero gluon mass $m^2$, which is sufficient for any of the loop integrals we encounter below.

For the first two terms we use \cref{eqn:T1,eqn:T1v} to obtain
\begin{subequations}
\begin{align}
  \Gamma_1 &\equiv \frac{c_\text{kin}}{2\s m_Q}\, g_s^2\, (d-1)\, C_F\, i\Tr \pqty{ \frac{1}{D^2 + m^2}\, \bar{h}_v\, h_v } \supset 0\,, \\[5pt]
  \Gamma_2 &\equiv -g_s^2\, C_F\, i\Tr \pqty{ \frac{1}{D^2 + m^2}\, \bar{h}_v\, \frac{1}{i\s v \cdot D}\, h_v }\notag\\[3pt]
           &\supset -2\,g_s^2\, C_F \int \dd[4]{x} \frac{1}{(4\s \pi)^2} \left(\ln\frac{\mu^2}{m^2}\right) \, \pqty{\bar{h}_v\, i\s v \cdot D\, h_v} \,.
\end{align}
\end{subequations}
For the third term, none of the results in~\cref{appsubsubsec:functionaltraces} directly apply, so we provide some explicit steps:
\begin{align}
  \Gamma_3 &\equiv g_s^2\, C_F\, \frac{c_\text{kin}}{2\s m_Q}\, i\Tr \bqty{ \frac{1}{D^2 + m^2}\, \bar{h}_v\, \frac{1}{i\s v \cdot D} \big(i\s D_\perp\big)^2 \frac{1}{i\s v \cdot D}\, h_v } \nonumber \\[5pt]
           &=  -g_s^2\, C_F\, \frac{c_\text{kin}}{2\s m_Q}\, i\int \ddx{x} \int \ddp{q} \tr \Bigg[ \frac{1}{(i\s D - q)^2 - m^2} \notag\\[-3pt]
           &\hspace{180pt} \times \bar{h}_v\, \frac{1}{v \cdot (i\s D - q)}(i\s D - q)^2 \frac{1}{v \cdot (i\s D - q)}\, h_v \Bigg]  \notag\\[5pt]
           &\supset g_s^2\, C_F\, \frac{c_\text{kin}}{2\s m_Q} \int \dd[4]{x} \Big(\eta^{\mu\nu}\, V_{1,2} - 4\,V_{2,2}^{\mu\nu}\Big) \Big(\bar{h}_v\, D_\mu\, D_\nu\, h_v\Big) \nonumber \\[5pt]
           &= \int \dd[4]{x} \frac{1}{(4\s\pi)^2} \ln\frac{\mu^2}{m^2} \Bqty{ -2\,g_s^2\, C_F\, \frac{c_\text{kin}}{2\s m_Q} \bqty{ \bar{h}_v\, (i\s D_\perp)^2\, h_v - 3\bar{h}_v (i\s v \cdot D)^2\, h_v } } \,.
\end{align}
In the above, the loop integrals $V_{1,2}$ and $V_{2,2}^{\mu\nu}$ can be found in \cref{appsubsubsec:HLIntegralsHQET}. For $\Gamma_4$ and $\Gamma_5$ we can use \cref{eqn:T2} and \cref{eqn:T1vv} to find
\begin{subequations}
\begin{align}
    \Gamma _4
      &\equiv g_s^2\, \pqty{C_F - \frac{1}{2}\,C_A}\, \frac{c_\text{mag}}{4\s m_Q}\,
                i\Tr \pqty{\frac{1}{D^2 + m^2}\, \bar{h}_v\, \frac{1}{i\s v \cdot D}\,
                             g_s \,\sigma^{\mu\nu}\, G_{\mu\nu}^a\, T^a\, \frac{1}{i\s v \cdot D}\,h_v }\notag \\[5pt]
      &\supset \int \dd[4]{x} \frac{1}{(4\s \pi)^2} \ln\frac{\mu^2}{m^2}
                 \bqty{ g_s^2\, \big(C_A - 2\,C_F\big)\, \frac{c_\text{mag}}{4\s m_Q} \,g_s \bar{h}_v \,\sigma^{\mu\nu}\, G_{\mu\nu}^a\, T^a\, h_v } \,, \\[12pt]
    \Gamma_5
      &\equiv 2\,g_s^2\, C_A\, \frac{c_\text{mag}}{4\s m_Q}
                i\Tr \pqty{ \frac{1}{D^2 + m^2}\, g_s\, G_{\mu\nu}^a\, \frac{1}{D^2 + m^2}\, \bar{h}_v\, \sigma^{\mu\nu}\, T^a\, h_v } \notag\\[5pt]
      &\supset \int \dd[4]{x} \frac{1}{(4\s \pi)^2} \ln\frac{\mu^2}{m^2}
                 \bqty{ -2\,g_s^2\, C_A \,\frac{c_\text{mag}}{4\s m_Q}\, g_s\, \bar{h}_v\, \sigma^{\mu\nu}\,G_{\mu\nu}^a\, T^a\, h_v}\,.
\end{align}
\end{subequations}
For the sixth term, we again provide the details:
\begin{align}
  \Gamma_6 &\equiv -2\,g_s^2\, C_A\, \frac{c_\text{mag}}{2\s m_Q}\, i\Tr \pqty{ \frac{1}{D^2 + m^2}\, g_s\, G_{\mu\nu}^a\, \frac{1}{D^2 + m^2} i\s D_\alpha \, \bar{h}_v\, v^\nu\, \sigma^{\mu\alpha}\, \frac{1}{i\s v \cdot D}\, T^a\, h_v } \nonumber \\[5pt]
   &= -2\,g_s^2\, C_A\, \frac{c_\text{mag}}{2\s m_Q} i\int \ddx{x} \int \ddp{q}
      \left[ \frac{1}{(i\s D - q)^2 - m^2}\, g_s\, G_{\mu\nu}^a\, \frac{1}{(i\s D - q)^2 - m^2} \right.\notag\\[5pt]
      &\hspace{165pt}\left. \times(i\s D - q)_\alpha\, \bar{h}_v\, v^\nu\, \sigma^{\mu\alpha}\, \frac{1}{v \cdot (i\s D - q)}\, T^a\, h_v \right] \nonumber \\[5pt]
   &\supset  -2\,g_s^2\, C_A \,\frac{c_\text{mag}}{2\s m_Q} i\int \ddx{x} \int \ddp{q}\, \frac{v^\nu \, q_\alpha\,v \cdot q}{\big(q^2 - m^2\big)^2} \,g_s\, \bar{h}_v\, \sigma^{\mu\alpha}\, G_{\mu\nu}^a\,T^a\, h_v \nonumber \\[5pt]
   &= \int \dd[4]{x} \bqty{ 2\,g_s^2\, C_A\, \frac{c_\text{mag}}{2\s m_Q}\, v^\nu\, \eta_{\alpha\beta} \,V_{2,1}^\beta\, \big( g_s\, \bar{h}_v\, \sigma^{\mu\alpha} G_{\mu\nu}^a \,T^a\, h_v\big) } \nonumber \\[5pt]
   &= \int \dd[4]{x} \frac{1}{(4\s\pi)^2} \ln\frac{\mu^2}{m^2} \,2\,g_s^2\, C_A \,\frac{c_\text{mag}}{2\s m_Q}\, v^\nu\, g_s\, \bar{h}_v\, v_\alpha \,\sigma^{\mu\alpha}\, G_{\mu\nu}^a\, T^a\, h_v = 0 \,.
\end{align}
Summing all the six $\Gamma_i$ yields the one-loop level effective action:
\begin{align}
      \hspace{-6pt}\Gamma_\text{HQET}^{(1)}
        &\supset \Gamma_1 + \Gamma_2 + \Gamma_3 + \Gamma_4 + \Gamma_5 + \Gamma_6\notag \\[5pt]
        &\supset \int \dd[4]{x} \frac{\alpha_s}{4\s\pi} \ln\frac{\mu^2}{m^2} \Bigg\{ - C_A\,\frac{c_\text{mag}}{4\s m_Q}\, g_s\, \bar{h}_v \,\sigma^{\mu\nu} \,G_{\mu\nu}^a\, T^a\, h_v \notag\\[-3pt]
        &\hspace{18pt} - 2\, C_F \left[ \bar{h}_v\, (i\s v \cdot D)\, h_v + \frac{c_\text{kin}}{2\s m_Q}\, \bar{h}_v\, (i\s D_\perp)^2\, h_v + \frac{c_\text{mag}}{4\s m_Q} \,g_s\, \bar{h}_v\, \sigma^{\mu\nu}\, G_{\mu\nu}^a\, T^a\, h_v \right] \Bigg\} \,.
    \label{eq:Gamma1RGE}
\end{align}
Note that in the last line, we have dropped the operator $\bar{h}_v \left(iv\cdot D\right)^2 h_v$ since it is redundant due to the equations of motion. Combining this result with the tree-level effective action $\Gamma_\text{HQET}^{\left(0\right)} = S_\text{HQET}$ yields
  \begin{align}
  \Gamma_\text{HQET} &\supset \int \dd[4]{x} \left\{
  \bqty{ 1 - \frac{2\,\alpha_s\, C_F}{4\s\pi} \ln\frac{\mu^2}{m^2} } \right.\notag\\[-3pt]
  &\hspace{60pt}\times \bqty{ \bar{h}_v\, (i\s v \cdot D)\, h_v + \frac{c_\text{kin}}{2\s m_Q} \bar{h}_v\,\big(i\s D_\perp\big)^2\, h_v + \frac{c_\text{mag}}{4\s m_Q}\, g_s\, \bar{h}_v \,\sigma^{\mu\nu}\, G_{\mu\nu}^a\,T^a\, h_v }\notag\\[-3pt]
&\left. \hspace{60pt} - \frac{\alpha_s}{4\s\pi} \ln\frac{\mu^2}{m^2}\,  C_A\, \frac{c_\text{mag}}{4\s m_Q}\,g_s\, \bar{h}_v\, \sigma^{\mu\nu}\, G_{\mu\nu}^a\, T^a\, h_v \right\} \notag\\[8pt]
    \longrightarrow\,\,&\hspace{12pt} \int \dd[4]{x} \Bigg[ \bar{h}_v (i\s v \cdot D)\, h_v
           + \frac{c_\text{kin}}{2\s m_Q}\, \bar{h}_v\, \big(i\s D_\perp\big)^2\, h_v\notag\\[-3pt]
        &\hspace{2cm}   +\left(1- \frac{\alpha_s}{4\s\pi}\, C_A \ln\frac{\mu^2}{m^2}\right) \frac{c_\text{mag}}{4\s m_Q}\,g_s\, \bar{h}_v\, \sigma^{\mu\nu} \,G_{\mu\nu}^a\, T^a\, h_v     \Bigg],
\end{align}
where in the second line, we have performed the field redefinition
\begin{equation}
  h_v \,\,\longrightarrow\,\, \bqty{ 1 - \frac{2\,C_F\, \aS}{4\s \pi} \ln\frac{\mu^2}{m^2} }^{-1/2}\, h_v\,,
\end{equation}
to canonically normalize the tree-level kinetic term $\bar{h}_v\, (i\s v\cdot D)\, h_v$. Finally, we can read off the RGE equations:
\begin{subequations}
  \begin{alignat}{3}
    0 &= \mu\dv{\mu} \frac{c_\text{kin}}{2\s m_Q} &\qquad\Longrightarrow\qquad&& \mu\dv{\mu} c_\text{kin} &= 0\,, \\[5pt]
    0 &= \mu\dv{\mu} \bqty{ \frac{c_\text{mag}}{4\s m_Q}\left(1 - \frac{\alpha_s}{4\s\pi}\,  C_A \ln\frac{\mu^2}{m^2}\right) } &\qquad\Longrightarrow\qquad&& \mu\dv{\mu} c_\text{mag} &= \frac{\alpha_s}{2\s\pi}\,C_A \,c_\text{mag} \,.
  \end{alignat}
\end{subequations}
These results agree with the literature as summarized in \cref{eq:RGESummary} above.

\section{Conclusions}
\label{sec:Conc}
This paper provides the first application of functional methods augmented by the covariant derivative expansion to a kinematic EFT. In particular, we studied HQET --- a description that emerges in the limit that a Dirac fermion is very heavy compared to the scale being probed by the propagating fluctuations. We focused on the particular example of heavy quarks coupled to QCD, and reproduced a variety of matching and running results that had been previously computed using Feynman diagrammatic methods. A critical component was the existence of an operator valued set of projectors that are used to derive an equation of motion for the long distance modes that is valid to all-orders in the heavy mass expansion. Furthermore, the point at which RPI becomes manifest was emphasized. The primary result of this work was the matching master formula given in \cref{eqn:SHQET1loop}.

There are two clear directions for future progress. First, one can now use these efficient methods to take one-loop matching calculations to higher order in the heavy mass limit. This would be relevant for high precision applications to either experiments such as LHCb and Belle~II, or to theory explorations comparing with lattice QCD calculations in the heavy mass limit~\cite{Heitger:2003nj, Blossier:2010jk}. Another interesting direction would be to understand how to use these methods for other kinematic EFTs, such as SCET, nrQCD, and others. Furthermore, these methods provide a compelling impetus to revisit the issue of higher-order functional integration directly in the path integral. If the relevant technology could be developed to the stage where two-loop (or higher) calculations could be implemented in a straightforward manner,\footnote{For promising work in this direction, see \Ref{Abbott:1980hw,Capper:1982tf}.} then these techniques could contribute to extend precision in $\aS$ as well. This paper makes the benefits of the functional approach to HQET clear, and furthermore opens the door to using these methods across a broader class of EFTs than had been previously explored.

\acknowledgments
We thank Zoltan Ligeti, Gil Paz, and Dean Robinson for useful discussions. TC and XL are supported by the U.S. Department of Energy (DOE), under grant DE-SC0011640.  MF is supported by the DOE under grant DE-SC0010008 and partially by the Zuckerman STEM Leadership Program. TC and MF performed some of this work at the Munich Institute for Astro- and Particle Physics (MIAPP) which is funded by the Deutsche Forschungsgemeinschaft (DFG, German Research Foundation) under Germany's Excellence Strategy -- EXC-2094 -- 390783311. MF thanks the Aspen Center for Physics, supported by National Science Foundation grant PHY-1607611, for hospitality while parts of this work were carried out.

\appendix
\section*{Appendices}
\addcontentsline{toc}{section}{\protect\numberline{}Appendices}%

\section{Loop Integrals}
\label{appsec:LoopIntegrals}
In this appendix, we provide a number of frequently encountered loop integrals that are relevant for the calculations performed in the main text. All loop integrals are regulated with dimensional regularization $d=4-2\eps$, and parameters are renormalized using the \MSbar scheme.

\subsection{Combining Denominators}
\label{appsubsec:IntegralReduction}
A general loop integral contains multiple propagators in the integrand. In order to facilitate the use of the general results presented in what follows, we use the Feynman parameter technique to combine propagators into a single term:
\begin{align}
\hspace{-3mm}
  \frac{1}{A_1^{m_1} \dotsm A_n^{m_n}}
    &= \int_0^1 \dd{x_1} \dotsi \dd{x_n}\, \delta(x_1 + \dotsb + x_n - 1)\notag\\[-3pt]
    &\hspace{40pt} \times \frac{x_1^{m_1 - 1} \dotsm x_n^{m_n - 1}}{(x_1 A_1 + \dotsb  + x_n A_n)^{m_1 + \dotsb + m_n}} \frac{\Gamma( m_1 + \dotsb + m_n)}{\Gamma(m_1) \dotsm \Gamma(m_n)} \notag\\[5pt]
    &= \int_0^\infty \dd{\lambda_2} \dotsi \dd{\lambda_n}
         \frac{\lambda_2^{m_2 - 1} \dotsm \lambda_n^{m_n - 1}}
              {(A_1 + \lambda_2 A_2 + \dotsb + \lambda_n A_n)^{m_1 + \dotsb + m_n}}
         \frac{\Gamma(m_1 + \dotsb + m_n)}{\Gamma(m_1) \cdots \Gamma(m_n)} \,.
\label{eqn:PropagatorCombine}
\end{align}
The first integral in the above expression is convenient for combining propagators that are quadratic in the loop momentum. For situations where propagators are linear in the loop momentum, which frequently occurs for HQET calculations, it is more convenient to use the second integral form of this trick. Note that it is straightforward to obtain this last line from the more commonly used form given in the first line by making the variable change $x_{k \ge 2} = x_1 \lambda_{k \ge 2}$ and then integrating over $x_1$ using the delta function.

\subsection{Single Scale Relativistic Integrals}\label{appsubsec:EssentialIntegrals}

After combining the propagators using \cref{eqn:PropagatorCombine}, if the numerator of the integrand has factors of the loop momenta with open Lorentz indices, one can convert them by relying on the symmetry of the integral:
\begin{subequations}
\label{eqn:NumeratorReduction}
\begin{align}
  p^\mu p^\nu &\to \frac{1}{d}\, p^2\, \eta^{\mu\nu} \,, \\
  p^\mu p^\nu p^\rho p^\sigma
    &\to \frac{1}{d(d+2)}\, p^4\, \big( \eta^{\mu\nu} \eta^{\rho\sigma} + \eta^{\mu\rho} \eta^{\nu\sigma} + \eta^{\mu\sigma} \eta^{\eta\rho} \big) \,.
\end{align}
\end{subequations}
This reduces all loop integrals (with at least one relativistic propagator) to the following essential form with a single scale $\Delta$:
\begin{align}
  I_n^m(\Delta)
    &\equiv  -i\mu^{4-d}  \int\ddp{p} \frac{{\left( p^2 \right)}^m}{\left( p^2 - \Delta \right)^n} \notag\\
    &= \frac{\mu^{4-d}}{(4\pi)^{d/2}} (-1)^{n-m} \frac{\Gamma(m + d/2)}{\Gamma(d/2)} \frac{\Gamma(n - m - d/2)}{\Gamma(n)} \frac{1}{\Delta^{n - m - d/2}} \,.
\label{eq:Inm}
\end{align}
This integral is straightforward to evaluate after performing a Wick rotation to Euclidean space. Then to derive \MSbar expressions, one expands these integrals for small $\eps$, identifies the $1/\eps$ poles and cancels all the occurrences of $1/\eps-\gamma_\text{E}+\ln 4\pi$ using counterterms, where $\gamma_\text{E}$ is the Euler-Mascheroni constant. For completeness, we provide explicit results that are frequently used here. For divergent integrals, we use an arrow to indicate that contributions from counterterms defined in the \MSbar scheme have been added. The integrals for $n - m = 1$ are quadratically UV divergent:
\begin{align}
  I_{m+1}^m(\Delta) &= \frac{1}{(4\pi)^2} \bqty{\frac{1}{m!} \frac{\Gamma(m + d/2)}{\Gamma(d/2)}} \,
                       \frac{\Delta}{d/2 - 1} \, \pqty{\frac{4\pi\mu^2}{\Delta}}^{2 - d/2}
                       \Gamma\pqty{2 - \frac{d}{2}} \notag \\
                    &\longrightarrow \frac{1}{(4\pi)^2} (m+1)\, \Delta \,
                                     \pqty{\ln\frac{\mu^2}{\Delta} + 2 - \sum_{r=1}^{m+1} \frac{1}{r} } \,,
\end{align}
For $n - m = 2$, the integrals are logarithmically UV divergent:
\begin{align}
  I_{m+2}^m(\Delta) &= \frac{1}{(4\pi)^2} \bqty{\frac{1}{(m+1)!} \frac{\Gamma(m + d/2)}{\Gamma(d/2)}} \,
                       \pqty{\frac{4\pi\mu^2}{\Delta}}^{2 - d/2}
                       \Gamma\pqty{2 - \frac{d}{2}} \notag \\
                    &\longrightarrow \frac{1}{(4\pi)^2}
                                     \pqty{\ln\frac{\mu^2}{\Delta} + 1 - \sum_{r=1}^{m+1} \frac{1}{r} } \,.
\end{align}
For $n - m = 3$ (and greater), the integrals are UV finite:
\begin{align}
  I_{m+3}^m(\Delta) &= -\frac{1}{(4\pi)^2} \frac{1}{m+2}\, \frac{1}{\Delta} \,, \\
  I_{m+4}^m(\Delta) &= \frac{1}{(4\pi)^2} \frac{1}{(m+3)(m+2)}\, \pqty{\frac{1}{\Delta}}^2 \,,
\end{align}
and so on.

\subsection{Two Scale Relativistic Integrals}
\label{appsubsec:HLIntegrals}
Here, we explicitly discuss a commonly appearing integral in relativistic theories that depends on two scales:
\begin{equation}\label{eqn:Inkr}
  I_{n,k}^r(m^2,M^2)
    \equiv -i\mu^{4-d} \int \ddp{p}
             \frac{\left( p^2 \right)^r}{\left(p^2 - m^2\right)^n \left( p^2 - M^2 \right)^k} \,.
\end{equation}
Note that when this integral contains a nontrivial numerator $p^{\mu_1}\cdots p^{\mu_s}$, one can use \cref{eqn:NumeratorReduction} before combining the propagators, to reduce it to the above form. We evaluate this integral as follows:
\begin{align}
  I_{n,k}^r(m^2,M^2)
    &= -i\mu^{4-d} \int_0^1 \dd{x} x^{n-1} (1-x)^{k-1} \frac{\Gamma(n+k)}{\Gamma(n) \Gamma(k)}
       \int \ddp{p} \frac{\left( p^2 \right)^r}{\left[ p^2 - xm^2 - (1-x)M^2 \right]^{n+k}} \notag\\[5pt]
    &= \frac{\mu^{4-d}}{(4\pi)^{d/2}} (-1)^{n+k-r} \frac{\Gamma(r + d/2)}{\Gamma(d/2)}
         \frac{\Gamma(n+k-r-d/2)}{\Gamma(n) \Gamma(k)} \notag\\
    &\qquad \times \int_0^1 \dd{x} \frac{x^{n-1} (1-x)^{k-1}}{\left[ xm^2 +  (1-x)M^2 \right]^{n+k-r-d/2}} \notag \\[5pt]
    &= \frac{\mu^{4-d}}{(4\pi)^{d/2}} \frac{(-1)^{n+k-r}}{\left( M^2 \right)^{n+k-r-d/2}}
         \frac{\Gamma(r + d/2)}{\Gamma(d/2)} \frac{\Gamma(n+k-r-d/2)}{\Gamma(n+k)} \notag \\
    &\qquad \times {}_2 F_1 \left( n + k - r - \frac{d}{2}, n; n+k; \frac{M^2 - m^2}{M^2} \right) \,.
\end{align}
In the first line above, we have combined the propagators using \cref{eqn:PropagatorCombine}. In the second line, we have evaluated the single propagator loop integral using the result in \cref{eq:Inm}. In the last line, we have performed the Feynman parameter integral which yields a Gaussian Hypergeometric function $_2F_1\left(a,b;c;z\right)$. If the arguments $a$, $b$, and $c$ are integers (or half integers), the Gaussian Hypergeometric function can be expressed in terms of elementary functions. A non-exhaustive set of explicit evaluations are listed below for some frequently encountered integrals. Note that $I_{n,k}^r(m^2,M^2) = I_{k,n}^r(M^2,m^2)$, so we will only provide results for integrals with $n \le k$. Taking $n+k-r=2$ yields the log divergent integral
\begin{align}
  I_{1,1}^0(m^2,M^2)
    &= \frac{1}{(4\pi)^2} \left(\frac{4\pi\mu^2}{M^2} \right)^{2 - d/2} \Gamma\left( 2 - \frac{d}{2} \right)
         {}_2 F_1 \left( 2 - \frac{d}{2}, 1; 2; z \right) \notag\\
    &= \frac{1}{(4\pi)^2} \left(\frac{4\pi\mu^2}{M^2} \right)^{2 - d/2} \Gamma\left( 2 - \frac{d}{2} \right)
         \frac{1 - (1-z)^{d/2 - 1}}{(d/2 - 1) z} \notag\\
    &= \frac{1}{(4\pi)^2} \frac{\left( M^2 \right)^{d/2 - 1} - \left( m^2 \right)^{d/2 - 1}}
                               {\left( M^2 - m^2 \right)}
         \frac{\left( 4\pi\mu^2 \right)^{2 - d/2}}{(d/2 - 1)} \Gamma \left( 2 - \frac{d}{2} \right) \notag\\
    &\longrightarrow \frac{1}{(4\pi)^2} \left[ \frac{1}{M^2 - m^2}
           \left( M^2 \ln\frac{\mu^2}{M^2} - m^2 \ln\frac{\mu^2}{m^2} \right) + 1 \right] \,.
\end{align}
where for convenience we have defined
\begin{align}
  z \equiv 1 - \frac{m^2}{M^2} \,.
\end{align}
For brevity, we will we simply state the results for the other evaluations of use here. Taking $n+k-r=3$ yields UV finite integrals proportional to $1/M^2$
\begin{subequations}
\begin{align}
I_{1,2}^0\left( {{m^2},{M^2}} \right) &=
\frac{1}{{{{\left( {4\pi } \right)}^2}}}\frac{1}{{{M^2}}}\frac{{ - 1}}{2}\frac{2}{{{z^2}}}\left[ {z + \left( {1 - z} \right)\ln \left( {1 - z} \right)} \right] \,, \\
I_{1,3}^1\left( {{m^2},{M^2}} \right) &=
 \frac{1}{{{{\left( {4\pi } \right)}^2}}}\frac{1}{{{M^2}}}\frac{{ - 1}}{3}\frac{{ - 3}}{{2{z^3}}}\left[ {z\left( {2 - 3z} \right) + 2{{\left( {1 - z} \right)}^2}\ln \left( {1 - z} \right)} \right] \,, \\
I_{2,2}^1\left( {{m^2},{M^2}} \right) &=
 \frac{1}{{{{\left( {4\pi } \right)}^2}}}\frac{1}{{{M^2}}}\frac{{ - 1}}{3}\frac{3}{{{z^3}}}\left[ {z\left( {2 - z} \right) + 2\left( {1 - z} \right)\ln \left( {1 - z} \right)} \right] \,.
\end{align}
\end{subequations}
Taking $n+k-r=4$ yields UV finite integrals proportional to $1/M^4$
\begingroup
\allowdisplaybreaks
\begin{subequations}\renewcommand\arraystretch{1.3}
\begin{align}
I_{1,3}^0\left( {{m^2},{M^2}} \right) &=
\frac{1}{{{{\left( {4\pi } \right)}^2}}}\frac{1}{{{M^4}}}\frac{1}{6}\frac{3}{{{z^3}}}\left[ {z\left( {2 - z} \right) + 2\left( {1 - z} \right)\ln \left( {1 - z} \right)} \right] \,, \\
I_{2,2}^0\left( {{m^2},{M^2}} \right) &=
\frac{1}{{{{\left( {4\pi } \right)}^2}}}\frac{1}{{{M^4}}}\frac{1}{6}\frac{{ - 6}}{{{z^3}}}\left[ {2z + \left( {2 - z} \right)\ln \left( {1 - z} \right)} \right] \,, \\
I_{1,4}^1\left( {{m^2},{M^2}} \right) &=
 \frac{1}{{{{\left( {4\pi } \right)}^2}}}\frac{1}{{{M^4}}}\frac{1}{{12}}\frac{{ - 2}}{{{z^4}}}\left[ {z\left( {6 - 9z + 2{z^2}} \right) + 6{{\left( {1 - z} \right)}^2}\ln \left( {1 - z} \right)} \right] \,, \\
I_{2,3}^1\left( {{m^2},{M^2}} \right) &=
\frac{1}{{{{\left( {4\pi } \right)}^2}}}\frac{1}{{{M^4}}}\frac{1}{{12}}\frac{6}{{{z^4}}}\left[ {z\left( {6 - 5z} \right) + 2\left( {3 - z} \right)\left( {1 - z} \right)\ln \left( {1 - z} \right)} \right] \,, \\
I_{1,5}^2\left( {{m^2},{M^2}} \right) &=
 \frac{1}{{{{\left( {4\pi } \right)}^2}}}\frac{1}{{{M^4}}}\frac{1}{{20}}\frac{5}{{3{z^5}}}\left[ \begin{array}{l}
 z\left( {12 - 30z + 22{z^2} - 3{z^3}} \right) \\
 + 12{{\left( {1 - z} \right)}^3}\ln \left( {1 - z} \right)
 \end{array} \right] \,, \\
I_{2,4}^2\left( {{m^2},{M^2}} \right) &=
 \frac{1}{{{{\left( {4\pi } \right)}^2}}}\frac{1}{{{M^4}}}\frac{1}{{20}}\frac{{ - 10}}{{3{z^5}}}\left[ \begin{array}{l}
 z\left( {24 - 42z + 17{z^2}} \right) \\
 + 6\left( {4 - z} \right){{\left( {1 - z} \right)}^2}\ln \left( {1 - z} \right)
 \end{array} \right] \,, \\
I_{3,3}^2\left( {{m^2},{M^2}} \right) &=
\frac{1}{{{{\left( {4\pi } \right)}^2}}}\frac{1}{{{M^4}}}\frac{1}{{20}}\frac{{10}}{{{z^5}}}\left[ {z\left( {12 - 12z + {z^2}} \right) + 6\left( {2 - z} \right)\left( {1 - z} \right)\ln \left( {1 - z} \right)} \right] \,.
\end{align}
\end{subequations}
\endgroup

\subsection{Integrals with Relativistic and Linear Propagators}\label{appsubsubsec:HLIntegralsHQET}
A class of integrals that commonly appear in HQET calculations are built from a relativistic massive propagator (the mass is often included to regulate the IR) and an HQET propagator:
\begin{equation}\label{eqn:Vnsr}
  V_{n,2s + r}^{\mu_1 \dotsm \mu_r} (m^2)
    \equiv -i\mu^{4-d} \int \ddp{p}
             \frac{p^{\mu_1} \dotsm p^{\mu_r}}{\left( p^2 - m^2 \right)^n (v \cdot p)^{2s + r}} \,.
\end{equation}
Note that for this integral, we cannot apply the numerator reduction formulas in \cref{eqn:NumeratorReduction} before combining the propagators, because more tensor structures are available due to the presence of $v^\mu$. However, the same logic obviously applies, and we simply need to contract the integrand with both $\eta^{\mu\nu}$ or $v^\mu$ to determine the possible forms. For example,
\begin{subequations}\label{eqn:VReplacements}
\begin{align}
{p^\mu } &\to {v^\mu }\left( {v \cdot p} \right) \,, \\
{p^\mu }{p^\nu } &\to \frac{1}{{d - 1}}\left[ {{p^2}\left( {{\eta ^{\mu \nu }} - {v^\mu }{v^\nu }} \right) - {{\left( {v \cdot p} \right)}^2}\left( {{\eta ^{\mu \nu }} - d{v^\mu }{v^\nu }} \right)} \right] \,.
\end{align}
\end{subequations}
Results for numerators with more factors of $p^\mu$ can be reduced in the same way. Applying these relations, we can eventually reduce the integral in \cref{eqn:Vnsr} to the trivial numerator case with $r=0$, which we can evaluate as follows:
\begin{align}\label{eqn:Vns}
{V_{n,2s}}\left( {{m^2}} \right) &\equiv  - i{\mu ^{4 - d}}\int {\ddp{p}\frac{1}{{{{\left( {{p^2} - {m^2}} \right)}^n}{{\left( {v \cdot p} \right)}^{2s}}}}} \nonumber \\
 &=  - i{\mu ^{4 - d}}\int {\ddp{p}\int_0^\infty  {\D\lambda \frac{{{{\left( {2m} \right)}^{2s}}{\lambda ^{2s - 1}}}}{{{{\left( {{p^2} - {m^2} + 2\lambda mv \cdot p} \right)}^{n + 2s}}}}\frac{{\Gamma \left( {n + 2s} \right)}}{{\Gamma \left( n \right)\Gamma \left( {2s} \right)}}} } \nonumber \\
 &=  - i{\mu ^{4 - d}}{2^{2s - 1}}\frac{{\Gamma \left( {n + 2s} \right)}}{{\Gamma \left( n \right)\Gamma \left( {2s} \right)}}\int {\ddl\int_0^\infty  {\D\Delta \frac{{{\Delta ^{s - 1}}}}{{{{\left( {{\ell^2} - {m^2} - \Delta } \right)}^{n + 2s}}}}} } \nonumber \\
 &=  - i{\mu ^{4 - d}}{2^{2s - 1}}\frac{{\Gamma \left( {n + s} \right)\Gamma \left( s \right)}}{{\Gamma \left( n \right)\Gamma \left( {2s} \right)}}\int {\ddl\frac{{{{\left( { - 1} \right)}^s}}}{{{{\left( {{\ell^2} - {m^2}} \right)}^{n + s}}}}} \nonumber \\
 &= \frac{{{\mu ^{4 - d}}}}{{{{\left( {4\pi } \right)}^{d/2}}}}{\left( { - 1} \right)^n}\frac{{{2^{2s}}\Gamma \left( {1 + s} \right)}}{{\Gamma \left( {1 + 2s} \right)}}\frac{{\Gamma \left( {n + s - d/2} \right)}}{{\Gamma \left( n \right)}}\frac{1}{{{{\left( {{m^2}} \right)}^{n + s - d/2}}}} \,,
\end{align}
where in the second line we combined the propagators using the $\lambda$ version of the Feynman parameters in \cref{eqn:PropagatorCombine}, and then replaced $\lambda \rightarrow 2m\lambda$ for convenience. In the third line, we made the replacements $\ell^\mu\equiv p^\mu+\lambda mv^\mu$ and $\Delta=\lambda^2 m^2$. In the fourth line, we integrated over $\Delta$, and to derive the final result, we used \cref{eq:Inm} to obtain the last line. As a cross check, we see that taking $s=0$ in this expression agrees with \cref{eq:Inm}:
\begin{equation}
{V_{n,0}}\left( {{m^2}} \right) = \frac{{{\mu ^{4 - d}}}}{{{{\left( {4\pi } \right)}^{d/2}}}}{\left( { - 1} \right)^n}\frac{{\Gamma \left( {n - d/2} \right)}}{{\Gamma \left( n \right)}}\frac{1}{{{{\left( {{m^2}} \right)}^{n - d/2}}}} = I_n^0\left( {{m^2}} \right) \,.
\end{equation}
For reference, we provide a non-exhaustive set of explicit formulas with non-zero values for $s$:
\begingroup
\allowdisplaybreaks
\begin{subequations}
\begin{align}
{V_{1,2}}\left( {{m^2}} \right) &= \frac{1}{{{{\left( {4\pi } \right)}^2}}}\left( { - 2} \right){\left( {\frac{{4\pi {\mu ^2}}}{{{m^2}}}} \right)^{2 - d/2}}\Gamma \left( {2 - \frac{d}{2}} \right) \longrightarrow \frac{1}{{{{\left( {4\pi } \right)}^2}}}\ln \frac{{{\mu ^2}}}{{{m^2}}}\left( { - 2} \right) \,, \label{eq:V12Eval} \\
{V_{2,2}}\left( {{m^2}} \right) &= \frac{1}{{{{\left( {4\pi } \right)}^2}}}\frac{1}{{{m^2}}}2 \,, \\
 {V_{3,2}}\left( {{m^2}} \right) &= \frac{1}{{{{\left( {4\pi } \right)}^2}}}\frac{1}{{{m^4}}}\left( { - 1} \right) \,, \\
  {V_{4,2}}\left( {{m^2}} \right) &= \frac{1}{{{{\left( {4\pi } \right)}^2}}}\frac{1}{{{m^6}}}\frac{2}{3} \,, \\
{V_{1,4}}\left( {{m^2}} \right) &= \frac{1}{{{{\left( {4\pi } \right)}^2}}}\frac{1}{{{m^2}}}\frac{{ - 4}}{3} \,, \\
 {V_{2,4}}\left( {{m^2}} \right) &= \frac{1}{{{{\left( {4\pi } \right)}^2}}}\frac{1}{{{m^4}}}\frac{4}{3} \,, \\
 {V_{3,4}}\left( {{m^2}} \right) &= \frac{1}{{{{\left( {4\pi } \right)}^2}}}\frac{1}{{{m^6}}}\frac{{ - 4}}{3} \,,
\end{align}
\end{subequations}
\endgroup
where as before, the arrow in \cref{eq:V12Eval} denotes that we have included the counterterm contributions using the $\overline{\text{MS}}$ scheme.

Combining \cref{eqn:VReplacements} with the results in \cref{eqn:Vns}, one can compute the integrals in \cref{eqn:Vnsr} with $r\ne0$. For example
\begin{subequations}
\begin{align}
V_{n,2s + 1}^\mu \left( {{m^2}} \right) &=  - i{\mu ^{4 - d}}\int {\ddp{p}\frac{{{p^\mu }}}{{{{\left( {{p^2} - {m^2}} \right)}^n}{{\left( {v \cdot p} \right)}^{2s + 1}}}}}  = {v^\mu }{V_{n,2s}}\left( {{m^2}} \right) \,, \\[7pt]
V_{n,2s + 2}^{\mu \nu }\left( {{m^2}} \right) &=  - i{\mu ^{4 - d}}\int {\ddp{p}\frac{{{p^\mu }{p^\nu }}}{{{{\left( {{p^2} - {m^2}} \right)}^n}{{\left( {v \cdot p} \right)}^{2s + 2}}}}} \notag\\
& =  - \frac{1}{{1 + 2s}}\left[ {{\eta ^{\mu \nu }} - 2\left( {1 + s} \right){v^\mu }{v^\nu }} \right]{V_{n,2s}}\left( {{m^2}} \right) \,.
\end{align}
\end{subequations}

\section{Covariant Derivative Expansion}
\label{appsec:CDE}

In this appendix, we give a comprehensive review of the algebraic approach to evaluating functional traces (or determinants) known as the Covariant Derivative Expansion (CDE). A typical object of interest is the one-loop correction to the 1PI effective action
\begin{align}
\Gamma^{(1)}[\phi] = \frac{i}{2} \ln \det \left( { - \frac{{{\delta ^2}S}}{{\delta {\phi ^2}}}} \right) = \Tr \ln \left[ {{D^2} + M^2 + U(\phi)} \right] \,,
\label{eq:Gamma1App}
\end{align}
which derives from a Lagrangian of the form
\begin{equation}
{\cal L}\left( {\phi } \right) = -{\phi ^\dag }\left( {  {D^2} + M^2 } \right)\phi - V(\phi) \,,
\end{equation}
where $\phi$ is a field, $D_\mu$ is a covariant derivative, $V(\phi)$ is the potential encoding additional interactions, and $U = \dd^2 V/\dd \phi^2$ for this explicit example. Here we will present the formalism capable of computing generalizations of \cref{eq:Gamma1App} such as\footnote{For results, see \cref{eqn:T0Tn,eqn:TraceSingle} below.}
\begin{equation}
{T_n}\left(M^2\right) \equiv i\Tr \left[ {\frac{1}{{{D^2} + {M^2}}}\, {U_1} \cdots \frac{1}{{{D^2} + {M^2}}}\, {U_n}} \right] \,,
\end{equation}
where $U_i(x)$ can be any functions of interest. This technique was originally invented in the 1980s~\cite{Gaillard:1985uh, Chan:1986jq, Cheyette:1987qz}, and recently reintroduced in the context of modern EFT calculations in \Ref{Henning:2014wua}, followed by the generalization given in \Ref{Henning:2016lyp}. This technology has been utilized to compute universal one-loop matching results for relativistic theories~\cite{Drozd:2015rsp,Ellis:2016enq,Zhang:2016pja,Ellis:2017jns,Summ:2018oko}.

Our review of the CDE methodology is organized as follows. In \cref{appsubsec:ToolsTricks}, we provide a number of useful tools and tricks for manipulating covariant derivatives and functional traces. In \cref{appsubsec:TraceAlgebraic}, we review three different algebraic approaches to evaluating functional traces, contrasting them against each other, and highlighting the features of two formulations of the CDE. Then in \cref{appsubsec:BackgroundField}, we show how  the background field method can be used to take functional variations with respect to gauge bosons in a way that manifestly preserves gauge covariance. Finally, we provide a compilation of useful results obtained by applying CDE in \cref{appsubsec:CDEApplications}.

\subsection{Tools and Tricks}\label{appsubsec:ToolsTricks}
In this subsection, we spell out many of the manipulations that are required to perform the calculations utilizing the CDE. This has the additional benefit that it allows us the opportunity to define notation and conventions. Particular emphasis is placed on subtleties that can arise.

\subsubsection{Baker-Campbell-Hausdorff Formula}
In deriving the CDE, we will make frequent use of the Baker-Campbell-Hausdorff formula (see \eg \cref{eqn:iDCDEpCDE,eqn:UCDE}):
\begin{equation}
  e^A B e^{-A} = \sum_{n=0}^\infty \frac{1}{n!} A^n [B] \,.
\label{eqn:Baker-Campbell-Hausdorff}
\end{equation}
Here we have introduced a successive commutator notation $A^n [B]$ defined as
\begin{equation}
  A^0[B] \equiv B \qand
  A^n[B] \equiv \comm{A}{A^{n-1}[B]} \qfor n \ge 1 \,.
\end{equation}

\subsubsection{Gauge Coupling}
As is well known, one can redefine the gauge field to move the gauge coupling from appearing within the covariant derivative to the coefficient of the gauge kinetic term. In this paper, we take the normalization such that the covariant derivative is
\begin{equation}
  D_\mu = \partial_\mu - ig G_\mu = \partial _\mu - ig G_\mu^a T^a \,,
\end{equation}
where the generator matrix $T^a$ depends on the representation of the field that it acts on:
\begin{equation}
  D_\mu \phi  = (\partial_\mu - igG_\mu^a T_\phi^a) \phi \,.
\end{equation}
We will mostly suppress the representation subscript $\phi$, as it should be clear from the context. Note that if there are successive covariant derivatives, their representation subscripts must all be the same due to the ``covariant'' nature of the covariant derivative, \ie, it preserves the representation:
\begin{equation}
  D_{\mu_1} \dotsm D_{\mu_n} \phi = (\partial_{\mu_1} - igG_{\mu_1}^{a_1} T_\phi^{a_1}) \dotsm (\partial_{\mu_n} - igG_{\mu_n}^{a_n} T_\phi ^{a_n}) \phi \,.
 \label{eqn:SuccessiveD}
\end{equation}

For convenience, \eg see \cref{eq:LQCDwithGD} below, we define a convenient form of the field strength from the commutator of two covariant derivatives
\begin{equation}
  G_{\mu\nu}^D \equiv \comm{D_\mu}{D_\nu} =  -ig G_{\mu\nu} =  -ig G_{\mu\nu}^a T^a \,,
\label{eq:defGD}
\end{equation}
where as usual the field strength is
\begin{equation}
  G_{\mu\nu}^a = \partial_\mu G_\nu^a - \partial_\nu G_\mu^a + g f^{abc} G_\mu^b G_\nu^c \,.
\end{equation}
Note that if there are multiple gauge sectors, $G_{\mu\nu}^D$ is the sum of the relevant field strengths.

\subsubsection{Functional Traces}
We use the notation ``$\tr$'' to denote a trace over the internal symmetries, and ``$\Tr$'' to denote a trace over both the internal symmetries and the functional space of the operators:
\begin{equation}
  \Tr_R\, (A) = \int \ddx{x} \mel{x}{\tr_R\, (A)}{x} = \int \ddp{p} \mel{p}{\tr_R\, (A)}{p} \,.
\label{eqn:TraceDef}
\end{equation}
Here the subscript $R$ denotes the internal symmetry representation that is being traced over, which we will often suppress when it is clear from the context. Throughout this appendix, we work in $d=4$ dimensional spacetime, except that when we confront divergent integrals we will use dimensional regularization with $d=4-2\eps$.

Another remarkable property is that the functional trace $\Tr$ is invariant under cyclic permutations of operators/functionals, but this is generically not the case for the internal trace $\tr$, because it only traces over a subspace of the functional. However, when all the objects in an internal trace $\tr$ are functions (\textit{i.e.} they are all diagonal in the basis $\ket{x}$, see more elaborations in \cref{appsubsubsec:Territory}), the cyclic permutation rule holds, see \eg \cref{eqn:trzero} below.

\subsubsection{Product Rule for Covariant Derivative}
We highlight an important feature of the covariant derivative $D_\mu$ --- the product rule takes the same form as for the partial derivative when it is acting on fields:
\begin{equation}
  D_\mu\, (A_i\, B_j) = (D_\mu A)_i\, B_j + A_i\, (D_\mu B)_j \,.
\label{eqn:ProductRuleD}
\end{equation}
This can be justified by noting that the direct product of the two fields $A_i \otimes B_j$ forms a tensor representation, for which we have the generator relation
\begin{equation}
  T_{A \otimes B}^a = T_A^a \otimes 1_B + 1_A \otimes T_B^a \,.
\end{equation}
Suppressing the representation subscript, this reads
\begin{equation}
  T^a\, (A_i\, B_j) = (T^a A)_i\, B_j + A_i\, (T^a B)_j \,,
\label{eqn:ProductRuleT}
\end{equation}
and leads to \cref{eqn:ProductRuleD}.

Although it might be clear, we emphasize that \cref{eqn:ProductRuleD} can be contracted with any non-field constants (that do not transform under the symmetry group in consideration). This leads to some familiar results such as
\begin{subequations}
\begin{align}
{D_\mu }\left( {{\phi ^\dag }\phi } \right) &= {D_\mu }\left( {{\delta _{ij}}\phi _i^*{\phi _j}} \right) = \left( {{D_\mu }{\phi ^\dag }} \right)\phi  + {\phi ^\dag }\left( {{D_\mu }\phi } \right) \,, \notag\\
{D_\mu }\left( {\bar \psi \psi } \right) &= {D_\mu }\left( {\psi _i^*\gamma _{ij}^0{\psi _j}} \right) = \left( {{D_\mu }\bar \psi } \right)\psi  + \bar \psi \left( {{D_\mu }\psi } \right) \,, \notag\\
{D_\mu }\left( {{\phi ^\dag }T_\phi ^a\phi } \right) &= {D_\mu }\left( {\phi _i^*T_{\phi ,ij}^a{\phi _j}} \right) = {\left( {{D_\mu }\phi } \right)^\dag }T_\phi ^a\phi  + {\phi ^\dag }T_\phi ^a\left( {{D_\mu }\phi } \right) = {\left( {{\partial _\mu } - igG_\mu ^bT_G^b} \right)^{ac}}\left( {{\phi ^\dag }T_\phi ^c\phi } \right) \,. \notag
\end{align}
\end{subequations}
For the last equation above, one can also check the non-trivial consistency between the last two expressions, where in the final form $D_\mu$ is written out explicitly noting that $\left( \phi^\dag T_\phi ^a \phi \right)$ forms an adjoint representation. A similar but maybe slightly less intuitive consequence is that the covariant derivative can be pulled out of a trace over the internal symmetries:
\begin{equation}
\tr \left( D_\mu A \right) = D_\mu \left[\tr (A)\right] \,.
\label{eqn:DPulltr}
\end{equation}

\subsubsection{Territory of Covariant Derivative}\label{appsubsubsec:Territory}
In what follows, we will be extensively manipulating functionals --- typically referred to as \emph{operators} in quantum mechanics. Note that a special class of functionals are just functions. For example, in quantum mechanics the Hamiltonian $\hat H$ is built out of two functionals, the kinetic term $\hat T = -\nabla^2/(2 m)$ and the potential $\hat V(x)$. This latter functional is of this special kind, in that it is just a function with the property $\hat V(x)|x\rangle = V(x)|x\rangle$.

Because of this, the territory of a covariant derivative $D_\mu$ could be ambiguous. When an object like $D_\mu U$ appears, where $U=U(x)$ is a function, it is not always clear whether this is a function with $D_\mu$ acting \emph{only} on $U(x)$, or a non-trivial functional with $D_\mu$ acting on everything that appears to the right. Usually, one would use a parenthesis (or a bracket of any kind) to remove this ambiguity. However, the calculations to be presented are cumbersome enough that brackets are often used for grouping functionals, as opposed to specifying the territory of $D_\mu$. As an explicit example, let us consider the following two toy Lagrangians
\begin{subequations}
\begin{align}
\mathcal{L}_1 &= \phi^\dag \left( - D^2 - M^2 \right)\phi  - \phi^\dag \left( D^\mu U_\mu^a T^a \right) \phi \,, \\
\mathcal{L}_2 &= \phi^\dag \left( - D^2 - M^2 \right)\phi  - \phi^\dag D^\mu \left( U_\mu^a T^a \phi \right) \,,
\end{align}
\end{subequations}
where we have introduced a vector field $U_\mu^a$ transforming in the adjoint representation. The two interactions are different. In the first Lagrangian, $D^\mu$ acts only on $U_\mu^a$ and the whole thing $\left(D^\mu U_\mu^a\right)$ is a function, which can be treated as a field (an adjoint scalar). But in the second Lagrangian, $D^\mu$ is acting on both $U_\mu^a$ and $\phi$. In the above, we successfully used the parenthesis to distinguish the two scenarios. However, in calculating the 1PI effective action using \cref{eq:Gamma1App}, we will get
\begin{subequations}
\begin{align}
\ln \det \left( - \frac{\delta^2 S_1}{\delta\phi^2} \right) &= \ln\det \left[ D^2 + M^2 + \left( D^\mu U_\mu^a T^a \right) \right] \to \ln\det \left[ D^2 + M^2 + \left( D^\mu U_\mu^a T^a \right)_x \right] \,, \notag\\
\ln \det \left( - \frac{\delta^2 S_2}{\delta\phi^2} \right) &= \ln \det \left( D^2 + M^2 + D^\mu U_\mu^a T^a \right) \,. \notag
\end{align}
\end{subequations}
Note that in the second line, a bracket has been introduced to group the three terms, but we do not mean to ``close'' the territory of $D^\mu$ (\textit{i.e.} to restrict the territory of $D_\mu$ to $U_\mu^a$ only). In fact, this is precisely an example of a term with an ``open'' covariant derivative, a universal evaluation of which we will provide in \cref{eqn:UniversalDUJ} up to mass dimension four. This is in contrast with the case of the first line, where the parenthesis is intended to indicate ``closing'' the territory of $D^\mu$. Needless to say, carefully tracking this distinction is critical to our ability to calculate with functional methods. Practically, when we evaluate functional traces with the CDE, an \emph{open} covariant derivative will get shifted due to $e^{ip\cdot x} iD_\mu e^{-ip\cdot x} = iD_\mu + p_\mu$ (see \cref{eqn:T0naiveCDE}), while a \emph{closed} covariant derivative does not. Therefore, to remove this ambiguity in the meaning of brackets, we put an additional subscript ``$x$'' on the brackets when they are used for specifying the territory of the covariant derivatives. This was explicitly done in the last expression of the first line above.\footnote{When there is no ambiguity in the meaning of brackets, we will drop the explicit subscript $x$ and go back to our usual way of addressing this issue. In particular, for Lagrangian expressions such as \cref{eqn:LQCD} and final results such as \cref{eq:EvalResInt}, the subscript $x$ is often dropped. However, note that we have carefully kept all the subscript $x$ explicit in the results presented in \cref{appsubsubsec:functionaltraces}.}

We emphasize that a covariant derivative $D_\mu = \partial_\mu - igG_\mu^a T^a$ has two parts --- the partial derivative and the gauge fields. Therefore, the meaning of closing its territory by $\left(D_\mu U\right)_x$ is twofold. First, the partial derivative is only acting on $U$; second, the generators $T^a$ are in the representation of $U$:
\begin{equation}
\tcboxmath{
(D_\mu U^i)_x \equiv \bigg[ \left(\partial_\mu U^i\right)\, \mathds{1}(x) - ig\, G_\mu^a \left(T_U^a\right)^{ij} U^j\, \mathds{1}(x)\, \bigg] \,,
}
\label{eqn:defDUix}
\end{equation}
where every functional is projected onto the identity function $\mathds{1}(x)$. Since this notation is non-standard, and the second aspect above can be easily overlooked, we provide a few explicit examples. First, when a $D_\mu$ is followed by a field strength $G_{\rho\sigma}^D$ (acting on a field $\phi(x)$), we have
\begin{subequations}
\begin{align}
 D_\mu G_{\rho\sigma}^D\, \phi &= \left( \partial_\mu - igG_\mu^a T_\phi^a \right) G_{\rho\sigma}^{D, b}\, T_\phi^b\, \phi \,, \\
 \left(D_\mu G_{\rho\sigma}^D\right)_x \phi &= \left\{ \left[ \partial_\mu \delta^{bc} - igG_\mu^a \left(T_G^a\right)^{bc} \right] G_{\rho\sigma}^{D, c} \right\} T_\phi^b\, \phi \,. \label{eqn:DGx}
\end{align}
\end{subequations}
Note the difference in the representations of $T^a$. Second, since the field strength is related to the covariant derivative through $G_{\mu\nu}^D = \comm{D_\mu}{D_\nu}$, it also has a notion of territory. So similar to the case above, for $G_{\mu\nu}^D$ we also have
\begin{subequations}
\begin{align}
 G_{\mu\nu}^D\, G_{\rho\sigma}^D\, \phi &= G_{\mu\nu}^{D, a}\, T_\phi^a\, G_{\rho\sigma}^{D, b}\, T_\phi^b\, \phi \,, \\
 \left(G_{\mu\nu}^D\, G_{\rho\sigma}^D\right)_x \phi &= \left[\, G_{\mu\nu}^{D, a}\, \left(T_G^a\right)^{bc}\, G_{\rho\sigma}^{D, c}\, \right] T_\phi^b\, \phi \,.
\end{align}
\end{subequations}

To further demystify this notation, we can use the product rule in \cref{eqn:ProductRuleD} to derive the following operator/functional equation for a generic functional $A$:
\begin{align}
 D_\mu U A = (D_\mu U)_x\, A + U D_\mu A \,.
\end{align}
This shows that our notation defined in \cref{eqn:defDUix} can be written as a commutator:
\begin{equation}
(D_\mu U)_x  =  D_\mu U - U D_\mu = \comm{D_\mu}{U} \,,
\label{eqn:DUcomm}
\end{equation}
In addition, one can repeatedly use this relation to derive expressions like
\begin{equation}
  (D_\mu D_\nu U)_x = \comm{D_\mu}{(D_\nu U)_x} = \comm{D_\mu}{\comm{D_\nu}{U}} \,.
  \label{eqn:DURepeat}
\end{equation}
This will be extensively used in our CDE derivations (see \eg \cref{eqn:iDCDEpCDE,eqn:UCDE}). Again, because $G_{\mu\nu}^D = \comm{D_\mu}{D_\nu}$, there is a similar expression for the field strength:
\begin{equation}
\left(G_{\mu\nu}^D U\right)_x  = \comm{G_{\mu\nu}^D}{U} \,.
\label{eqn:GUcomm}
\end{equation}
A potentially confusing consequence is
\begin{subequations}
\begin{align}
\tr \left[ \left(G_{\mu\nu}^D U\right)_x \right] &= \tr \comm{G_{\mu\nu}^D}{U} = 0 \,, \label{eqn:trzero} \\
\tr \left(G_{\mu\nu}^D U\right) &\ne 0 \,. \label{eqn:trnonzero}
\end{align}
\end{subequations}
Note that in the first line above, we are allowed to use the cyclic permutation property for the internal trace, because both objects $G_{\mu\nu}^D(x)$ and $U(x)$ are functions. On the other hand, the trace in the second line is generically non-zero. For example, taking $U(x)$ to be another field strength, we have
\begin{equation}
\tr_R \left( G_{\mu\nu}^D G^{D, \mu\nu} \right) = -g^2 G_{\mu\nu}^a G^{b, \mu\nu} \tr_R \left(T^a T^b\right) = -g^2 G_{\mu\nu}^a G^{a, \mu\nu} d_R \,,
\end{equation}
where $d_R$ is defined as $\tr_R\left(T^a T^b\right) = d_R \delta^{ab}$. This term will be used in \cref{appsubsec:TraceAlgebraic} as a benchmark result for contrasting the three algebraic approaches to evaluating functional traces (see in particular \cref{eqn:T0result}). Some additional non-intuitive manipulations are
\begin{subequations}
\begin{align}
  G_{\mu\nu}^D G_{\rho\sigma}^D &= (G_{\mu\nu}^D G_{\rho\sigma}^D)_x + G_{\rho\sigma}^D G_{\mu\nu}^D \,, \\[3pt]
  (D^\mu D^\nu G_{\mu\nu}^D)_x &= \frac{1}{2} (G^{D, \mu\nu} G_{\mu\nu}^D)_x
    = \frac{1}{2} \comm{G^{D, \mu\nu}}{G_{\mu\nu}^D} = 0 \,.
\label{eqn:DDG0}
\end{align}
\end{subequations}

\subsubsection{Integration by Parts for Covariant Derivative}
Another useful property of the covariant derivative is integration by parts:
\begin{equation}
  S \supset \int \ddx{x}\, \tr\bqty{ (D_\mu U)_x } = \int \ddx{x} \Bqty{ D_\mu [\tr(U)] }_x
    = \int \ddx{x} \Bqty{ \partial_\mu [\tr(U)] }_x = 0 \,.
    \label{eqn:DIBP}
\end{equation}
To derive this relation, we first pulled the covariant derivative out of the trace using \cref{eqn:DPulltr}, and then used the fact that $\tr(U)$ must be a group singlet if this term appears in the Lagrangian. \cref{eqn:DIBP} implies
\begin{equation}
  \int\ddx{x}\, \tr \left[ U_1 (D_\mu U_2)_x \right] = \int\ddx{x}\, \tr \left[ -(D_\mu U_1)_x U_2 \right] \,.
\end{equation}
For contrast, we emphasize that integration by parts cannot be used for an open covariant derivative
\begin{equation}
  \int\ddx{x}\, \tr(D_\mu U) \ne 0 .
\end{equation}

\subsection{Algebraic Evaluation of Functional Traces}\label{appsubsec:TraceAlgebraic}
In this subsection, we present three algebraic approaches to evaluating functional traces: the Partial Derivative Expansion (PDE), the Simplified CDE, and the Original CDE. Presenting the PDE approach will highlight the benefits of CDE. Since both versions of CDE are used in the text (matching uses simplified CDE and running uses original CDE), we will present them both in detail here. For concreteness, we will contrast the three approaches by evaluating the simplest functional trace $T_0$ in each framework up to mass dimension four:
\begin{equation}
 T_0(M^2) \equiv i\Tr_R\pqty{\frac{1}{D^2 + M^2}} \supset
   \int \dd^4 x \frac{1}{(4\pi)^2} \frac{1}{M^2} \frac{-1}{12} \tr_R( G_{\mu\nu}^D G^{D,\mu\nu}) \,.
\label{eqn:T0result}
\end{equation}
For notational simplicity, we will suppress the representation subscript $R$, with the understanding that the derivation holds for arbitrary representations. A catalog of results for more involved functional traces are given in \cref{appsubsubsec:functionaltraces}.

\subsubsection{Partial Derivative Expansion}
To begin, we will derive the Partial Derivative Expansion (PDE). This is the brute force approach to evaluating functional traces. The idea is to ignore the covariant derivative structure, \emph{i.e.}, by treating partial derivative and the gauge fields independently, and to perform an expansion in terms of the gauge fields. This yields a large number of terms that are not gauge invariant individually, but that all combine into gauge invariant quantities at the end. In fact, PDE is nothing but the Feynman diagram approach represented algebraically. In particular, since each term generated in the PDE step corresponds to the contribution of a Feynman diagram, these approaches share the disadvantage that they lack manifest gauge invariance, which requires dealing with a large number of terms at intermediate steps. We review this approach to establish that any functional traces can be evaluated purely algebraically, without invoking any of the additional tricks required to derive the CDE. This already demonstrates the benefit of functional methods: calculations are organized into mindless algebraic expansions, which automatically takes care of all the Feynman rules, relative signs, symmetric factors, \textit{etc.} In later subsections, we will derive the CDE approaches, which are dramatically streamlined by comparison.

As a simple example, consider a gauge theory with an associated coupling $g$. This implies there is a covariant derivative:
\begin{align}
  D_\mu &= \partial_\mu - ig G_\mu \,,
\end{align}
so that
\begin{align}
  D^2 + M^2 &= \partial^2 + M^2 - \bqty{ ig(\partial^\mu G_\mu + G_\mu \partial^\mu) + g^2 G_\mu G^\mu } \,.
\end{align}
In order to compute $T_0$ up to mass dimension four, we should expand the argument of the trace up to four powers in $G_\mu$:
\begin{align}
  T_0  \equiv i\Tr\pqty{ \frac{1}{D^2 + M^2} }
      &= -i\Tr\pqty{ \frac{1}{-\partial^2 - M^2
                              + \bqty{ig (\partial^\mu G_\mu + G_\mu \partial^\mu) + g^2 G_\mu G^\mu} } } \notag\\[5pt]
      &\supset T_0^{1G} + T_0^{2G} + T_0^{3G} + T_0^{4G} \,,
\end{align}
with\footnote{Here we have used
\[
  \frac{1}{A+B} = A^{-1} - A^{-1}BA^{-1} + A^{-1}BA^{-1}BA^{-1} - A^{-1}BA^{-1}BA^{-1}BA^{-1} + \dotsb \,.
\]
}
\begingroup
\allowdisplaybreaks
\begin{subequations}
\begin{align}
  T_0^{1G} &\equiv -g \Tr\bqty{ \frac{1}{-\partial^2 - M^2} (\partial^\mu G_\mu + G_\mu \partial^\mu) \frac{1}{-\partial^2 - M^2} } \,, \\[5pt]
  T_0^{2G} &\equiv ig^2 \Tr \Bigg[ \dfrac{1}{{ - {\partial ^2} - {M^2}}}{G_\mu }{G^\mu }\dfrac{1}{{ - {\partial ^2} - {M^2}}} \notag\\[-7pt]
           &\hspace{30pt} + \dfrac{1}{{ - {\partial ^2} - {M^2}}}\left( {{\partial ^\mu }{G_\mu } + {G_\mu }{\partial ^\mu }} \right)\dfrac{1}{{ - {\partial ^2} - {M^2}}}\left( {{\partial ^\nu }{G_\nu } + {G_\nu }{\partial ^\nu }} \right)\dfrac{1}{{ - {\partial ^2} - {M^2}}} \Bigg] \,, \\[5pt]
  T_0^{3G} &\equiv g^3 \Tr \Bigg\{ \dfrac{1}{{ - {\partial ^2} - {M^2}}}{G_\mu }{G^\mu }\dfrac{1}{{ - {\partial ^2} - {M^2}}}\left( {{\partial ^\nu }{G_\nu } + {G_\nu }{\partial ^\nu }} \right)\dfrac{1}{{ - {\partial ^2} - {M^2}}} \notag\\[-3pt]
           &\hspace{50pt} + \dfrac{1}{{ - {\partial ^2} - {M^2}}}\left( {{\partial ^\nu }{G_\nu } + {G_\nu }{\partial ^\nu }} \right)\dfrac{1}{{ - {\partial ^2} - {M^2}}}{G_\mu }{G^\mu }\dfrac{1}{{ - {\partial ^2} - {M^2}}} \notag\\[-3pt]
           &\hspace{50pt} + \Bigg[ \dfrac{1}{{ - {\partial ^2} - {M^2}}}\left( {{\partial ^\mu }{G_\mu } + {G_\mu }{\partial ^\mu }} \right)\dfrac{1}{{ - {\partial ^2} - {M^2}}}\left( {{\partial ^\nu }{G_\nu } + {G_\nu }{\partial ^\nu }} \right) \notag\\[-7pt]
           &\hspace{150pt} \times \dfrac{1}{{ - {\partial ^2} - {M^2}}}\left( {{\partial ^\rho }{G_\rho } + {G_\rho }{\partial ^\rho }} \right)\dfrac{1}{{ - {\partial ^2} - {M^2}}} \Bigg] \Bigg\} \,, \\[5pt]
  T_0^{4G} &\equiv -ig^4 \Tr \Bigg\{ \dfrac{1}{{ - {\partial ^2} - {M^2}}}{G_\mu }{G^\mu }\dfrac{1}{{ - {\partial ^2} - {M^2}}}{G_\nu }{G^\nu }\dfrac{1}{{ - {\partial ^2} - {M^2}}} \notag\\[-3pt]
  &\hspace{50pt} + \Bigg[ \dfrac{1}{{ - {\partial ^2} - {M^2}}}{G_\mu }{G^\mu }\dfrac{1}{{ - {\partial ^2} - {M^2}}}\left( {{\partial ^\nu }{G_\nu } + {G_\nu }{\partial ^\nu }} \right) \notag\\[-7pt]
  &\hspace{120pt} \times\dfrac{1}{{ - {\partial ^2} - {M^2}}}\left( {{\partial ^\rho }{G_\rho } + {G_\rho }{\partial ^\rho }} \right)\dfrac{1}{{ - {\partial ^2} - {M^2}}} \Bigg] \notag\\[-3pt]
  &\hspace{50pt} + \Bigg[ \dfrac{1}{{ - {\partial ^2} - {M^2}}}\left( {{\partial ^\mu }{G_\mu } + {G_\mu }{\partial ^\mu }} \right)\dfrac{1}{{ - {\partial ^2} - {M^2}}}{G_\nu }{G^\nu } \notag\\[-7pt]
  &\hspace{120pt} \times\dfrac{1}{{ - {\partial ^2} - {M^2}}}\left( {{\partial ^\rho }{G_\rho } + {G_\rho }{\partial ^\rho }} \right)\dfrac{1}{{ - {\partial ^2} - {M^2}}} \Bigg] \notag\\[-3pt]
  &\hspace{50pt} + \Bigg[ \dfrac{1}{{ - {\partial ^2} - {M^2}}}\left( {{\partial ^\mu }{G_\mu } + {G_\mu }{\partial ^\mu }} \right)\dfrac{1}{{ - {\partial ^2} - {M^2}}} \notag\\[-7pt]
  &\hspace{120pt} \times\left( {{\partial ^\nu }{G_\nu } + {G_\nu }{\partial ^\nu }} \right)\dfrac{1}{{ - {\partial ^2} - {M^2}}}{G_\rho }{G^\rho }\dfrac{1}{{ - {\partial ^2} - {M^2}}} \Bigg] \notag\\[-3pt]
  &\hspace{50pt} + \Bigg[ \dfrac{1}{{ - {\partial ^2} - {M^2}}}\left( {{\partial ^\mu }{G_\mu } + {G_\mu }{\partial ^\mu }} \right)\dfrac{1}{{ - {\partial ^2} - {M^2}}} \left( {{\partial ^\nu }{G_\nu } + {G_\nu}{\partial^\nu }} \right) \dfrac{1}{{ - {\partial ^2} - {M^2}}} \notag\\[-5pt]
  &\hspace{80pt} \times \left( {{\partial ^\rho }{G_\rho } + {G_\rho }{\partial ^\rho }} \right) \dfrac{1}{{ - {\partial ^2} - {M^2}}}\left( {{\partial ^\sigma }{G_\sigma } + {G_\sigma }{\partial ^\sigma }} \right)\dfrac{1}{{ - {\partial ^2} - {M^2}}} \Bigg] \Bigg\} \,.
\end{align}
\end{subequations}
\endgroup
We see that even for this simplest functional trace $T_0$, the PDE generates many terms:\\

\begin{minipage}{0.9 \textwidth}
\centering
\begin{tabular}{c | c}
Operator & Terms\\
\hline\\[-10pt]
$T_0^{1G}$& 2 \\[4pt]
$T_0^{2G}$&  5 \\[4pt]
$T_0^{3G}$& 12 \\[4pt]
$T_0^{4G}$&  29
\end{tabular}
\end{minipage}
\\

It is straightforward to evaluate each term through the repeated insertion of the functional identity element
\begin{equation}
  1 = \int \ddx{x} \op{x}  = \int \ddp{p} \op{p} \,.
\end{equation}
For concreteness, we will provide some details for calculating part of $T_0^{2G}$. It contains five terms\footnote{Note that the gauge field $G_\mu(x)$ does not commute with the partial derivative, as it depends on the spacetime coordinate $x$. Therefore, the last four terms are generically inequivalent. On the other hand, sometimes one assumes that the field is independent of the spacetime coordinate $x$, \eg when computing the Coleman-Weinberg potential. In this case that the background field is constant, it is a lot easier to evaluate such terms. See \eg Eqs.~(11.71)--(11.73) in Peskin and Schroeder~\cite{Peskin:1995ev}.}
\begingroup
\allowdisplaybreaks
\begin{subequations}
\begin{align}
  T_0^{2G} &= T_0^{2G0} + T_0^{2G1} + T_0^{2G2} + T_0^{2G3} + T_0^{2G4} \,, \\
  T_0^{2G0} &\equiv ig^2 \Tr\pqty{ \frac{1}{-\partial^2 - M^2} G_\mu G^\mu \frac{1}{-\partial^2 - M^2} } \,, \\
  T_0^{2G1} &\equiv ig^2 \Tr\pqty{ \frac{1}{-\partial^2 - M^2} \partial^\mu G_\mu \frac{1}{-\partial^2 - M^2}
    \partial^\nu G_\nu \frac{1}{-\partial^2 - M^2} } \,, \\
  T_0^{2G2} &\equiv ig^2 \Tr\pqty{ \frac{1}{-\partial^2 - M^2} \partial^\mu G_\mu \frac{1}{-\partial^2 - M^2}
    G_\nu \partial^\nu \frac{1}{-\partial^2 - M^2} } \,, \\
  T_0^{2G3} &\equiv ig^2 \Tr\pqty{ \frac{1}{-\partial^2 - M^2} G_\mu \partial^\mu \frac{1}{-\partial^2 - M^2}
    \partial^\nu G_\nu \frac{1}{-\partial^2 - M^2} } \,, \\
  T_0^{2G4} &\equiv ig^2 \Tr\pqty{ \frac{1}{-\partial^2 - M^2} G_\mu \partial^\mu \frac{1}{-\partial^2 - M^2}
    G_\nu \partial^\nu \frac{1}{-\partial^2 - M^2} } \,.
\end{align}
\end{subequations}
\endgroup
In order to demonstrate every detail, we provide the evaluation of $T_0^{2G1}$:
\begingroup
\allowdisplaybreaks
\begin{align}
  T_0^{2G1} &\equiv ig^2 \Tr\pqty{ \dfrac{1}{-\partial^2 - M^2} \partial^\mu G_\mu \dfrac{1}{-\partial^2 - M^2}
      \partial^\nu G_\nu \dfrac{1}{-\partial^2 - M^2} } \notag\\
    &= ig^2 \int \dfp{p_1}
         \tr\mel**{p_1}{\dfrac{1}{-\partial^2 - M^2} \partial^\mu G_\mu \dfrac{1}{-\partial^2 - M^2}
                                 \partial^\nu G_\nu \dfrac{1}{-\partial^2 - M^2}}{p_1} \notag\\
    &= ig^2 \int \dfx{x_1} \dfx{x_2} \int \dfp{p_1} \dfp{p_2} \tr \Bigg[ \left\langle {{p_1}\left| {\dfrac{1}{{ - {\partial ^2} - {M^2}}}{\partial ^\mu }} \right|{x_1}} \right\rangle \left\langle {{x_1}\left| {{G_\mu }} \right|{p_2}} \right\rangle \notag\\[-5pt]
         &\hspace{150pt} \times \left\langle {{p_2}\left| {\dfrac{1}{{ - {\partial ^2} - {M^2}}}{\partial ^\nu }} \right|{x_2}} \right\rangle \left\langle {{x_2}\left| {{G_\nu }\dfrac{1}{{ - {\partial ^2} - {M^2}}}} \right|{p_1}} \right\rangle \Bigg] \notag\\
    &= ig^2 \int \dfx{x_1} \dfx{x_2} \int \dfp{p_1} \dfp{p_2} \tr \Bigg[ \dfrac{ip_1^\mu}{p_1^2 - M^2} G_\mu(x_1) \dfrac{ip_2^\nu}{p_2^2 - M^2} G_\nu(x_2) \notag\\[-5pt]
         &\hspace{200pt} \times \dfrac{1}{p_1^2 - M^2} e^{i(p_1 - p_2)\cdot x_1} e^{i(p_2 - p_1)\cdot x_2} \Bigg] \notag\\
    &= g^2 \int \ddx{x_1} \ddx{x_2} \tr \left[ G_\mu(x_1) G_\nu(x_2) \right] \notag\\
        &\hspace{40pt}\times \int \ddp{k}\, e^{-ik\cdot(x_1 - x_2)}
           \bqty{ -i\mu^{4-d} \int \ddp{p_1}
                  \dfrac{p_1^\mu}{(p_1^2 - M^2)^2} \dfrac{(p_1 + k)^\nu}{(p_1 + k)^2 - M^2} } \,.
 \label{eq:T02G1intermediate}
\end{align}
\endgroup
To evaluate this expression we used the definition of functional trace in \cref{eqn:TraceDef} to derive the second line, followed by insertions of the unit operator to derive the third line. In the fourth line, we extracted the eigenvalues, and used the fact that $\ip{x}{p} = e^{-ip\cdot x}$. In the last line, we make a convenient change of variables $p_2 = p_1+k$. We are then left with a ``loop integral'' to evaluate. This integral is divergent, so we use dim.\ reg.\ (which explains the appearance of the $\mu^{4-d}$ factor):
\begingroup
\allowdisplaybreaks
\begin{align}
  &-i\mu^{4-d} \int \ddp{p_1}
                \frac{p_1^\mu}{(p_1^2 - M^2)^2} \frac{(p_1 + k)^\nu}{(p_1 + k)^2 - M^2}\notag \\
    &\hspace{40pt} = -i\mu^{4-d} \int_0^1 \dd{x} 2x \int \ddl
        \bqty{ \frac{\eta^{\mu\nu}\ell^2}{d(\ell^2 - \Delta)^3} - \frac{x(1-x) k^\mu k^\nu}{(\ell^2 - \Delta)^3} } \notag\\
    &\hspace{40pt} = \int_0^1 \dd{x} 2x \bqty{ \eta^{\mu\nu} \frac{1}{(4\pi)^2} \frac{1}{4}
         \pqty{\frac{4\pi\mu^2}{\Delta}}^{2 - \frac{d}{2}} \Gamma\pqty{2 - \frac{d}{2}}
         + x(1-x) \frac{1}{(4\pi)^2} \frac{1}{2} \frac{k^\mu k^\nu}{\Delta} } \notag\\
    &\hspace{40pt} \supset \frac{1}{(4\pi)^2} \frac{1}{2} \int_0^1 \dd{x} x \bqty{ \eta^{\mu\nu} \ln\frac{\mu^2}{M^2}
               + x(1-x) \frac{\eta^{\mu\nu} k^2 + 2k^\mu k^\nu}{M^2}}\notag \\
    &\hspace{40pt} = \frac{1}{(4\pi)^2} \pqty{ \frac{1}{4} \eta^{\mu\nu} \ln\frac{\mu^2}{M^2}
                                 + \frac{1}{24} \frac{\eta^{\mu\nu}{k^2} + 2 k^\mu k^\nu}{M^2} } \,.
\end{align}
\endgroup
where we used \cref{eqn:PropagatorCombine} to combine denominators, substituted $\ell^\mu \equiv p_1^\mu+(1-x)k^\mu$ and $\Delta\equiv M^2-x(1-x)k^2$, and used \cref{eq:Inm} to evaluate the integral. In the fourth line, we have expanded in the external momentum $k^\mu$ up to quadratic order, which is sufficient to capture all effective operators with mass dimension four. Plugging the integrated result into \cref{eq:T02G1intermediate} yields
\begingroup
\allowdisplaybreaks
\begin{align}
T_0^{2G1} &= {g^2}\int {\D^4}{x_1}{\D^4}{x_2} \tr\left[ {{G_\mu }\left( {{x_1}} \right){G_\nu }\left( {{x_2}} \right)} \right]\notag\\
&\hspace{40pt}\times\int {\frac{{{\D^4}k}}{{{{\left( {2\s\pi } \right)}^4}}}\,{e^{ - ik\left( {{x_1} - {x_2}} \right)}}\frac{1}{{{{\left( {4\pi } \right)}^2}}}\left( {\frac{1}{4}{\eta ^{\mu \nu }}\ln \frac{{{\mu ^2}}}{{{M^2}}} + \frac{1}{{24}}\frac{{{\eta ^{\mu \nu }}{k^2} + 2{k^\mu }{k^\nu }}}{{{M^2}}}} \right)}  \nonumber \\
 &= \int {\D^4}{x_1}{\D^4}{x_2} \tr \left[ {{G_\mu }\left( {{x_1}} \right){G_\nu }\left( {{x_2}} \right)} \right]\notag\\
 &\hspace{40pt}\times\frac{1}{{{{\left( {4\pi } \right)}^2}}}\left[ {\ln \frac{{{\mu ^2}}}{{{M^2}}}\frac{{{g^2}}}{4}{\eta ^{\mu \nu }} + \frac{1}{{{M^2}}}\frac{{ - {g^2}}}{{24}}\left( {{\eta ^{\mu \nu }}\partial _1^2 + 2\partial _1^\mu \partial _1^\nu } \right)} \right]\delta^4 \left( {{x_1} - {x_2}} \right) \nonumber \\
 &= \int {{\D^4}x\,\frac{1}{{{{\left( {4\pi } \right)}^2}}}\left\{ {\ln \frac{{{\mu ^2}}}{{{M^2}}}\frac{{{g^2}}}{4} \tr\left( {{G_\mu }{G^\mu }} \right) + \frac{1}{{{M^2}}}\frac{{ - {g^2}}}{{24}} \tr\left[ {{G_\mu }\left( {{\eta ^{\mu \nu }}{\partial ^2} + 2{\partial ^\mu }{\partial ^\nu }} \right){G_\nu }} \right]} \right\}} \,.
\end{align}
\endgroup
Note that $k$ has become a derivative starting with the second equality. Unsurprisingly, this object is not gauge invariant since this is only part of $T_0$. Deriving a gauge invariant final result is therefore one of the consistency checks of the calculation.

All the five terms in $T_0^{2G}$ can be evaluated in the same way:
\begingroup
\allowdisplaybreaks
\begin{subequations}
\begin{align}
\hspace{-8pt} T_0^{2G0} &\supset \int {{\D^4}x\,\frac{1}{{{{\left( {4\pi } \right)}^2}}}\ln \frac{{{\mu ^2}}}{{{M^2}}}\left( { - {g^2}} \right) \tr \left( {{G_\mu }{G^\mu }} \right)} \,, \\
\hspace{-8pt} T_0^{2G1} &\supset \int {{\D^4}x\,\frac{1}{{{{\left( {4\pi } \right)}^2}}}\left\{ {\ln \frac{{{\mu ^2}}}{{{M^2}}}\frac{{{g^2}}}{4} \tr \left( {{G_\mu }{G^\mu }} \right) + \frac{1}{{{M^2}}}\frac{{ - {g^2}}}{{24}} \tr \left[ {{G_\mu }\left( {{\eta ^{\mu \nu }}{\partial ^2} + 2{\partial ^\mu }{\partial ^\nu }} \right){G_\nu }} \right]} \right\}} \,, \\
\hspace{-8pt} T_0^{2G2} &\supset \int {{\D^4}x\,\frac{1}{{{{\left( {4\pi } \right)}^2}}}\left\{ {\ln \frac{{{\mu ^2}}}{{{M^2}}}\frac{{{g^2}}}{4} \tr \left( {{G_\mu }{G^\mu }} \right) + \frac{1}{{{M^2}}}\frac{{ - {g^2}}}{{24}} \tr \left[ {{G_\mu }\left( {{\eta ^{\mu \nu }}{\partial ^2} - 2{\partial ^\mu }{\partial ^\nu }} \right){G_\nu }} \right]} \right\}} \,, \\
\hspace{-8pt} T_0^{2G3} &\supset \int {{\D^4}x\,\frac{1}{{{{\left( {4\pi } \right)}^2}}}\left\{ {\ln \frac{{{\mu ^2}}}{{{M^2}}}\frac{{{g^2}}}{4} \tr \left( {{G_\mu }{G^\mu }} \right) + \frac{1}{{{M^2}}}\frac{{ - {g^2}}}{{24}} \tr \left[ {{G_\mu }\left( {{\eta ^{\mu \nu }}{\partial ^2} - 6{\partial ^\mu }{\partial ^\nu }} \right){G_\nu }} \right]} \right\}} \,, \\
\hspace{-8pt} T_0^{2G4} &\supset \int {{\D^4}x\,\frac{1}{{{{\left( {4\pi } \right)}^2}}}\left\{ {\ln \frac{{{\mu ^2}}}{{{M^2}}}\frac{{{g^2}}}{4} \tr \left( {{G_\mu }{G^\mu }} \right) + \frac{1}{{{M^2}}}\frac{{ - {g^2}}}{{24}} \tr \left[ {{G_\mu }\left( {{\eta ^{\mu \nu }}{\partial ^2} + 2{\partial ^\mu }{\partial ^\nu }} \right){G_\nu }} \right]} \right\}} \,,
\end{align}
\end{subequations}
\endgroup
which gives
\begin{align}
T_0^{2G} &= T_0^{2G0} + T_0^{2G1} + T_0^{2G2} + T_0^{2G3} + T_0^{2G4} \nonumber \\
& \supset \int {{\D^4}x\,\frac{1}{{{{\left( {4\pi } \right)}^2}}}\frac{1}{{{M^2}}}\frac{{ - {g^2}}}{6} \tr \left[ {{G_\mu }\left( {{\eta ^{\mu \nu }}{\partial ^2} - {\partial ^\mu }{\partial ^\nu }} \right){G_\nu }} \right]} \,.
\end{align}
Note that $T_0^{2G}$ is still not gauge invariant. We need terms with higher powers of the gauge fields $T_0^{3G}$ and $T_0^{4G}$, which can be evaluated using the same procedure. A quite tedious evaluation yields
\begin{subequations}
\begin{align}
T_0^{1G} &= 0 \,, \\
T_0^{2G} &\supset \int {{\D^4}x\,\frac{1}{{{{\left( {4\pi } \right)}^2}}}\frac{1}{{{M^2}}}\frac{{ - {g^2}}}{6} \tr \left[ {{G_\mu }\left( {{\eta ^{\mu \nu }}{\partial ^2} - {\partial ^\mu }{\partial ^\nu }} \right){G_\nu }} \right]} \,, \\
T_0^{3G} &\supset \int {{\D^4}x\,\frac{1}{{{{\left( {4\pi } \right)}^2}}}\frac{1}{{{M^2}}}\frac{{ - i{g^3}}}{3} \tr \left[ {\left( {{\partial ^\mu }{G^\nu }} \right)\left( {{G_\mu }{G_\nu } - {G_\nu }{G_\mu }} \right)} \right]} \,, \\
T_0^{4G} &\supset \int {{\D^4}x\,\frac{1}{{{{\left( {4\pi } \right)}^2}}}\frac{1}{{{M^2}}}\frac{{ - {g^4}}}{6} \tr \left[ {\left( {{G_\mu }{G_\nu } - {G_\nu }{G_\mu }} \right){G^\mu }{G^\nu }} \right]} \,.
\end{align}
\end{subequations}
Putting everything together, we obtain the final gauge invariant result in Eq.~\eqref{eqn:T0result}
\begin{align}
{T_0} &\supset T_0^{1G} + T_0^{2G} + T_0^{3G} + T_0^{4G} \nonumber \\
 &\supset \int {\D^4}x\,\dfrac{1}{{{{\left( {4\pi } \right)}^2}}}\dfrac{1}{{{M^2}}} \bigg\{ - \dfrac{{{g^2}}}{6} \tr \left[ {{G_\mu }\left( {{\partial ^2}{\eta ^{\mu \nu }} - {\partial ^\mu }{\partial ^\nu }} \right){G_\nu }} \right] - \dfrac{{i{g^3}}}{3} \tr \left[ {\left( {{G_\mu }{G_\nu } - {G_\nu }{G_\mu }} \right)\left( {{\partial ^\mu }{G^\nu }} \right)} \right] \notag\\[-3pt]
 &\hspace{100pt} - \dfrac{{{g^4}}}{6} \tr \left[ {\left( {{G_\mu }{G_\nu } - {G_\nu }{G_\mu }} \right){G^\mu }{G^\nu }} \right] \bigg\} \nonumber \\
 &= \int {{\D^4}x\,\dfrac{1}{{{{\left( {4\pi } \right)}^2}}}\dfrac{1}{{{M^2}}}\dfrac{{{g^2}}}{{12}} \tr \left( {{G_{\mu \nu }}{G^{\mu \nu }}} \right)} \nonumber \\
 &= \int {{\D^4}x\,\dfrac{1}{{{{\left( {4\pi } \right)}^2}}}\dfrac{1}{{{M^2}}}\dfrac{{ - 1}}{{12}} \tr \left( {G_{\mu \nu }^D{G^{D,\mu \nu }}} \right)} \,,
 \label{eq:T0PDEfinal}
\end{align}
where we used \cref{eq:defGD} in the last line.

This completes our example for using the PDE to evaluate a functional trace. It should now be clear that this approach is very tedious, due to the many terms that are generated at intermediate steps. This can be traced back to the fact that we split the covariant derivative into a partial derivative and the gauge fields to perform the calculation, so gauge invariance appeared broken until all the pieces were finally assembled. Seeing that all the terms ultimately combined together to yield a very simple gauge invariant answer motivates finding a method for evaluating the functional traces without splitting the covariant derivative --- this is the covariant derivative expansion.

\subsubsection{Simplified CDE}\label{appsubsubsec:naiveCDE}
Having motivated the desire for a covariant approach to evaluating functional traces, we now turn to the most straightforward implementation of a covariant derivative expansion, which we call ``simplified CDE.'' This is the CDE method introduced in \cite{Henning:2016lyp}, which is used to derive the matching results in the main text of this paper. This will also set the stage for a modified approach --- ``original CDE'' to be presented in the next subsection.

Again, we evaluate $T_0$ as a concrete example. The CDE relies on manipulations like the following:
\begin{align}
{T_0}\left(M^2\right) &\equiv i\Tr\left( {\frac{1}{{{D^2} + {M^2}}}} \right) = i\int {\ddx{x} \int {\ddp{p}\left\langle {p}
 \mathrel{\left | {\vphantom {p x}}
 \right. \kern-\nulldelimiterspace}
 {x} \right\rangle \left\langle {x\left| {\tr\left( {\frac{1}{{{D^2} + {M^2}}}} \right)} \right|p} \right\rangle } } \notag\\
 &=  - i\int {\ddx{x}\int {\ddp{p}{e^{ip \cdot x}}\tr\left[ {\frac{1}{{{{\left( {iD} \right)}^2} - {M^2}}}} \right]{e^{ -ip\cdot x}}} }  \notag\\
 &=  - i\int {\ddx{x}\int {\ddp{p}\tr\left[ {\frac{1}{{{{\left( {iD + p} \right)}^2} - {M^2}}}} \right]} } \notag\\
 &=  - i\int {\ddx{x}\int {\ddp{p}\tr\left[ {\frac{1}{{{{\left( {iD - p} \right)}^2} - {M^2}}}} \right]} } \,. \label{eqn:T0naiveCDE}
\end{align}
For this derivation, the key step is going from the first line to the second line, which relies on the fact that $\left\langle x\left| f\left({\hat x}^\mu, {\hat p}^\nu\right) \right| p \right\rangle = f \left(x^\mu, i\partial^\nu\right) \left\langle x | p \right\rangle$. After this step, the integrand in the second line is a function of $x$ and $p$, and the partial derivative in $D_\mu=\partial_\mu - igG_\mu^a\left(x\right) T^a$ is interpreted as an operator that acts on any function of $x$ to its right. In this example, this function of $x$ is simply $e^{-ip\cdot x}$. We then use the fact $e^{ip\cdot x} iD_\mu e^{-ip\cdot x} = iD_\mu + p_\mu$ to obtain the third line. Then in the last line, we have flipped the sign of the loop momentum for future convenience. Again we emphasize that the integrands in lines three and four are no more mysterious than simply being functions of $x$ and $p$; the territory of the partial derivative $\partial_\mu$ (contained in $D_\mu$) is \emph{closed} by the square bracket, to the end of which there is an implicit identity function $\mathds{1}(x)$. However, this is not the same as the action implied by our notation defined in \cref{eqn:defDUix}, because we are not closing the territory of the generator $T^a$ contained in the covariant derivative $D_\mu$ --- its representation is determined by the representation $R$ (which is suppressed in the expressions above) of the trace $\tr$ being evaluated.

Next, one can express the integrand as an expansion in the covariant derivative $D_\mu$:
\begin{equation}
\frac{1}{{{{\left( {iD - p} \right)}^2} - {M^2}}} = \frac{1}{{p^2 - M^2 - \left[ {2p\cdot iD - {{\left(iD\right)}^2}} \right]}} = \sum\limits_{n = 0}^\infty  {\frac{1}{{{{\left( {{p^2} - {M^2}} \right)}^{n+1}}}}{{\left[ {2p\cdot iD - {{\left(iD\right)}^2}} \right]}^n}} \,. \notag
\end{equation}
Truncating \cref{eqn:T0naiveCDE} up to the fourth power of $D_\mu$, we obtain
\begingroup
\allowdisplaybreaks
\begin{align}
{T_0} &\supset - i\int\ddx{x} \int\ddp{p} \tr \left[ \dfrac{1}{{{p^2} - {M^2}}} + \dfrac{{2i{p^{{\alpha _1}}}}}{{{{\left( {{p^2} - {M^2}} \right)}^2}}}{D_{{\alpha _1}}} + \dfrac{1}{{{{\left( {{p^2} - {M^2}} \right)}^2}}}{D^2} \right. \notag\\
                &\hspace{120pt} - \dfrac{{4{p^{{\alpha _1}}}{p^{{\alpha _2}}}}}{{{{\left( {{p^2} - {M^2}} \right)}^3}}}{D_{{\alpha _1}}}{D_{{\alpha _2}}} + \dfrac{{2i{p^{{\alpha _1}}}}}{{{{\left( {{p^2} - {M^2}} \right)}^3}}}\left( {{D_{{\alpha _1}}}{D^2} + {D^2}{D_{{\alpha _1}}}} \right) \notag\\
                &\hspace{120pt} - \dfrac{{8i{p^{{\alpha _1}}}{p^{{\alpha _2}}}{p^{{\alpha _3}}}}}{{{{\left( {{p^2} - {M^2}} \right)}^4}}}{D_{{\alpha _1}}}{D_{{\alpha _2}}}{D_{{\alpha _3}}} + \dfrac{1}{{{{\left( {{p^2} - {M^2}} \right)}^3}}}{D^4} \notag\\
                &\hspace{120pt} - \dfrac{{4{p^{{\alpha _1}}}{p^{{\alpha _2}}}}}{{{{\left( {{p^2} - {M^2}} \right)}^4}}}\left( {{D_{{\alpha _1}}}{D_{{\alpha _2}}}{D^2} + {D_{{\alpha _1}}}{D^2}{D_{{\alpha _2}}} + {D^2}{D_{{\alpha _1}}}{D_{{\alpha _2}}}} \right) \notag\\[-3pt]
                &\hspace{120pt} \left. + \dfrac{{16{p^{{\alpha _1}}}{p^{{\alpha _2}}}{p^{{\alpha _3}}}{p^{{\alpha _4}}}}}{{{{\left( {{p^2} - {M^2}} \right)}^5}}}{D_{{\alpha _1}}}{D_{{\alpha _2}}}{D_{{\alpha _3}}}{D_{{\alpha _4}}} \right]  \notag\\[5pt]
      &\supset - i\int\ddx{x} \int\ddp{p} \tr \left[ \dfrac{1}{{{{\left( {{p^2} - {M^2}} \right)}^2}}}{D^2} - \dfrac{{4{p^{{\alpha _1}}}{p^{{\alpha _2}}}}}{{{{\left( {{p^2} - {M^2}} \right)}^3}}}{D_{{\alpha _1}}}{D_{{\alpha _2}}} + \dfrac{1}{{{{\left( {{p^2} - {M^2}} \right)}^3}}}{D^4} \right. \notag\\
                &\hspace{120pt} - \dfrac{{4{p^{{\alpha _1}}}{p^{{\alpha _2}}}}}{{{{\left( {{p^2} - {M^2}} \right)}^4}}}\left( {{D_{{\alpha _1}}}{D_{{\alpha _2}}}{D^2} + {D_{{\alpha _1}}}{D^2}{D_{{\alpha _2}}} + {D^2}{D_{{\alpha _1}}}{D_{{\alpha _2}}}} \right) \notag\\
                &\hspace{120pt} \left. + \dfrac{{16{p^{{\alpha _1}}}{p^{{\alpha _2}}}{p^{{\alpha _3}}}{p^{{\alpha _4}}}}}{{{{\left( {{p^2} - {M^2}} \right)}^5}}}{D_{{\alpha _1}}}{D_{{\alpha _2}}}{D_{{\alpha _3}}}{D_{{\alpha _4}}} \right] \,. \label{eqn:DString}
\end{align}
\endgroup
Note that the string of covariant derivatives, \eg $D_{\alpha_1} D_{\alpha_2} D_{\alpha_3} D_{\alpha_4}$, are acting on the identity, which projects out the gauge fields
\begin{equation}
\left( {i{D_{\alpha_4} }} \right)\mathds{1}_x = \left( {i{\partial_{\alpha_4} }} \right)\mathds{1}_x + gG_{\alpha_4} ^a\left( x \right){T^a} \mathds{1}_x = gG_{\alpha_4} ^a\left( x \right){T^a} \,, \label{eqn:CDEbreak}
\end{equation}
thereby breaking the manifest gauge covariance. If one were to carry this out for every term in the first line of \cref{eqn:DString}, the resulting manipulations would simply revert to the PDE. Instead, the CDE approach treats these terms abstractly as $D_\mu$, and manipulate the covariant derivatives directly. This yields significant organizational improvements. Not only does it group many terms together, it also allows us to drop quite a few terms before evaluating the loop integral (another way of seeing that they would all conspire to cancel). Take for example the second line of \cref{eqn:DString}, where we have dropped terms that do not involve any covariant derivatives, as well as terms that are odd in the loop momentum $p$ since these vanish once the loop integral is carried out. Additionally, simplifying the numerator using \cref{eqn:NumeratorReduction}, we can further reduce the remaining seven terms down to only four:
\begin{align}
{T_0} &\supset - i\int\ddx{x} \int\ddp{p} \tr \left\{ \left[ {\dfrac{1}{{{{\left( {{p^2} - {M^2}} \right)}^2}}} - \dfrac{{4{p^2}}}{{d{{\left( {{p^2} - {M^2}} \right)}^3}}}} \right]{D^2} \right. \notag\\
                &\hspace{120pt} + \left[ {\dfrac{1}{{{{\left( {{p^2} - {M^2}} \right)}^3}}} - \dfrac{{8{p^2}}}{{d{{\left( {{p^2} - {M^2}} \right)}^4}}} + \dfrac{{16{p^4}}}{{d\left( {d + 2} \right){{\left( {{p^2} - {M^2}} \right)}^5}}}} \right]{D^4} \notag\\
                &\hspace{120pt} - \left[ {\dfrac{{4{p^2}}}{{d{{\left( {{p^2} - {M^2}} \right)}^4}}} - \dfrac{{16{p^4}}}{{d\left( {d + 2} \right){{\left( {{p^2} - {M^2}} \right)}^5}}}} \right]{D_\mu }{D^2}{D^\mu } \notag\\
                &\hspace{120pt} \left. + \dfrac{{16{p^4}}}{{d\left( {d + 2} \right){{\left( {{p^2} - {M^2}} \right)}^5}}}{D_\mu }{D_\nu }{D^\mu }{D^\nu} \right\}  \,.
\end{align}
Rewriting the last term using
\begin{equation}
{D_\mu }{D_\nu }{D^\mu }{D^\nu } = {D_\mu }{D_\nu }\left[ {{D^\mu },{D^\nu }} \right] + {D_\mu }{D^2}{D^\mu } = \frac{1}{2}G_{\mu \nu }^D{G^{D,\mu \nu }} + {D_\mu }{D^2}{D^\mu } \,,
\end{equation}
we get
\begin{align}
{T_0} &\supset - i\int\ddx{x} \int\ddp{p} \tr \left\{ \left[ {\dfrac{1}{{{{\left( {{p^2} - {M^2}} \right)}^2}}} - \dfrac{{4{p^2}}}{{d{{\left( {{p^2} - {M^2}} \right)}^3}}}} \right]{D^2} \right. \notag\\
                &\hspace{120pt} + \left[ {\dfrac{1}{{{{\left( {{p^2} - {M^2}} \right)}^3}}} - \dfrac{{8{p^2}}}{{d{{\left( {{p^2} - {M^2}} \right)}^4}}} + \dfrac{{16{p^4}}}{{d\left( {d + 2} \right){{\left( {{p^2} - {M^2}} \right)}^5}}}} \right]{D^4} \notag\\
                &\hspace{120pt} - \left[ {\dfrac{{4{p^2}}}{{d{{\left( {{p^2} - {M^2}} \right)}^4}}} - \dfrac{{32{p^4}}}{{d\left( {d + 2} \right){{\left( {{p^2} - {M^2}} \right)}^5}}}} \right]{D_\mu }{D^2}{D^\mu } \notag\\
                &\hspace{120pt} \left. + \dfrac{{8{p^4}}}{{d\left( {d + 2} \right){{\left( {{p^2} - {M^2}} \right)}^5}}}\, G_{\mu \nu }^D{G^{D,\mu \nu }} \right\}  \,.
 \label{eqn:T0FourTerms}
\end{align}
Now we are ready to evaluate the loop integrals. Note that the $D^2$, $D^4$, and $D_\mu D^2 D^\mu$ terms do not yield gauge invariant objects after acting on the identity function $\mathds{1}(x)$, so their coefficients must vanish. We can see this magic happen explicitly by evaluating
\begingroup
\allowdisplaybreaks
\begin{subequations}\label{eqn:vanishingloopintegral}
\begin{align}
& \int \ddp{p} \left[ \frac{1}{{{{\left( {{p^2} - {M^2}} \right)}^2}}} - \frac{{4{p^2}}}{{d{{\left( {{p^2} - {M^2}} \right)}^3}}} \right] \notag\\[3pt]
&\hspace{40pt} = I_2^0 - \frac{4}{d}I_3^1 = 0 \,, \\[8pt]
& \int \ddp{p} \left[ \frac{1}{{{{\left( {{p^2} - {M^2}} \right)}^3}}} - \frac{{8{p^2}}}{{d{{\left( {{p^2} - {M^2}} \right)}^4}}} + \frac{{16{p^4}}}{{d\left( {d + 2} \right){{\left( {{p^2} - {M^2}} \right)}^5}}} \right] \notag\\[3pt]
&\hspace{40pt} = I_3^0 - \frac{8}{d}I_4^1 + \frac{{16}}{{d\left( {d + 2} \right)}}I_5^2 = 0 \,, \\[8pt]
& \int \ddp{p} \left[ \frac{{4{p^2}}}{{d{{\left( {{p^2} - {M^2}} \right)}^4}}} - \frac{{32{p^4}}}{{d\left( {d + 2} \right){{\left( {{p^2} - {M^2}} \right)}^5}}} \right] \notag\\[3pt]
&\hspace{40pt} = \frac{4}{d}I_4^1 - \frac{{32}}{{d\left( {d + 2} \right)}}I_5^2 = 0 \,.
\end{align}
\end{subequations}
\endgroup
Here we have used the loop integral notation defined in \cref{eq:Inm} and integrated results given in \cref{appsubsec:EssentialIntegrals}. So the only term in \cref{eqn:T0FourTerms} that survives is the last line, which gives
\begin{equation}
{T_0} \supset \int \dd^4 x \tr \left( {G_{\mu \nu }^D{G^{D,\mu \nu }}} \right) \frac{1}{3} I_5^2 \left(M^2\right) = \int {\dd^4}x\, \frac{1}{{{{\left( {4\pi } \right)}^2}}} \frac{1}{{{M^2}}}\frac{{ - 1}}{{12}} \tr \left( {G_{\mu \nu }^D{G^{D,\mu \nu }}} \right) \,.
\end{equation}
This reproduces \cref{eqn:T0result}.

This shows how to perform calculation using the most naive implementation of a CDE. The benefit of this approach is that it is very straightforward. In particular, if we are not interested in operators that involve a field strength, we can discard many terms at intermediate steps, which dramatically simplifies the calculations. The downside is that it does generate terms that are not manifestly gauge invariant, but these do not ultimately contribute. As we saw, checking that the coefficients are zero requires tracking cancellations among various loop integrals, \textit{e.g.} \cref{eqn:vanishingloopintegral}. It would be desirable if this could be avoided, which is the purpose of the original CDE method presented next.

\subsubsection{Original CDE}\label{appsubsubsec:elegantCDE}
Having worked through the PDE and simplified CDE approaches, we will now explain the ``original CDE'' proposal for performing functional traces that was developed in \cite{Gaillard:1985uh,Chan:1986jq,Cheyette:1987qz} and recently reviewed in~\cite{Henning:2014wua}. The simplified CDE generates terms that are not manifestly gauge invariant, which can be shown to not contribute. One can understand the origin of this attribute by studying the origin of the loop integral cancellations in \cref{eqn:vanishingloopintegral}. We notice that integration by parts lets us organize the integrand into a total derivative of the loop momentum. For example,
\begin{equation}
\frac{1}{{{{\left( {{p^2} - {M^2}} \right)}^2}}} - \frac{{4{p^2}}}{{d{{\left( {{p^2} - {M^2}} \right)}^3}}} = \frac{\partial }{{\partial {p^\mu }}}\left[ {\frac{{{p^\mu }}}{{d{{\left( {{p^2} - {M^2}} \right)}^2}}}} \right] \,,
\end{equation}
which is a surface term that integrates to zero. This is why the non-gauge-invariant terms in the naive CDE approach do not contribute. This CDE approach discussed next is formulated to ensure that these contributions do not appear at intermediate steps, thereby further streamlining functional calculations. Practically, it is more efficient for extracting the matching coefficients at higher mass dimensions, where multiple insertions of the field strength could appear in many of the terms.

We start by performing the same steps that led to \cref{eqn:T0naiveCDE}, followed by the pair of operator insertions that are marked in red:
\begin{align}
{T_0}\left(M^2\right) &\equiv i \Tr \left( {\frac{1}{{{D^2} + {M^2}}}} \right) =  - i\int\ddx{x} \int\ddp{p} \tr \left[ {\frac{1}{{{{\left( {iD - p} \right)}^2} - {M^2}}}} \right]  \notag\\
 &= - i\int\ddx{x} \int\ddp{p} \tr \left[ \textcolor{red}{e^{iD\cdot\frac{\partial}{\partial p}}} \frac{1}{{{{\left( {iD - p} \right)}^2} - {M^2}}} \textcolor{red}{e^{-iD\cdot\frac{\partial}{\partial p}}} \right] \,.
\label{eq:T0CDEbegin}
\end{align}
These expressions are equal. The factor inserted on the right $e^{-iD\cdot\frac{\partial}{\partial p}}$ acts on the constant unit function $\mathds{1}_p$ thereby evaluating to the identity. The factor on the left $e^{iD\cdot\frac{\partial}{\partial p}}$ also evaluates to the identity since Taylor expanding it only generates total derivative terms in $p$ that vanish upon integration. As we will see, sandwiching our integrand between these two factors will eliminate all the non-gauge-invariant terms that appeared when using the simplified CDE.

To understand how this works in detail, we note that the operator being sandwiched is a function of the combination $f\left( iD_\mu - p_\mu \right)$. As the two inserted factors are inverse of each other, they can be brought inside this function such that they act directly on the argument:
\begin{equation}
{e^{iD \cdot \frac{\partial }{{\partial p}}}}f\left( {i{D_\mu } - {p_\mu }} \right){e^{ - iD \cdot \frac{\partial }{{\partial p}}}} = f\left[ {{e^{iD \cdot \frac{\partial }{{\partial p}}}}\left( {i{D_\mu } - {p_\mu }} \right){e^{ - iD \cdot \frac{\partial }{{\partial p}}}}} \right] \,.
\end{equation}
Next, we manipulate this into a more useful form using the Baker-Campbell-Hausdorff formula given in \cref{eqn:Baker-Campbell-Hausdorff}:
\begin{subequations}\label{eqn:iDCDEpCDE}
\begin{align}
{e^{iD \cdot \frac{\partial }{{\partial p}}}}\left( {i{D_\mu }} \right){e^{ - iD \cdot \frac{\partial }{{\partial p}}}} &= \sum\limits_{n = 0}^\infty  {\frac{1}{{n!}}{{\left( {iD \cdot \frac{\partial }{{\partial p}}} \right)}^n}\left[ {i{D_\mu }} \right]} \notag\\
 &= i{D_\mu } + \sum\limits_{n = 0}^\infty  {\frac{1}{{\left( {n + 1} \right)!}}{{\left( {iD \cdot \partial } \right)}^n}\left[ {i{D_\nu }{\partial ^\nu },i{D_\mu }} \right]} \notag\\
 &= i{D_\mu } + \sum\limits_{n = 0}^\infty  {\frac{{{i^n}}}{{\left( {n + 1} \right)!}}{{\left( {D \cdot \partial } \right)}^n}\left[ {G_{\mu \nu }^D} \right]{\partial ^\nu }} \notag\\
 &= i{D_\mu } + \sum\limits_{n = 0}^\infty  {\frac{{{i^n}}}{{\left( {n + 1} \right)!}}{{\left( {{D_{{\alpha _1}}} \cdots {D_{{\alpha _n}}}G_{\mu \nu }^D} \right)}_x}{\partial ^{{\alpha _1}}} \cdots {\partial ^{{\alpha _n}}}{\partial ^\nu }} \,, \\[10pt]
{e^{iD \cdot \frac{\partial }{{\partial p}}}}\left( {{p_\mu }} \right){e^{ - iD \cdot \frac{\partial }{{\partial p}}}} &= \sum\limits_{n = 0}^\infty  {\frac{1}{{n!}}{{\left( {iD \cdot \frac{\partial }{{\partial p}}} \right)}^n}\left[ {{p_\mu }} \right]} \notag\\
 &= {p_\mu } + i{D_\mu } + \sum\limits_{n = 0}^\infty  {\frac{1}{{\left( {n + 2} \right)!}}{{\left( {iD \cdot \partial } \right)}^n}\left[ {i{D_\nu }{\partial ^\nu },i{D_\mu }} \right]} \notag\\
 &= {p_\mu } + i{D_\mu } + \sum\limits_{n = 0}^\infty  {\frac{{{i^n}}}{{\left( {n + 2} \right)!}}{{\left( {D \cdot \partial } \right)}^n}\left[ {G_{\mu \nu }^D} \right]{\partial ^\nu }} \notag\\
 &= {p_\mu } + i{D_\mu } + \sum\limits_{n = 0}^\infty  {\frac{{{i^n}}}{{\left( {n + 2} \right)!}}{{\left( {{D_{{\alpha _1}}} \cdots {D_{{\alpha _n}}}G_{\mu \nu }^D} \right)}_x}{\partial ^{{\alpha _1}}} \cdots {\partial ^{{\alpha _n}}}{\partial ^\nu }} \,.
\end{align}
\end{subequations}
In the above, we have introduced $\partial^\alpha$ as a shorthand for $\frac{\partial}{\partial p_\alpha}$. We will adopt this notation from now on unless otherwise specified, since the position partial derivative $\frac{\partial}{\partial x_\alpha}$ will always be contained in $D^\alpha$ and as such will never explicitly appear. Note that to obtain the last lines in the above two equations, we have used $\left[ {{D_\alpha },G_{\mu \nu }^D} \right] = {\left( {{D_\alpha }G_{\mu \nu }^D} \right)_x}$ repeatedly, see \cref{eqn:DGx,eqn:DURepeat}. Combining these two equations, we get
\begin{align}
{e^{iD \cdot \frac{\partial }{{\partial p}}}}\left( {i{D_\mu } - {p_\mu }} \right){e^{ - iD \cdot \frac{\partial }{{\partial p}}}} &=  - {p_\mu } + \sum\limits_{n = 0}^\infty  {\frac{{\left( {n + 1} \right){i^n}}}{{\left( {n + 2} \right)!}}{{\left( {{D_{{\alpha _1}}} \cdots {D_{{\alpha _n}}}G_{\mu \nu }^D} \right)}_x}{\partial ^{{\alpha _1}}} \cdots {\partial ^{{\alpha _n}}}{\partial ^\nu }} \notag\\[3pt]
 &\equiv  - {p_\mu } + G_{\mu \nu }^{{\text{CDE}}}\left( x \right){\partial ^\nu } \,,
\label{eqn:iDpCDE}
\end{align}
with
\begin{align}
G_{\mu \nu }^{{\text{CDE}}}\left( x \right) &\equiv \sum\limits_{n = 0}^\infty  {\frac{{\left( {n + 1} \right){i^n}}}{{\left( {n + 2} \right)!}}{{\left( {{D_{{\alpha _1}}} \cdots {D_{{\alpha _n}}}G_{\mu \nu }^D} \right)}_x}{\partial ^{{\alpha _1}}} \cdots {\partial ^{{\alpha _n}}}} \,. \label{eqn:GCDE}
\end{align}
Here we have introduced the ``CDE field strength'' function $G_{\mu\nu}^\text{CDE}\left(x\right)$. The benefit of the additional insertions in \cref{eq:T0CDEbegin} is that they systematically converted all covariant derivatives into commutators, $G_{\mu\nu}^\text{CDE}\left(x\right)$. Furthermore, since this object is an explicit function of $x$, there are no gauge covariance breaking terms appearing due to $D_\mu$ acting on the unit function. The price is that we now have to keep track of the momentum derivatives that appear in \cref{eqn:GCDE}. In other words, we have systematically traded the position derivatives for momentum derivatives, which have the feature that they do not have any impact on the gauge transformation properties.

Putting this all together, \cref{eq:T0CDEbegin} becomes
\begin{equation}
{T_0} =  - i\int\ddx{x} \int\ddp{p} \tr \left[ \frac{1}{\left( p_\mu - G_{\mu\nu}^\text{CDE} \partial^\nu \right)^2 - M^2} \right] \,.
\end{equation}
The next step is to perform the CDE expansion, which requires performing two simultaneous expansions. First, we expand the ``CDE propagator''
\begin{align}
{\left( {{p_\mu } - G_{\mu \nu }^{{\text{CDE}}}{\partial ^\nu }} \right)^2} - {M^2} = {p^2} - {M^2} - \left[ {{\eta ^{\mu \alpha }}\left( {{p_\alpha }G_{\mu \nu }^{{\text{CDE}}} + G_{\mu \nu }^{{\text{CDE}}}{p_\alpha }} \right){\partial ^\nu } - {\eta ^{\mu \nu }}G_{\mu \alpha }^{{\text{CDE}}}G_{\nu \beta }^{{\text{CDE}}}{\partial ^\alpha }{\partial ^\beta }} \right] \,, \notag
\end{align}
which is a Taylor expansion in small $G_{\mu\nu}^\text{CDE}\left(x\right)$. This expansion is performed up to the mass dimension of interest, noting that each $G_{\mu\nu}^\text{CDE}\left(x\right)$ contributes mass dimension two or more. Second, $G_{\mu\nu}^\text{CDE}\left(x\right)$ itself should be expressed as a series using \cref{eqn:GCDE}, which should also be truncated up to the desired mass dimension.

For our example, we keep all terms up to mass dimension four:
\begin{align}
\dfrac{1}{{{{\left( {{p_\mu } - G_{\mu \nu }^{{\text{CDE}}}{\partial ^\nu }} \right)}^2} - {M^2}}} &\supset \left[\, \dfrac{1}{{{p^2} - {M^2}}} + \dfrac{1}{{{p^2} - {M^2}}}{\eta ^{\mu \alpha }}\left( {{p_\alpha }G_{\mu \nu }^{{\text{CDE}}} + G_{\mu \nu }^{{\text{CDE}}}{p_\alpha }} \right){\partial ^\nu }\dfrac{1}{{{p^2} - {M^2}}} \right. \notag\\[-3pt]
                &\hspace{20pt} - \dfrac{1}{{{p^2} - {M^2}}}{\eta ^{\mu \nu }}G_{\mu \alpha }^{{\text{CDE}}}G_{\nu \beta }^{{\text{CDE}}}{\partial ^\alpha }{\partial ^\beta }\dfrac{1}{{{p^2} - {M^2}}} \notag\\
                &\hspace{20pt} + \dfrac{1}{{{p^2} - {M^2}}}{\eta ^{\mu \alpha }}\left( {{p_\alpha }G_{\mu \nu }^{{\text{CDE}}} + G_{\mu \nu }^{{\text{CDE}}}{p_\alpha }} \right) \partial^\nu \dfrac{1}{{{p^2} - {M^2}}} \notag\\
                &\hspace{70pt} \left. \times {\eta ^{\rho \beta }}\left( {{p_\beta }G_{\rho \sigma }^{{\text{CDE}}} + G_{\rho \sigma }^{{\text{CDE}}}{p_\beta }} \right){\partial ^\sigma }\dfrac{1}{{{p^2} - {M^2}}} \right] \,,
\end{align}
with
\begin{align}
G_{\mu \nu }^{{\text{CDE}}} &\supset {\dfrac{1}{2}G_{\mu \nu }^D + \dfrac{i}{3}{{\left( {{D_{{\alpha _1}}}G_{\mu \nu }^D} \right)}_x}{\partial ^{{\alpha _1}}} - \dfrac{1}{8}{{\left( {{D_{{\alpha _1}}}{D_{{\alpha _2}}}G_{\mu \nu }^D} \right)}_x}{\partial ^{{\alpha _1}}}{\partial ^{{\alpha _2}}}} \,,
\end{align}
which gives
\begin{align}
\dfrac{1}{{{{\left( {{p_\mu } - G_{\mu \nu }^{{\text{CDE}}}{\partial ^\nu }} \right)}^2} - {M^2}}} &\supset \left\{\, \dfrac{1}{{{p^2} - {M^2}}} + G_{\mu \nu }^D\dfrac{1}{{{p^2} - {M^2}}}\left( {{p^\mu }{\partial ^\nu }} \right)\dfrac{1}{{{p^2} - {M^2}}} \right. \notag\\
                &\hspace{-20pt} + \dfrac{1}{3}\dfrac{1}{{{p^2} - {M^2}}}\left[ {{{\left( {{iD^\mu }G_{\mu \nu }^D} \right)}_x}{\partial ^\nu } + 2{{\left( {{iD_\alpha }G_{\mu \nu }^D} \right)}_x}{p^\mu }{\partial ^\nu }{\partial ^\alpha }} \right]\dfrac{1}{{{p^2} - {M^2}}} \notag\\
                &\hspace{-20pt} - \dfrac{1}{4}\dfrac{1}{{{p^2} - {M^2}}}\left[ {{{\left( {{D_\alpha }{D^\mu }G_{\mu \nu }^D} \right)}_x}{\partial ^\nu }{\partial ^\alpha } + {{\left( {{D_\alpha }{D_\beta }G_{\mu \nu }^D} \right)}_x}{p^\mu }{\partial ^\nu }{\partial ^\alpha }{\partial ^\beta }} \right]\dfrac{1}{{{p^2} - {M^2}}} \notag\\
                &\hspace{-20pt} - \dfrac{1}{4}G_{\mu \alpha }^DG_{\nu \beta }^D\left( {\eta ^{\mu \nu }}\dfrac{1}{{{p^2} - {M^2}}}{\partial ^\alpha }{\partial ^\beta }\dfrac{1}{{{p^2} - {M^2}}}\right. \notag\\
                &\hspace{50pt} \left. \left.  - 4\dfrac{1}{{{p^2} - {M^2}}}{p^\mu }{\partial ^\alpha }\dfrac{1}{{{p^2} - {M^2}}}{p^\nu }{\partial ^\beta }\dfrac{1}{{{p^2} - {M^2}}} \right) \right\} \,.
 \label{eqn:eCDEpropagator}
\end{align}
Dropping the term without $G_{\mu\nu}^D$, we integrate over the loop momentum and get (note that terms odd in the loop momentum $p$ vanish upon integration)
\begingroup
\allowdisplaybreaks
\begin{align}
{T_0} &\supset - i\int\ddx{x} \int\ddp{p} \tr \Bigg\{ G_{\mu \nu }^D\dfrac{1}{{{p^2} - {M^2}}}\left( {{p^\mu }{\partial ^\nu }} \right)\dfrac{1}{{{p^2} - {M^2}}} \notag\\
                &\hspace{60pt} - \dfrac{1}{4}\dfrac{1}{{{p^2} - {M^2}}}\left[ {{{\left( {{D_\alpha }{D^\mu }G_{\mu \nu }^D} \right)}_x}{\partial ^\nu }{\partial ^\alpha } + {{\left( {{D_\alpha }{D_\beta }G_{\mu \nu }^D} \right)}_x}{p^\mu }{\partial ^\nu }{\partial ^\alpha }{\partial ^\beta }} \right]\dfrac{1}{{{p^2} - {M^2}}} \notag\\
                &\hspace{60pt} - \dfrac{1}{4}G_{\mu \alpha }^DG_{\nu \beta }^D\left( {\eta ^{\mu \nu }}\dfrac{1}{{{p^2} - {M^2}}}{\partial ^\alpha }{\partial ^\beta }\dfrac{1}{{{p^2} - {M^2}}} \right. \notag\\[-5pt]
                &\hspace{140pt} \left.- 4\dfrac{1}{{{p^2} - {M^2}}}{p^\mu }{\partial ^\alpha }\dfrac{1}{{{p^2} - {M^2}}}{p^\nu }{\partial ^\beta }\dfrac{1}{{{p^2} - {M^2}}} \right) \Bigg\} \notag\\[5pt]
 &=  - i\int\ddx{x} \int\ddp{p} \tr \left[ { - \dfrac{1}{4}{\eta ^{\mu \nu }}G_{\mu \alpha }^DG_{\nu \beta }^D\left( {\dfrac{1}{{{p^2} - {M^2}}}{\partial ^\alpha }{\partial ^\beta }\dfrac{1}{{{p^2} - {M^2}}}} \right)} \right]  \notag\\[5pt]
 &=  - i\int\ddx{x} \int\ddp{p} \tr \left\{ {\dfrac{1}{2}{\eta ^{\mu \nu }}{\eta ^{\alpha \beta }}G_{\mu \alpha }^DG_{\nu \beta }^D\left[ {\dfrac{1}{{{{\left( {{p^2} - {M^2}} \right)}^3}}} - \dfrac{{4{p^2}}}{{d{{\left( {{p^2} - {M^2}} \right)}^4}}}} \right]} \right\}  \notag\\[5pt]
 &= \int{\dd^4}x\,\frac{1}{{{{\left( {4\pi } \right)}^2}}}\frac{1}{{{M^2}}}\frac{{ - 1}}{{12}} \tr \left( {G_{\mu \nu }^D{G^{D,\mu \nu }}} \right) \,, \label{eqn:DDGused}
\end{align}
\endgroup
in agreement with the previous two approaches. To obtain the second line, we further dropped terms that vanish due to symmetries, \emph{i.e.}, the term in the first line of the curly bracket is zero since $p^\mu\partial^\nu$ is symmetric under the exchange of $\mu\leftrightarrow\nu$ while $G_{\mu\nu}^D$ is antisymmetric. The same argument holds also for the second term in the second line (after using integration by parts) and the second term in the third line of the curly bracket. The first term in the second line of the curly bracket vanishes because the loop integral will yield $\eta^{\nu\alpha}$, which contracts with the field strength to give $\left(D^\mu D^\nu G_{\mu\nu}^D\right)_x=0$, see \cref{eqn:DDG0}. This leaves only one term, which can be evaluated using the loop integrals tabulated in~\cref{appsubsec:EssentialIntegrals}.

Having worked out this simple example in detail, we will briefly comment on evaluating more involved functional traces. For example, we frequently encounter functional traces of the following form
\begin{equation}
{T_n}\left(M^2\right) \equiv i \Tr \left[ {\frac{1}{{{D^2} + {M^2}}}\, {U_1}\cdots \frac{1}{{{D^2} + {M^2}}}\, {U_n}} \right] \,,
\label{eq:Tn}
\end{equation}
First performing the simplified CDE step gives
\begin{equation}
{T_n}\left(M^2\right) = {\left( { - 1} \right)^n}i\int\ddx{x} \int\ddp{p} \tr \left[ {\frac{1}{{{{\left( {iD - p} \right)}^2} - {M^2}}}\, {U_1} \cdots \frac{1}{{{{\left( {iD - p} \right)}^2} - {M^2}}}\, {U_n}} \right] \,.
\end{equation}
Then converting this into an original CDE expression using the additional insertions as in \cref{eq:T0CDEbegin}, we have
\begin{align}
{T_n}\left(M^2\right) &= {\left( { - 1} \right)^n} i \int\ddx{x} \int\ddp{p} \tr \left[ \frac{1}{{{{\left( {{p_\mu } - G_{\mu \nu }^{{\text{CDE}}}{\partial ^\nu }} \right)}^2} - {M^2}}}\, U_1^{{\text{CDE}}}\right. \notag\\
&\hspace{150pt}\left.\times\cdots\times \frac{1}{{{{\left( {{p_\mu } - G_{\mu \nu }^{{\text{CDE}}}{\partial ^\nu }} \right)}^2} - {M^2}}}\, U_n^{{\text{CDE}}} \right] \,.
\end{align}
Note that the function $U_k(x)$ should also be sandwiched by the insertions, which defines the series
\begin{equation}
U_k^\text{CDE} \equiv {e^{iD \cdot \frac{\partial }{{\partial p}}}}\, U_k\, {e^{ - iD \cdot \frac{\partial }{{\partial p}}}} = \sum\limits_{n = 0}^\infty  {\frac{{{i^n}}}{{n!}}{{\left( {{D_{{\alpha _1}}} \cdots {D_{{\alpha _n}}}U_k} \right)}_x}{\partial ^{{\alpha _1}}} \cdots {\partial ^{{\alpha _n}}}} \,.
\label{eqn:UCDE}
\end{equation}
The evaluation of several frequently used generic functional traces is given in~\cref{appsubsubsec:functionaltraces}.

\subsection{Functional Variations for Gauge Bosons}\label{appsubsec:BackgroundField}
In the previous section, we showed how to use CDE to evaluate functional traces while maintaining manifest gauge covariance. In doing so, we assumed that the functional trace is already expressed in terms of the covariant derivative $D_\mu$. This is the case if the variation is being taken with respect to a field that is not a gauge boson. In HQET, we are typically interested in loops involving gluons. In order to take the functional variation with respect to the gauge fields, we would naively be forced to split up the covariant derivative, which implies a loss of the manifest gauge covariance. By using the background field method (see \textit{e.g.}~\cite{Abbott:1980hw,Abbott:1981ke,Peskin:1995ev}) we can avoid this issue.

The first step is to decompose the gauge field into a background field component $G_{B,\mu}^a$ plus fluctuations $A_\mu$:
\begin{align}
{G_\mu } &= {G_{B,\mu }} + {A_\mu } \,,
\end{align}
One defines a background covariant derivative $D_{B,\mu}$:
\begin{align}
{D_{B,\mu }} &\equiv {\partial _\mu } - ig{G_{B,\mu }} \,,
\end{align}
such that the full covariant derivative is
\begin{equation}
{D_\mu } \equiv {D_{B,\mu }} - ig{A_\mu } \,.
\end{equation}
Then up to quadratic terms in $A_\mu$, the original field strength and the gauge field kinetic term are written as
\begin{align}
G_{\mu\nu}^a &= G_{B,\mu \nu }^a + \left(D_{B,\mu} A_\nu\right)^a - \left(D_{B,\nu} A_\mu\right)^a + g f^{abc} A_\mu^b A_\nu^c \,,
\end{align}
and
\begin{align}
 - \frac{1}{4}G_{\mu \nu }^a{G^{\mu \nu ,a}} \supset& - \frac{1}{4}G_{B,\mu \nu }^aG_B^{\mu \nu ,a} - G_B^{\mu\nu,a}\left( D_{B,\mu} A_\nu \right)^a \notag\\
 &+ \frac{1}{2}A_\mu^a \left[ \left( \eta^{\mu \nu} D_B^2 - D_B^\mu D_B^\nu \right)^{ab} -2g f^{abc} G_B^{\mu\nu,c} \right] A_\nu^b \,.
\end{align}
Everything is manifestly gauge covariant with respect to the background gauge field $G_{B,\mu}$, \emph{i.e.}, every term is composed of either $D_{B,\mu}$ or $G_{B,\mu\nu}^a$.

Factors of $A_\mu$ appear at intermediate steps. However, they will be integrated over perturbatively, which requires gauge fixing. The key to the background field method is that we can implement a gauge fixing condition for $A_\mu$ that is manifestly gauge covariant with respect to the background gauge field $G_{B,\mu}$:
\begin{equation}
{\Lag_\text{gf}} =  - \frac{1}{{2\xi }}{\left[ {{{\left( {D_B^\mu {A_\mu }} \right)}^a}} \right]^2} = \frac{1}{2}A_\mu ^a{\left( {D_B^\mu D_B^\nu } \right)^{ab}}A_\nu ^b \,,
\label{eq:gaugefix}
\end{equation}
where we have specified to the Feynman gauge $\xi=1$ in the last expression. The resultant ghost term is also manifestly background gauge covariant
\begin{equation}
{\Lag_\text{gh}} = {\left( {D_B^\mu \bar c} \right)^a}\left[ {{{\left( {{D_{B,\mu }}c} \right)}^a} - g{f^{abc}}A_\mu ^c{c^b}} \right] \supset {{\bar c}^a}{\left( { - D_B^2} \right)^{ab}}{c^b} \,.
\label{eq:ghost}
\end{equation}
Note that both the fluctuating part $A_\mu^a$ and the ghosts $c^a$ and ${\bar c}^a$ transform under the adjoint representation of the background gauge symmetry. Once we have preformed a functional variation, we will drop the subscript $B$ from $D_{B,\mu}$ and $G_{B,\mu\nu}^a$.

\subsection{Applications}\label{appsubsec:CDEApplications}
In this section, we apply the CDE to evaluate functional traces.

\subsubsection{Expansions up to Mass Dimension Six}\label{appsubsubsec:Expansions}
This section collects some intermediate expansion results that are frequently needed for evaluating functional traces with the original CDE. First, the CDE field strength $G_{\mu\nu}^\text{CDE}$ and $U^\text{CDE}$ expanded up to mass dimension six are
\begingroup
\allowdisplaybreaks
\begin{subequations}
\begin{align}
G_{\mu \nu }^{{\text{CDE}}} &\equiv \sum\limits_{n = 0}^\infty  {\frac{{\left( {n + 1} \right){i^n}}}{{\left( {n + 2} \right)!}}{{\left( {{D_{{\alpha _1}}} \cdots {D_{{\alpha _n}}}G_{\mu \nu }^D} \right)}_x}{\partial ^{{\alpha _1}}} \cdots {\partial ^{{\alpha _n}}}} \nonumber \\
 &\supset \Bigg[ \frac{1}{2}G_{\mu \nu }^D + \frac{1}{3}{\left( {i{D_{{\alpha _1}}}G_{\mu \nu }^D} \right)_x}{\partial ^{{\alpha _1}}} - \frac{1}{8}{\left( {{D_{{\alpha _1}}}{D_{{\alpha _2}}}G_{\mu \nu }^D} \right)_x}{\partial ^{{\alpha _1}}}{\partial ^{{\alpha _2}}} \notag\\
 &\hspace{20pt} - \frac{1}{{30}}{\left( {i{D_{{\alpha _1}}}{D_{{\alpha _2}}}{D_{{\alpha _3}}}G_{\mu \nu }^D} \right)_x}{\partial ^{{\alpha _1}}}{\partial ^{{\alpha _2}}}{\partial ^{{\alpha _3}}} \notag\\[-3pt]
 &\hspace{20pt} + \frac{1}{{144}}{\left( {{D_{{\alpha _1}}}{D_{{\alpha _2}}}{D_{{\alpha _3}}}{D_{{\alpha _4}}}G_{\mu \nu }^D} \right)_x}{\partial ^{{\alpha_1}}}{\partial ^{{\alpha _2}}}{\partial ^{{\alpha _3}}}{\partial ^{{\alpha _4}}} \Bigg] \,, \\[8pt]
U^{{\text{CDE}}} &\equiv \sum\limits_{n = 0}^\infty  {\frac{{{i^n}}}{{n!}}{{\left( {{D_{{\alpha _1}}} \cdots {D_{{\alpha _n}}}U} \right)}_x}{\partial ^{{\alpha _1}}} \cdots {\partial ^{{\alpha _n}}}} \notag\\
 &\supset \Bigg[ U + {\left( {i{D_{{\alpha _1}}}U} \right)_x}{\partial ^{{\alpha _1}}} - \frac{1}{2}{\left( {{D_{{\alpha _1}}}{D_{{\alpha _2}}}U} \right)_x}{\partial ^{{\alpha _1}}}{\partial ^{{\alpha _2}}} \notag\\
 &\hspace{20pt} - \frac{1}{6}{\left( {i{D_{{\alpha _1}}}{D_{{\alpha _2}}}{D_{{\alpha _3}}}U} \right)_x}{\partial ^{{\alpha _1}}}{\partial ^{{\alpha _2}}}{\partial ^{{\alpha _3}}} + \frac{1}{{24}}{\left( {{D_{{\alpha _1}}}{D_{{\alpha _2}}}{D_{{\alpha _3}}}{D_{{\alpha _4}}}U} \right)_x}{\partial ^{{\alpha _1}}}{\partial ^{{\alpha _2}}}{\partial ^{{\alpha _3}}}{\partial ^{{\alpha _4}}} \notag\\
 &\hspace{20pt} + \frac{1}{{120}}{\left( {i{D_{{\alpha _1}}}{D_{{\alpha _2}}}{D_{{\alpha _3}}}{D_{{\alpha _4}}}{D_{{\alpha _5}}}U} \right)_x}{\partial ^{{\alpha _1}}}{\partial ^{{\alpha _2}}}{\partial ^{{\alpha _3}}}{\partial ^{{\alpha _4}}}{\partial ^{{\alpha _5}}} \Bigg] \,.
\end{align}
\end{subequations}
\endgroup
Note that $G_{\mu\nu}^D = \comm{D_\mu}{D_\nu}$ always contributes mass dimension two to the effective operator. On the other hand, although $U$ is also an object with mass dimension two, it might contain a dimensionful coupling. So $U$ is guaranteed to contribute only mass dimension one to the effective operator. Next, we expand the ``CDE propagator'' up to mass dimension six
\begingroup
\allowdisplaybreaks
\begin{align}
&\hspace{-10pt}\dfrac{1}{{{{\left( {{p_\mu } - G_{\mu \nu }^{{\text{CDE}}}{\partial ^\nu }} \right)}^2} - {M^2}}} \notag\\
=\,\,& \dfrac{1}{{{p^2} - {M^2} - \left[ {{\eta ^{\mu \alpha }}\left( {{p_\alpha }G_{\mu \nu }^{{\text{CDE}}} + G_{\mu \nu }^{{\text{CDE}}}{p_\alpha }} \right){\partial ^\nu } - {\eta ^{\mu \nu }}G_{\mu \alpha }^{{\text{CDE}}}G_{\nu \beta }^{{\text{CDE}}}{\partial ^\alpha }{\partial ^\beta }} \right]}} \notag\\
\supset\,\, & \dfrac{1}{{{p^2} - {M^2}}} + G_{\mu \nu }^D\, \dfrac{1}{{{p^2} - {M^2}}}\, {p^\mu }{\partial ^\nu }\dfrac{1}{{{p^2} - {M^2}}} \notag\\
    & + \dfrac{1}{3}{\left( {i{D_{{\alpha _1}}}G_{\mu \nu }^D} \right)_x}\dfrac{1}{{{p^2} - {M^2}}}\left( {2{p^\mu }{\partial ^{{\alpha _1}}} + {\eta ^{\mu {\alpha _1}}}} \right){\partial ^\nu }\dfrac{1}{{{p^2} - {M^2}}} \notag\\
    & - \dfrac{1}{4}{\left( {{D_{{\alpha _1}}}{D_{{\alpha _2}}}G_{\mu \nu }^D} \right)_x}\, \dfrac{1}{{{p^2} - {M^2}}}\left( {{p^\mu }{\partial ^{{\alpha _2}}} + {\eta ^{\mu {\alpha _2}}}} \right){\partial ^\nu }{\partial ^{{\alpha _1}}}\dfrac{1}{{{p^2} - {M^2}}} \notag\\
    & - \dfrac{1}{4}G_{\mu \nu }^D G_{\rho \sigma }^D \left( {{\eta ^{\mu \rho }}\dfrac{1}{{{p^2} - {M^2}}}\,{\partial ^\nu }{\partial ^\sigma }\dfrac{1}{{{p^2} - {M^2}}} - 4\dfrac{1}{{{p^2} - {M^2}}}\,{p^\mu }{\partial ^\nu }\dfrac{1}{{{p^2} - {M^2}}}\,{p^\rho }{\partial ^\sigma }\dfrac{1}{{{p^2} - {M^2}}}} \right) \notag\\
    & - \dfrac{1}{{30}}\dfrac{1}{{{p^2} - {M^2}}} \Bigg[ {\left( {i{D^\mu }{D_{{\alpha _1}}}{D_{{\alpha _2}}}G_{\mu \nu }^D + i{D_{{\alpha _1}}}{D^\mu }{D_{{\alpha _2}}}G_{\mu \nu }^D + i{D_{{\alpha _1}}}{D_{{\alpha _2}}}{D^\mu }G_{\mu \nu }^D} \right)_x} \notag\\[-8pt]
          &\hspace{150pt} + 2{\left( {i{D_{{\alpha _1}}}{D_{{\alpha _2}}}{D_{{\alpha _3}}}G_{\mu \nu }^D} \right)_x}{p^\mu }{\partial ^{{\alpha _3}}} \Bigg] {\partial ^\nu }{\partial ^{{\alpha _1}}}{\partial ^{{\alpha _2}}}\dfrac{1}{{{p^2} - {M^2}}}\notag\\
    & - \dfrac{1}{6}G_{\mu \nu }^D{\left( {i{D_{{\alpha _1}}}G_{\rho \sigma }^D} \right)_x} \Bigg[ {\eta ^{\mu \rho }}\dfrac{1}{{{p^2} - {M^2}}}\,{\partial ^\nu }{\partial ^\sigma }{\partial ^{{\alpha _1}}}\dfrac{1}{{{p^2} - {M^2}}} \notag\\[-8pt]
          &\hspace{105pt} - 2\dfrac{1}{{{p^2} - {M^2}}}\,{p^\mu }{\partial ^\nu }\dfrac{1}{{{p^2} - {M^2}}}\left( {2{p^\rho }{\partial ^{{\alpha _1}}} + {\eta ^{\rho {\alpha _1}}}} \right){\partial ^\sigma }\dfrac{1}{{{p^2} - {M^2}}} \Bigg] \notag\\
    & - \dfrac{1}{6}{\left( {i{D_{{\alpha _1}}}G_{\mu \nu }^D} \right)_x}G_{\rho \sigma }^D \Bigg[ {\eta ^{\mu \rho }}\dfrac{1}{{{p^2} - {M^2}}}\,{\partial ^\nu }{\partial ^\sigma }{\partial ^{{\alpha _1}}}\dfrac{1}{{{p^2} - {M^2}}} \notag\\[-8pt]
          &\hspace{105pt} - 2\dfrac{1}{{{p^2} - {M^2}}}\left( {2{p^\mu }{\partial ^{{\alpha _1}}} + {\eta ^{\mu {\alpha _1}}}} \right){\partial ^\nu }\dfrac{1}{{{p^2} - {M^2}}}\,{p^\rho }{\partial ^\sigma }\dfrac{1}{{{p^2} - {M^2}}} \Bigg] \notag\\
    & + \dfrac{1}{{144}}\dfrac{1}{{{p^2} - {M^2}}} \left[ \renewcommand\arraystretch{1.2} \mqty( {D^\mu }{D_{{\alpha _1}}}{D_{{\alpha _2}}}{D_{{\alpha _3}}}G_{\mu \nu }^D + {D_{{\alpha _1}}}{D^\mu }{D_{{\alpha _2}}}{D_{{\alpha _3}}}G_{\mu \nu }^D \\ + {D_{{\alpha _1}}}{D_{{\alpha _2}}}{D^\mu }{D_{{\alpha _3}}}G_{\mu \nu }^D + {D_{{\alpha _1}}}{D_{{\alpha _2}}}{D_{{\alpha _3}}}{D^\mu }G_{\mu \nu }^D )_x \right. \notag\\[-5pt]
          &\hspace{125pt} + 2{\left( {{D_{{\alpha _1}}}{D_{{\alpha _2}}}{D_{{\alpha _3}}}{D_{{\alpha _4}}}G_{\mu \nu }^D} \right)_x}{p^\mu }{\partial ^{{\alpha _4}}} \Bigg] {\partial ^\nu }{\partial ^{{\alpha _1}}}{\partial ^{{\alpha _2}}}{\partial ^{{\alpha _3}}}\dfrac{1}{{{p^2} - {M^2}}}\notag\\
    & + \dfrac{1}{9}{\left( {{D_{{\alpha _1}}}G_{\mu \nu }^D} \right)_x}{\left( {{D_{{\alpha _2}}}G_{\rho \sigma }^D} \right)_x} \Bigg[ {\eta ^{\mu \rho }}\dfrac{1}{{{p^2} - {M^2}}}\,{\partial ^{{\alpha _1}}}{\partial ^{{\alpha _2}}}{\partial ^\nu }{\partial ^\sigma }\dfrac{1}{{{p^2} - {M^2}}} \notag\\[-5pt]
          &\hspace{90pt} - \dfrac{1}{{{p^2} - {M^2}}}\left( {2{p^\mu }{\partial ^{{\alpha _1}}} + {\eta ^{\mu {\alpha _1}}}} \right){\partial ^\nu }\dfrac{1}{{{p^2} - {M^2}}}\left( {2{p^\rho }{\partial ^{{\alpha _2}}} + {\eta ^{\rho {\alpha _2}}}} \right){\partial ^\sigma }\dfrac{1}{{{p^2} - {M^2}}} \Bigg] \notag\\
    & + \dfrac{1}{{16}}G_{\mu \nu }^D{\left( {{D_{{\alpha _1}}}{D_{{\alpha _2}}}G_{\rho \sigma }^D} \right)_x} \Bigg[ {\eta ^{\mu \rho }}\dfrac{1}{{{p^2} - {M^2}}}\,{\partial ^{{\alpha _1}}}{\partial ^{{\alpha _2}}}{\partial ^\nu }{\partial ^\sigma }\dfrac{1}{{{p^2} - {M^2}}} \notag\\[-8pt]
          &\hspace{125pt} - 4\dfrac{1}{{{p^2} - {M^2}}}\,{p^\mu }{\partial ^\nu }\dfrac{1}{{{p^2} - {M^2}}}\left( {{p^\rho }{\partial ^{{\alpha _2}}} + {\eta ^{\rho {\alpha _2}}}} \right){\partial ^\sigma }{\partial ^{{\alpha _1}}}\dfrac{1}{{{p^2} - {M^2}}} \Bigg] \notag\\
    & + \dfrac{1}{{16}}{\left( {{D_{{\alpha _1}}}{D_{{\alpha _2}}}G_{\mu \nu }^D} \right)_x}G_{\rho \sigma }^D \Bigg[ {\eta ^{\mu \rho }}\dfrac{1}{{{p^2} - {M^2}}}\,{\partial ^{{\alpha _1}}}{\partial ^{{\alpha _2}}}{\partial ^\nu }{\partial ^\sigma }\dfrac{1}{{{p^2} - {M^2}}} \notag\\[-8pt]
          &\hspace{125pt} - 4\dfrac{1}{{{p^2} - {M^2}}}\left( {{p^\mu }{\partial ^{{\alpha _2}}} + {\eta ^{\mu {\alpha _2}}}} \right){\partial ^\nu }{\partial ^{{\alpha _1}}}\dfrac{1}{{{p^2} - {M^2}}}\,{p^\rho }{\partial ^\sigma }\dfrac{1}{{{p^2} - {M^2}}} \Bigg] \notag\\
    & - \dfrac{1}{4}G_{\mu \nu }^DG_{\rho \sigma }^DG_{\lambda \eta }^D \Bigg[ {\eta^{\rho \lambda}} \dfrac{1}{{{p^2} - {M^2}}}\,{p^\mu }{\partial ^\nu }\dfrac{1}{{{p^2} - {M^2}}}\,{\partial ^\sigma }{\partial ^\eta }\dfrac{1}{{{p^2} - {M^2}}} \notag\\[-3pt]
          &\hspace{85pt} + {\eta ^{\mu \rho }}\dfrac{1}{{{p^2} - {M^2}}}\,{\partial ^\nu }{\partial ^\sigma }\dfrac{1}{{{p^2} - {M^2}}}\,{p^\lambda }{\partial ^\eta }\dfrac{1}{{{p^2} - {M^2}}} \notag\\[-5pt]
          &\hspace{85pt} - 4\dfrac{1}{{{p^2} - {M^2}}}\,{p^\mu }{\partial ^\nu }\dfrac{1}{{{p^2} - {M^2}}}\,{p^\rho }{\partial ^\sigma }\dfrac{1}{{{p^2} - {M^2}}}\,{p^\lambda }{\partial ^\eta }\dfrac{1}{{{p^2} - {M^2}}} \Bigg] \,.
\label{eq:ExpUpToDim6}
\end{align}
\endgroup
Evaluating this expression requires taking momentum derivatives of the free propagators:
\begin{subequations}
\begin{align}
{\partial ^{{\alpha _1}}}\,\dfrac{1}{{{p^2} - {M^2}}} &=  - \dfrac{2}{{{{\left( {{p^2} - {M^2}} \right)}^2}}}\,{p^{{\alpha _1}}} \,, \\[5pt]
{\partial ^{{\alpha _1}}}{\partial ^{{\alpha _2}}}\,\dfrac{1}{{{p^2} - {M^2}}} &= \dfrac{8}{{{{\left( {{p^2} - {M^2}} \right)}^3}}}\,{p^{{\alpha _1}}}{p^{{\alpha _2}}} - \dfrac{2}{{{{\left( {{p^2} - {M^2}} \right)}^2}}}\,{\eta ^{{\alpha _1}{\alpha _2}}} \,, \\[5pt]
{\partial ^{{\alpha _1}}}{\partial ^{{\alpha _2}}}{\partial ^{{\alpha _3}}}\,\dfrac{1}{{{p^2} - {M^2}}} &=  - \dfrac{{48}}{{{{\left( {{p^2} - {M^2}} \right)}^4}}}{p^{{\alpha _1}}}{p^{{\alpha _2}}}{p^{{\alpha _3}}} \notag\\
        &\hspace{10pt} + \dfrac{8}{{{{\left( {{p^2} - {M^2}} \right)}^3}}}\Big( {{p^{{\alpha _1}}}{\eta ^{{\alpha _2}{\alpha _3}}} + {p^{{\alpha _2}}}{\eta ^{{\alpha _1}{\alpha _3}}} + {p^{{\alpha _3}}}{\eta ^{{\alpha _1}{\alpha _2}}}} \Big) \,, \\[5pt]
{\partial ^{{\alpha _1}}}{\partial ^{{\alpha _2}}}{\partial ^{{\alpha _3}}}{\partial ^{{\alpha _4}}}\,\dfrac{1}{{{p^2} - {M^2}}} &=
\dfrac{{384}}{{{{\left( {{p^2} - {M^2}} \right)}^5}}}\,{p^{{\alpha _1}}}{p^{{\alpha _2}}}{p^{{\alpha _3}}}{p^{{\alpha _4}}} \notag\\
        &\hspace{10pt} - \dfrac{{48}}{{{{\left( {{p^2} - {M^2}} \right)}^4}}} {\renewcommand\arraystretch{1.2}\mqty( {p^{{\alpha _1}}}{p^{{\alpha _2}}}{\eta ^{{\alpha _3}{\alpha _4}}} + {p^{{\alpha _1}}}{p^{{\alpha _3}}}{\eta ^{{\alpha _2}{\alpha _4}}} + {p^{{\alpha _1}}}{p^{{\alpha _4}}}{\eta ^{{\alpha _2}{\alpha _3}}} \\ + {p^{{\alpha _2}}}{p^{{\alpha _3}}}{\eta ^{{\alpha _1}{\alpha _4}}} + {p^{{\alpha _2}}}{p^{{\alpha _4}}}{\eta ^{{\alpha _1}{\alpha _3}}} + {p^{{\alpha _3}}}{p^{{\alpha _4}}}{\eta ^{{\alpha _1}{\alpha _2}}} )} \notag\\
        &\hspace{10pt} + \dfrac{8}{{{{\left( {{p^2} - {M^2}} \right)}^3}}}\Big( {{\eta ^{{\alpha _1}{\alpha _2}}}{\eta ^{{\alpha _3}{\alpha _4}}} + {\eta ^{{\alpha _1}{\alpha _3}}}{\eta ^{{\alpha _2}{\alpha _4}}} + {\eta ^{{\alpha _1}{\alpha _4}}}{\eta ^{{\alpha _2}{\alpha _3}}}} \Big) \,.
\end{align}
\end{subequations}
Finally, using \cref{eqn:iDpCDE}, we can also get the ``HQET CDE propagator,'' which expanded up to mass dimension six is
\begingroup
\allowdisplaybreaks
\begin{align}
& - e^{iD\cdot\frac{\partial}{\partial p}} \, \frac{1}{v\cdot\left(iD-p\right)}\, e^{-iD\cdot\frac{\partial}{\partial p}} = \dfrac{1}{{v \cdot p - {v^\mu }G_{\mu \nu }^{{\text{CDE}}}{\partial ^\nu }}} \notag\\[5pt]
        &\hspace{40pt} \supset \dfrac{1}{{v \cdot p}} + {v^\mu }\dfrac{1}{{v \cdot p}}G_{\mu \nu }^{{\text{CDE}}}{\partial ^\nu }\dfrac{1}{{v \cdot p}} + {v^\mu }{v^\rho }\dfrac{1}{{v \cdot p}}G_{\mu \nu }^{{\text{CDE}}}{\partial ^\nu }\dfrac{1}{{v \cdot p}}G_{\rho \sigma }^{{\text{CDE}}}{\partial ^\sigma }\dfrac{1}{{v \cdot p}} \notag\\
        &\hspace{80pt} + {v^\mu }{v^\rho }{v^\lambda }\dfrac{1}{{v \cdot p}}G_{\mu \nu }^{{\text{CDE}}}{\partial ^\nu }\dfrac{1}{{v \cdot p}}G_{\rho \sigma }^{{\text{CDE}}}{\partial ^\sigma }\dfrac{1}{{v \cdot p}}G_{\lambda \eta }^{{\text{CDE}}}{\partial ^\eta }\dfrac{1}{{v \cdot p}} \notag\\[5pt]
        &\hspace{40pt} \supset \dfrac{1}{{v \cdot p}} + \dfrac{1}{2}G_{\mu \nu }^D\left( {{v^\mu }\dfrac{1}{{v \cdot p}}{\partial ^\nu }\dfrac{1}{{v \cdot p}}} \right) + \dfrac{1}{3}{\left( {i{D_{{\alpha _1}}}G_{\mu \nu }^D} \right)_x}\left( {{v^\mu }\dfrac{1}{{v \cdot p}}{\partial ^\nu }{\partial ^{{\alpha _1}}}\dfrac{1}{{v \cdot p}}} \right) \notag\\
        &\hspace{80pt} - \dfrac{1}{8}{\left( {{D_{{\alpha _1}}}{D_{{\alpha _2}}}G_{\mu \nu }^D} \right)_x}\left( {{v^\mu }\dfrac{1}{{v \cdot p}}{\partial ^\nu }{\partial ^{{\alpha _1}}}{\partial ^{{\alpha _2}}}\dfrac{1}{{v \cdot p}}} \right) \notag\\
        &\hspace{80pt} + \dfrac{1}{4}G_{\mu \nu }^DG_{\rho \sigma }^D\left( {{v^\mu }{v^\rho }\dfrac{1}{{v \cdot p}}{\partial ^\nu }\dfrac{1}{{v \cdot p}}{\partial ^\sigma }\dfrac{1}{{v \cdot p}}} \right) \notag\\
        &\hspace{80pt} - \dfrac{1}{{30}}{\left( {i{D_{{\alpha _1}}}{D_{{\alpha _2}}}{D_{{\alpha _3}}}G_{\mu \nu }^D} \right)_x}\left( {{v^\mu }\dfrac{1}{{v \cdot p}}{\partial ^\nu }{\partial ^{{\alpha _1}}}{\partial ^{{\alpha _2}}}{\partial ^{{\alpha _3}}}\dfrac{1}{{v \cdot p}}} \right) \notag\\
        &\hspace{80pt} + \dfrac{1}{6}G_{\mu \nu }^D{\left( {i{D_{{\alpha _1}}}G_{\rho \sigma }^D} \right)_x}\left( {{v^\mu }{v^\rho }\dfrac{1}{{v \cdot p}}{\partial ^\nu }\dfrac{1}{{v \cdot p}}{\partial ^\sigma }{\partial ^{{\alpha _1}}}\dfrac{1}{{v \cdot p}}} \right) \notag\\
        &\hspace{80pt} + \dfrac{1}{6}{\left( {i{D_{{\alpha _1}}}G_{\mu \nu }^D} \right)_x}G_{\rho \sigma }^D\left( {{v^\mu }{v^\rho }\dfrac{1}{{v \cdot p}}{\partial ^\nu }{\partial ^{{\alpha _1}}}\dfrac{1}{{v \cdot p}}{\partial ^\sigma }\dfrac{1}{{v \cdot p}}} \right) \notag\\
        &\hspace{80pt} + \dfrac{1}{{144}}{\left( {{D_{{\alpha _1}}}{D_{{\alpha _2}}}{D_{{\alpha _3}}}{D_{{\alpha _4}}}G_{\mu \nu }^D} \right)_x}\left( {{v^\mu }\dfrac{1}{{v \cdot p}}{\partial ^\nu }{\partial ^{{\alpha _1}}}{\partial ^{{\alpha _2}}}{\partial ^{{\alpha _3}}}{\partial ^{{\alpha _4}}}\dfrac{1}{{v \cdot p}}} \right) \notag\\
        &\hspace{80pt} - \dfrac{1}{9}{\left( {{D_{{\alpha _1}}}G_{\mu \nu }^D} \right)_x}{\left( {{D_{{\alpha _2}}}G_{\rho \sigma }^D} \right)_x}\left( {{v^\mu }{v^\rho }\dfrac{1}{{v \cdot p}}{\partial ^\nu }{\partial ^{{\alpha _1}}}\dfrac{1}{{v \cdot p}}{\partial ^\sigma }{\partial ^{{\alpha _2}}}\dfrac{1}{{v \cdot p}}} \right) \notag\\
        &\hspace{80pt} - \dfrac{1}{{16}}G_{\mu \nu }^D{\left( {{D_{{\alpha _1}}}{D_{{\alpha _2}}}G_{\rho \sigma }^D} \right)_x}\left( {{v^\mu }{v^\rho }\dfrac{1}{{v \cdot p}}{\partial ^\nu }\dfrac{1}{{v \cdot p}}{\partial ^\sigma }{\partial ^{{\alpha _1}}}{\partial ^{{\alpha _2}}}\dfrac{1}{{v \cdot p}}} \right) \notag\\
        &\hspace{80pt} - \dfrac{1}{{16}}{\left( {{D_{{\alpha _1}}}{D_{{\alpha _2}}}G_{\mu \nu }^D} \right)_x}G_{\rho \sigma }^D\left( {{v^\mu }{v^\rho }\dfrac{1}{{v \cdot p}}{\partial ^\nu }{\partial ^{{\alpha _1}}}{\partial ^{{\alpha _2}}}\dfrac{1}{{v \cdot p}}{\partial ^\sigma }\dfrac{1}{{v \cdot p}}} \right) \notag\\
        &\hspace{80pt} + \dfrac{1}{8}G_{\mu \nu }^DG_{\rho \sigma }^DG_{\lambda \eta }^D\left( {{v^\mu }{v^\rho }{v^\lambda }\dfrac{1}{{v \cdot p}}{\partial ^\nu }\dfrac{1}{{v \cdot p}}{\partial ^\sigma }\dfrac{1}{{v \cdot p}}{\partial ^\eta }\dfrac{1}{{v \cdot p}}} \right) \,.
\end{align}
\endgroup

\subsubsection{Functional Traces and Determinants}\label{appsubsubsec:functionaltraces}
Using the expressions provided in \cref{appsec:LoopIntegrals} and~\cref{appsubsubsec:Expansions}, we have the tools we need to evaluate various functional traces using the original CDE. In what follows, we provide a few explicit results.

\subsubsection*{Single Relativistic Propagator}
For functional traces with $n$ factors of the same relativistic propagator, we adopt the following notation
\begin{subequations}\label{eqn:T0Tn}
\begin{align}
{T_0}\left( {{M^2}} \right) &\equiv i \Tr \left( {\frac{1}{{{D^2} + {M^2}}}} \right) \,, \\
{T_n}\left( {{M^2}} \right) &\equiv i \Tr \left[ {\frac{1}{{{D^2} + {M^2}}}\, {U_1} \cdots \frac{1}{{{D^2} + {M^2}}}\, {U_n}} \right] \,,
\end{align}
\end{subequations}
where $U_i(x)$ are purely functions of $x$. Assuming that each factor of $U_i$ contributes at least dimension one to the effective operators, the results up to dimension six are
\begingroup
\allowdisplaybreaks
\begin{subequations} \label{eqn:TraceSingle}
\begin{align}
{T_0}\left( {{M^2}} \right) &\supset \int {\dd^4}x\, \dfrac{1}{{{{\left( {4\pi } \right)}^2}}} \tr \Bigg\{ \dfrac{1}{{{M^2}}}\dfrac{{ - 1}}{{12}}G_{\mu \nu }^D{G^{D,\mu \nu }} \notag\\[-5pt]
          &\hspace{20pt} + \dfrac{1}{{{M^4}}}\dfrac{1}{{180}}\Bigg[ 2{\eta ^{\nu \rho }}{\eta ^{\sigma \lambda }}{\eta ^{\eta \mu }}G_{\mu \nu }^DG_{\rho \sigma }^DG_{\lambda \eta }^D - 3{\eta ^{\nu \sigma }}{\left( {{D^\mu }G_{\mu \nu }^D} \right)}_x {\left( {{D^\rho }G_{\rho \sigma }^D} \right)}_x \Bigg] \Bigg\} \,, \label{eqn:T0} \\[5pt]
{T_1}\left( {{M^2}} \right) &\supset \int {{\dd^4}x\,\dfrac{1}{{{{\left( {4\pi } \right)}^2}}} \tr \left[ {{M^2}\left( {\ln \dfrac{{{\mu ^2}}}{{{M^2}}} + 1} \right){U_1} + \dfrac{1}{{{M^2}}}\dfrac{{ - 1}}{{12}}{U_1}G_{\mu \nu }^D{G^{D,\mu \nu }}} \right]} \,, \label{eqn:T1} \\[5pt]
{T_2}\left( {{M^2}} \right) &\supset \int {\dd^4}x\,\dfrac{1}{{{{\left( {4\pi } \right)}^2}}} \tr \Bigg\{ \left( {\ln \dfrac{{{\mu ^2}}}{{{M^2}}}} \right)\left( { - {U_1}{U_2}} \right) + \dfrac{1}{{{M^2}}}\dfrac{1}{6}{\left( {{U_1}{D^2}{U_2}} \right)_x} \notag\\[-3pt]
          &\hspace{40pt} + \dfrac{1}{{{M^4}}}\dfrac{{ - 1}}{{60}} \Bigg[ 2{\left( {{D^2}{U_1}} \right)_x}{\left( {{D^2}{U_2}} \right)_x} - {\left( {{D_\mu }{D_\nu }{U_1}} \right)_x}{\left( {{D^\mu }{D^\nu }{U_2}} \right)_x} \notag\\[-3pt]
          &\hspace{70pt} + {U_1}G_{\mu \nu }^D{G^{D,\mu \nu }}{U_2} + 3{U_1}G_{\mu \nu }^D{U_2}{G^{D,\mu \nu }} + {U_1}{U_2}G_{\mu \nu }^D{G^{D,\mu\nu}} \Bigg] \Bigg\}  \,, \label{eqn:T2} \\[5pt]
{T_3}\left( {{M^2}} \right) &\supset \int {\dd^4}x\,\dfrac{1}{{{{\left( {4\pi } \right)}^2}}} \tr \Bigg\{ \dfrac{1}{{{M^2}}}\dfrac{{ - 1}}{2}{U_1}{U_2}{U_3} \notag\\[-8pt]
          &\hspace{100pt} + \dfrac{1}{{{M^4}}}\dfrac{{ - 1}}{{12}}\left[ {{{\left( {{D^\mu }{U_1}} \right)}_x}{{\left( {{D_\mu }{U_2}} \right)}_x}{U_3} + {\text{cyclic}}} \right] \Bigg\}\,, \label{eqn:T3} \\[5pt]
{T_4}\left( {{M^2}} \right) &\supset \int {\dd^4}x\,\dfrac{1}{{{{\left( {4\pi } \right)}^2}}} \tr \Bigg\{ \dfrac{1}{{{M^4}}}\dfrac{{ - 1}}{6}{U_1}{U_2}{U_3}{U_4} \notag\\[-5pt]
          & + \dfrac{1}{{{M^6}}}\dfrac{{ - 1}}{{60}}\left[ {3{{\left( {{D^\mu }{U_1}} \right)}_x}{{\left( {{D_\mu }{U_2}} \right)}_x}{U_3}{U_4} + 2{{\left( {{D^\mu }{U_1}} \right)}_x}{U_2}{{\left( {{D_\mu }{U_3}} \right)}_x}{U_4} + {\text{cyclic}}} \right] \Bigg\}  \,, \label{eqn:T4} \\[5pt]
{T_5}\left( {{M^2}} \right) &\supset \int {{\dd^4}x\,\dfrac{1}{{{{\left( {4\pi } \right)}^2}}} \tr \left( {\dfrac{1}{{{M^6}}}\dfrac{{ - 1}}{{12}}{U_1}{U_2}{U_3}{U_4}{U_5}} \right)} \,, \label{eqn:T5} \\[5pt]
{T_6}\left( {{M^2}} \right) &\supset \int {{\dd^4}x\,\dfrac{1}{{{{\left( {4\pi } \right)}^2}}} \tr \left( {\dfrac{1}{{{M^8}}}\dfrac{{ - 1}}{{20}}{U_1}{U_2}{U_3}{U_4}{U_5}{U_6}} \right)} \,. \label{eqn:T6}
\end{align}
\end{subequations}
\endgroup

\subsubsection*{Mixed Relativistic Propagators}
The following functional trace with two different relativistic propagators is often useful:
\begingroup
\allowdisplaybreaks
\begin{align}
{T_{1,1}}\left( {{m^2},{M^2}} \right) & \equiv i \Tr \Bigg[ \dfrac{1}{{{D^2} + {m^2}}}\, U_1 \dfrac{1}{{{D^2} + {M^2}}}\, U_2 \Bigg] \notag \\
 &\supset \int{\dd^4}x \tr \Bigg[ - I_{1,1}^0{U_1}{U_2} - \dfrac{1}{2}I_{2,2}^1{\left( {{U_1}{D^2}{U_2}} \right)_x} \notag\\
          &\hspace{30pt} - \dfrac{1}{9}\left( {3I_{2,3}^1 + I_{3,3}^2} \right){\left( {{D^2}{U_1}} \right)_x}{\left( {{D^2}{U_2}} \right)_x} \notag\\
          &\hspace{30pt} + \dfrac{1}{9}\left( {3I_{2,3}^1 - 2I_{3,3}^2} \right){\left( {{D_\mu }{D_\nu }{U_1}} \right)_x}{\left( {{D^\mu }{D^\nu }{U_2}} \right)_x} \notag\\
          &\hspace{30pt} + \dfrac{1}{3}\left( {I_{2,3}^1 - I_{3,2}^1} \right){\left( {{D^\mu }{U_1}} \right)_x}{\left( {{D^\nu }{U_2}} \right)_x}G_{\mu \nu }^D \notag\\
          &\hspace{30pt} - \dfrac{1}{{18}}\left( {9I_{1,3}^0 - 9I_{1,4}^1 - 6I_{2,3}^1 + I_{3,3}^2} \right){U_1}G_{\mu \nu }^D{G^{D,\mu \nu }}{U_2} \notag\\
          &\hspace{30pt} - \dfrac{1}{{18}}{U_1}G_{\mu \nu }^D{U_2}{G^{D,\mu \nu }}\left( {9I_{2,3}^1 + 3I_{3,2}^1 - 2I_{3,3}^2} \right) \notag\\
          &\hspace{30pt} - \dfrac{1}{{18}}{U_1}{U_2}G_{\mu \nu }^D{G^{D,\mu \nu }}\left( {9I_{3,1}^0 - 3I_{2,3}^1 - 3I_{3,2}^1 - 9I_{4,1}^1 + I_{3,3}^2} \right) \Bigg] \,.
\label{eqn:TraceMixed}
\end{align}
\endgroup
The definitions and values of the loop integrals in this expression can be found in~\cref{appsubsec:HLIntegrals}.

\subsubsection*{Relativistic and Linear Propagators}
A few functional traces containing both relativistic propagators and HQET propagators are important for the calculations in the main text.
\begingroup
\allowdisplaybreaks
\begin{subequations}\label{eqn:TraceHQET}
\begin{align}
{T_{1v}}\left( {{m^2}} \right) &\equiv i \Tr \left( {\dfrac{1}{D^2+m^2}\, {U_1}\dfrac{1}{{v \cdot iD}}\, {U_2}} \right) \notag\\
 &\supset \int {\dd^4}x\, \dfrac{1}{{{{\left( {4\pi } \right)}^2}}} \tr \Bigg\{ \left( {\ln \dfrac{{{\mu ^2}}}{{{m^2}}}} \right)2{{\left( {{U_1}v \cdot iD{U_2}} \right)}_x} + \dfrac{1}{{{m^2}}}\dfrac{{4}}{3}{{\left[ {{U_1}{{\left( {v \cdot iD} \right)}^3}{U_2}} \right]}_x} \notag\\[-3pt]
          &\hspace{190pt} + \dfrac{1}{{{m^2}}}\dfrac{{1}}{3}{U_1}{U_2}{{\left( {{v^\nu }i{D^\mu }G_{\mu \nu }^D} \right)}_x} \Bigg\} \,, \label{eqn:T1v} \\[5pt]
{T_{1vv}}\left( {{m^2}} \right) &\equiv i \Tr \left( {\dfrac{1}{D^2+m^2}\, {U_1}\dfrac{1}{{v \cdot iD}}\, {U_2}\dfrac{1}{{v \cdot iD}}\, {U_3}} \right) \notag\\
 &\supset \int {\dd^4}x\, \dfrac{1}{{{{\left( {4\pi } \right)}^2}}} \tr \Bigg\{ \left( {\ln \dfrac{{{\mu ^2}}}{{{m^2}}}} \right)\left(-2\right){U_1}{U_2}{U_3} + \dfrac{1}{{{m^2}}}\dfrac{{4}}{3}{v^{{\alpha _1}}}{v^{{\alpha _2}}} \Bigg[ {U_1}{\left( {{D_{{\alpha _1}}}{D_{{\alpha _2}}}{U_2}} \right)_x}{U_3} \notag\\[-3pt]
          &\hspace{80pt} + 3{U_1}{\left( {{D_{{\alpha _1}}}{U_2}} \right)_x}{\left( {{D_{{\alpha _2}}}{U_3}} \right)_x} + 3{U_1}{U_2}{\left( {{D_{{\alpha _1}}}{D_{{\alpha _2}}}{U_3}} \right)_x} \Bigg] \Bigg\} \,, \label{eqn:T1vv} \\[5pt]
{T_{2v}}\left( {{m^2}} \right) &\equiv i \Tr \left( {\dfrac{1}{D^2+m^2}\, {U_1}\dfrac{1}{D^2+m^2}\, {U_2}\dfrac{1}{{v \cdot iD}}\, {U_3}} \right) \notag\\
 &\supset \int {\dd^4}x\, \dfrac{1}{{{{\left( {4\pi } \right)}^2}}} \tr \Bigg\{ \dfrac{1}{{{m^2}}}\left[ {2{U_1}{{\left( {v \cdot iD{U_2}} \right)}_x}{U_3} + {{\left( {v \cdot iD{U_1}} \right)}_x}{U_2}{U_3}} \right] \notag\\
          &\hspace{20pt} - \dfrac{1}{{{m^4}}}\dfrac{1}{{18}} \Bigg[ 12{v^{{\alpha _1}}}{v^{{\alpha _2}}}{v^{{\alpha _3}}} \Bigg( 3{\left( {i{D_{{\alpha _1}}}{U_1}} \right)_x}{\left( {{D_{{\alpha _2}}}{D_{{\alpha _3}}}{U_2}} \right)_x}{U_3} \notag\\[-5pt]
          &\hspace{150pt} + 2{U_1}{\left( {i{D_{{\alpha _1}}}{D_{{\alpha _2}}}{D_{{\alpha _3}}}{U_2}} \right)_x}{U_3} \Bigg) \notag\\
          &\hspace{70pt} + {\renewcommand\arraystretch{1.2}\mqty( {\eta ^{{\alpha _1}{\alpha _2}}}{v^{{\alpha _3}}} + {\eta ^{{\alpha _1}{\alpha _3}}}{v^{{\alpha _2}}} \\ + {\eta ^{{\alpha _2}{\alpha _3}}}{v^{{\alpha _1}}} + 6{v^{{\alpha _1}}}{v^{{\alpha _2}}}{v^{{\alpha _3}}} ) }
          {\left( {i{D_{{\alpha _1}}}{D_{{\alpha _2}}}{D_{{\alpha _3}}}{U_1}} \right)_x}{U_2}{U_3} \notag\\
          &\hspace{70pt} + 6\left( {{\eta ^{{\alpha _1}{\alpha _2}}} + 4{v^{{\alpha _1}}}{v^{{\alpha _2}}}} \right){v^{{\alpha _3}}}{\left( {{D_{{\alpha _1}}}{D_{{\alpha _2}}}{U_1}} \right)_x}{\left( {i{D_{{\alpha _3}}}{U_2}} \right)_x}{U_3} \notag\\
          &\hspace{70pt} + 2{v^\nu} \Bigg( 3{\left( {i{D^\mu }{U_1}} \right)_x}G_{\mu \nu }^D{U_2}{U_3} + {\left( {i{D^\mu }G_{\mu \nu }^D} \right)_x}{U_1}{U_2}{U_3} \notag\\[-5pt]
          &\hspace{200pt} + 2{U_1}{\left( {i{D^\mu }G_{\mu \nu }^D} \right)_x}{U_2}{U_3} \Bigg) \Bigg] \Bigg\} \,. \label{eqn:T2v}
\end{align}
\end{subequations}
\endgroup

\subsubsection*{Universal Functional Determinants}
Finally, with the previous results in hand, the following functional determinants can be evaluated:
\begingroup
\allowdisplaybreaks
\begin{subequations}\label{eqn:UniversalDeterminants}
\begin{align}
&i\ln \det \left( {{D^2} + {M^2} + U} \right) \supset \int{\dd^4}x\, \dfrac{1}{{{{\left( {4\pi } \right)}^2}}} \tr \Bigg\{
{M^2}\left( {\ln \dfrac{{{\mu ^2}}}{{{M^2}}} + 1} \right)U \notag\\
          &\hspace{50pt} + \left( {\ln \dfrac{{{\mu ^2}}}{{{M^2}}}} \right)\left( {\dfrac{1}{2}{U^2} + \dfrac{1}{{12}}G_{\mu \nu }^D{G^{D,\mu \nu }}} \right) \nonumber \\
          &\hspace{50pt} + \dfrac{1}{{{M^2}}} \Bigg[ -\dfrac{{1}}{{90}}{\eta ^{\nu \rho }}{\eta ^{\sigma \lambda }}{\eta ^{\eta \mu }}G_{\mu \nu }^DG_{\rho \sigma }^DG_{\lambda \eta }^D + \dfrac{1}{{60}}{\eta ^{\nu \sigma }}{\left( {{D^\mu }G_{\mu \nu }^D} \right)_x}{\left( {{D^\rho }G_{\rho \sigma }^D} \right)_x} \notag\\[-5pt]
          &\hspace{90pt} - \dfrac{1}{6}{U^3} + \dfrac{1}{{12}}{\left( {{D^\mu }U} \right)_x}{\left( {{D_\mu }U} \right)_x} - \dfrac{1}{{12}}UG_{\mu \nu }^D{G^{D,\mu \nu }} \Bigg] \notag\\
          &\hspace{50pt} + \dfrac{1}{{{M^4}}} \Bigg[ \dfrac{1}{{24}}{U^4} - \dfrac{1}{{12}}U{\left( {{D^\mu }U} \right)_x}{\left( {{D_\mu }U} \right)_x} + \dfrac{1}{{120}}{\left( {{D^2}U} \right)_x}{\left( {{D^2}U} \right)_x} \notag\\[-5pt]
          &\hspace{90pt} + \dfrac{1}{{60}}{\left( {{D^\mu }U} \right)_x}{\left( {{D^\nu }U} \right)_x}G_{\mu \nu }^D + \dfrac{1}{{60}}UG_{\mu \nu }^DU{G^{D,\mu \nu }} + \dfrac{1}{{40}}{U^2}G_{\mu \nu }^D{G^{D,\mu \nu }} \Bigg] \notag\\
          &\hspace{50pt} + \dfrac{1}{{{M^6}}}\left[ { - \dfrac{1}{{60}}{U^5} + \dfrac{1}{{20}}{U^2}{{\left( {{D^\mu }U} \right)}_x}{{\left( {{D_\mu }U} \right)}_x} + \dfrac{1}{{30}}U{{\left( {{D^\mu }U} \right)}_x}U{{\left( {{D_\mu }U} \right)}_x}} \right] \notag\\
          &\hspace{50pt} + \dfrac{1}{{{M^8}}}\dfrac{1}{{120}}{U^6} \Bigg\} \,, \label{eqn:UniversalDU} \\
&i\ln \det \left( {{D^2} + {M^2} + U + i{D_\mu }{J^\mu }} \right) \supset \int{\dd^4}x\, \dfrac{1}{{{{\left( {4\pi } \right)}^2}}} \tr \Bigg\{
{M^2}\left( {\ln \dfrac{{{\mu ^2}}}{{{M^2}}} + 1} \right)\left( {U + \dfrac{1}{4}{J^\mu }{J_\mu }} \right) \notag\\
          &\hspace{50pt} + \left( {\ln \dfrac{{{\mu ^2}}}{{{M^2}}}} \right) \Bigg[ \dfrac{1}{2}{U^2} + \dfrac{1}{{12}}G_{\mu \nu }^D{G^{D,\mu \nu }} + \dfrac{1}{4}U{J^2} + \dfrac{1}{2}U{\left( {i{D^\mu }{J_\mu }} \right)_x} \notag\\[-5pt]
          &\hspace{115pt} - \dfrac{1}{{12}}{J^\mu }{J^\nu }G_{\mu \nu }^D + \dfrac{1}{6}{\left( {i{D^\mu }{J^\nu }} \right)_x}G_{\mu \nu }^D - \dfrac{1}{{12}}{\left( {i{D_\mu }{J_\nu }} \right)_x}\left( {2{J^\mu }{J^\nu } + {J^\nu }{J^\mu }} \right) \notag\\
          &\hspace{115pt} - \dfrac{1}{{24}}\left( {{\eta ^{\mu \nu }}{\eta ^{\rho \sigma }} + {\eta ^{\mu \rho }}{\eta ^{\nu \sigma }} + {\eta ^{\mu \sigma }}{\eta ^{\nu \rho }}} \right){\left( {{D_\mu }{J_\nu }} \right)_x}{\left( {{D_\rho }{J_\sigma }} \right)_x} \notag\\
          &\hspace{115pt} + \dfrac{1}{{96}}\left( {{\eta ^{\mu \nu }}{\eta ^{\rho \sigma }} + {\eta ^{\mu \rho }}{\eta ^{\nu \sigma }} + {\eta ^{\mu \sigma }}{\eta ^{\nu \rho }}} \right){J_\mu }{J_\nu }{J_\rho }{J_\sigma } \Bigg] \notag\\
          &\hspace{50pt} + \dfrac{1}{{{M^2}}} \Bigg[ \dfrac{1}{{12}}{\left( {{D^\mu }U} \right)_x}{\left( {{D_\mu }U} \right)_x} - \dfrac{1}{6}{U^3} + \dfrac{1}{6}U{\left( {i{D^\mu }U} \right)_x}{J_\mu } \notag\\[-8pt]
          &\hspace{90pt} - \dfrac{1}{6}{U^2}{\left( {i{D^\mu }{J_\mu }} \right)_x} - \dfrac{1}{{12}}{U^2}{J^2} - \dfrac{1}{{24}}U{J_\mu }U{J^\mu } \Bigg] \notag\\
          &\hspace{50pt} + \dfrac{1}{{{M^4}}}\dfrac{1}{{24}}{U^4} \Bigg\} \,. \label{eqn:UniversalDUJ}
\end{align}
\end{subequations}
\endgroup
In expanding the above, we have assumed that $U$ and $J_\mu$ are purely functions of $x$, and that they contribute at least mass dimension one to the effective operator. The first determinant is truncated up to dimension six. This is the prototype of the UOLEA, which has now been studied extensively~\cite{Drozd:2015rsp, Ellis:2016enq, Zhang:2016pja, Ellis:2017jns, Summ:2018oko, Kramer:2019fwz}. Note that the second determinant contains an open covariant derivative, and we have provided the evaluation truncated at dimension four. This result was previously unknown as emphasized in~\cite{Ellis:2017jns, Brivio:2017vri, Kramer:2019fwz}.

\section{More on RGEs with CDE}
\label{appsec:ExamplesRGE}
In \cref{subsec:ReviewFunctionalRunning}, we discussed how one can obtain RGEs using functional methods, first introduced in \Ref{Henning:2016lyp}. The purpose of this appendix is to provide a pedagogical introduction to the methodology as applied to RGEs. First, we discuss a minor complication when computing the RGEs for gauge couplings in \cref{appsubsec:RGEgauge}, and emphasize how to maintain the manifest gauge invariance in this case. Then in \cref{sec:RGEExamples}, we provide detailed calculations that reproduce the RGEs for some classic examples: the real scalar $\phi^4$ theory in \cref{appsubsec:RGEphi}, a theory with two real scalars $\phi$ and $\Phi$ in \cref{appsubsec:RGEphiPhi}, and gauge theories in \cref{appsubsec:RGEQCDQED}.

\subsection{RGEs for Gauge Couplings}\label{appsubsec:RGEgauge}
The procedure described in \cref{subsec:ReviewFunctionalRunning} applies to any coupling $\lambda$. However, there is a minor complication when considering a gauge coupling $g$. If the 1PI effective action is written in terms of gauge invariant operators,\footnote{Note that this is not generically guaranteed. For example, in the traditional computation of the 1PI effective action for a non-Abelian gauge theory, the result will not be composed of gauge invariant operators due to the gauge fixing procedure, \eg see Chapter 16.5 of Peskin and Schroeder~\cite{Peskin:1995ev}. However, when using functional methods, we can benefit from background field gauge fixing \cref{eq:gaugefix} described in \cref{appsubsec:BackgroundField}. With this choice of gauge fixing, the 1PI effective action will be expressed in terms of gauge invariant operators, as emphasized in \eg Chapter 16.6 in Peskin and Schroeder~\cite{Peskin:1995ev}.} such as what we will obtain from a CDE procedure, there will be no explicit operator $\oper_g$ --- the gauge couplings are all hidden in the covariant derivatives. A brute force approach would be to abandon the manifest gauge invariance by splitting a gauge invariant term and isolating a part of it as a candidate $\oper_g$. In doing so, one typically gets multiple terms that can serve as $\oper_g$. For example, the QCD Lagrangian is
\begin{align}
  \Lag_\text{QCD} &= -\frac{1}{4} G_{\mu\nu}^a G^{\mu\nu,a} + \bar{\psi}\, i\slashed{D}\, \psi \nonumber \\
    &= \frac{1}{2} G_\mu^a (\eta^{\mu\nu}\partial^2 - \partial^\mu \partial^\nu) G_\nu^a
       - g_s f^{abc} (\partial_\mu G_\nu^a) G_\mu^b G_\nu^c
       - \frac{1}{4} g_s^2 f^{abc} f^{ade} G_\mu^b G_\nu^c  G^{\mu,d} G^{\nu,e} \nonumber \\
 &\qquad + \bar{\psi} i\slashed{\partial} \psi  + g_s \bar{\psi} \slashed{G} \psi \,.
 \label{eqn:LQCDtermSplit}
\end{align}
There are no explicit factors of the gauge coupling in the first line, while expanding the gluon field strength and covariant derivatives exposes the $g_s$ dependence. Clearly, there are three terms that could be identified as $\oper_g$. This is not a problem --- one can choose to work with any of these three operators since gauge invariance will ensure that the results are the same. However, there is another approach, which avoids breaking up the gauge invariant terms. This is additionally convenient here since the CDE approach yields a form of the 1PI effective action that is already packaged in this way.

Again taking QCD as the example, we focus on the gauge invariant kinetic term for the gauge fields:
\begin{equation}
  \Lag_\text{QCD} \supset  -\frac{1}{4} G_{\mu\nu}^a G^{\mu\nu,a}
    = \frac{1}{4 g_s^2} G_{\mu\nu}^{D,a} G^{D,a,\mu\nu} \,,
\label{eq:LQCDwithGD}
\end{equation}
where we are using a notation defined in \cref{eq:defGD} that is convenient when doing CDE calculations: $\comm{D_\mu}{D_\nu} \equiv G_{\mu\nu}^D = G_{\mu\nu}^{D,a} T^a = -i g_s G_{\mu\nu}^a T^a$. (See \cref{appsubsec:ToolsTricks} for a list of conventions.) Then the 1PI effective action takes the form
\begin{equation}
  \Gamma[\phi] \supset \int  \dd[4]{x} \pqty{ \frac{1}{4g_s^2} + a_g} G_{\mu\nu}^{D,a} G^{D,a,\mu\nu} \,,
\label{eqn:GammaQCDg}
\end{equation}
where $a_g$ represents the loop corrections.

Note that $D_\mu$ does not run as long as gauge invariance is maintained at the perturbative level. Since $G_{\mu\nu}^D$ is defined as the commutator of the covariant derivative, this term does not run either. Therefore, the RG invariance of the coefficient of this term directly yields the RGE for the gauge coupling:
\begin{equation}
  \mu \dv{\mu} \pqty{ \frac{1}{4g_s^2} + a_g } = 0 \,.
\label{eqn:RGEQCDgCDE}
\end{equation}
This is the approach that will be used in what follows for computing RGEs of gauge couplings.

For completeness, we will quickly show that the result derived by working with an individual non-gauge invariant term would yield an equivalent expression. Expanding the gauge invariant kinetic term in \cref{eqn:GammaQCDg} gives
\begin{align}
  \pqty{ \frac{1}{4g_s^2} + a_g} G_{\mu\nu}^{D,a} G^{D,a,\mu\nu}
    &= (1 + 4g_s^2 a_g)
        \bigg[ \frac{1}{2} G_\mu^a \pqty{ \eta^{\mu\nu}\partial^2 - \partial^\mu \partial^\nu } G_\nu^a- g_s f^{abc} (\partial^\mu G^{\nu,a}) G_\mu^b G_\nu^c \notag\\
      &   \hspace{80pt}     - \frac{1}{4} g_s^2 f^{abc} f^{ade} G_\mu^b G_\nu^c G^{\mu,d} G^{\nu,e} \bigg] \,.
\end{align}
Next, we rescale the gauge fields to renormalize the kinetic term
\begin{equation}
  G_\mu^a \to (1 + 4g_s^2 a_g)^{-1/2} G_\mu^a \,.
\label{eqn:Grescale}
\end{equation}
After this rescaling, the 1PI effective action in \cref{eqn:GammaQCDg} becomes
\begin{align}
  \Gamma[\phi] \to \int \dd[4]{x}
    \bigg[ \frac{1}{2} G_\mu^a \pqty{ \eta^{\mu\nu}\partial^2 - \partial^\mu \partial^\nu } G_\nu^a
    &- (1 + 4 g_s^2 a_g)^{-1/2} g_s f^{abc} (\partial^\mu G^{\nu,a}) G_\mu^b G_\nu^c \notag\\
    &- (1 + 4 g_s^2 a_g)^{-1} \frac{1}{4} g_s^2 f^{abc} f^{ade} G_\mu^b G_\nu^c G^{\mu,d} G^{\nu,e}
\bigg] \,.
\end{align}
Taking either the cubic or the quartic interaction term as $\oper_g$, we can use \cref{eqn:RGE1PIgeneral} to derive \cref{eqn:RGEQCDgCDE}. One can also perform a similar analysis on the fermion kinetic term $\bar{\psi} i\slashed{D} \psi$ (being careful to include both the rescaling of $\psi$ and $G_\mu$) as an alternative derivation that yields the same result.

\subsection{Example RGE Calculations}
\label{sec:RGEExamples}
In this section, we provide detailed calculations of RGEs for a number of well known examples.

\subsubsection{\boldmath Scalar $\phi^4$ Theory}\label{appsubsec:RGEphi}
We start with the simplest example, $\phi^4$ theory for a real scalar field $\phi$. The Lagrangian is
\begin{equation}
{{\cal L}_\phi } \supset \frac{1}{2}{\left( {\partial \phi } \right)^2} - \frac{1}{2}{m^2}{\phi ^2} - \frac{1}{{4!}}\lambda {\phi ^4} = \frac{1}{2}\phi \left( { - {D^2} - {m^2}} \right)\phi  - \frac{1}{{4!}}\lambda {\phi ^4} \,.
\label{eq:Lphi4}
\end{equation}
We will demonstrate how to compute the RGEs for the couplings $m^2$ and $\lambda$. Starting with the Lagrangian in \cref{eq:Lphi4}, we take the second variation with respect to $\phi$:
\begin{equation}
\frac{\delta^2 S_\phi}{\left(\delta\phi\right)^2} =  - {D^2} - {m^2} - \frac{1}{2}\lambda {\phi ^2} \,.
\end{equation}
Making use of the universal result \cref{eqn:UniversalDU} with $U=U_\phi=\frac{1}{2}\lambda\phi^2$, we see that the one-loop 1PI effective action contains
\begin{align}
\Gamma _\phi ^{\left( 1 \right)} &= \frac{i}{2}\ln {\textstyle\det_\phi }\left[ { - \frac{{{\delta ^2}{S_\phi }}}{{{{\left( {\delta \phi } \right)}^2}}}} \right] = \frac{i}{2}\ln {\textstyle\det_\phi }\left( {{D^2} + {m^2} + \frac{1}{2}\lambda {\phi ^2}} \right) \nonumber \\
 &\supset \frac{1}{2}\int {{\dd^4}x\frac{1}{{{{\left( {4\pi } \right)}^2}}}{\tr_\phi }\left[ {{m^2}\left( {\ln \frac{{{\mu ^2}}}{{{m^2}}} + 1} \right){U_\phi } + \left( {\ln \frac{{{\mu ^2}}}{{{m^2}}}} \right)\frac{1}{2}U_\phi ^2} \right]} \nonumber \\
 &= \int {{\dd^4}x\frac{1}{{{{\left( {4\pi } \right)}^2}}}\left[ {{m^2}\left( {\ln \frac{{{\mu ^2}}}{{{m^2}}} + 1} \right)\frac{1}{4}\lambda {\phi ^2} + \left( {\ln \frac{{{\mu ^2}}}{{{m^2}}}} \right)\frac{1}{{16}}{\lambda ^2}{\phi ^4}} \right]} \,.
\end{align}
In the second line above, we have only kept terms up to mass dimension four. Combining it with the tree-level 1PI effective action $\Gamma_\phi^{\left( 0 \right)} = S_\phi = \int \dd^4x\, {\cal L}_\phi$, we get
\begin{align}
\Gamma _\phi ^{\left( 0 \right)} + \Gamma _\phi ^{\left( 1 \right)} \supset \int {\dd^4}x\left\{ \frac{1}{2}\phi \left( { - {D^2}} \right)\phi  - \frac{1}{2}\left[ {{m^2} - \frac{\lambda }{{2{{\left( {4\pi } \right)}^2}}}{m^2}\left( {\ln \frac{{{\mu ^2}}}{{{m^2}}} + 1} \right)} \right]{\phi ^2}\right. \notag\\
\left.- \frac{1}{{4!}}\left[ {\lambda  - \frac{{3{\lambda ^2}}}{{2{{\left( {4\pi } \right)}^2}}}\left( {\ln \frac{{{\mu ^2}}}{{{m^2}}}} \right)} \right]{\phi ^4} \right\} \,.
\end{align}
The kinetic term is already canonically normalized, so no additional field redefinition is required. We then simply use \cref{eqn:RGE1PIgeneral} to find the RGEs:
\begin{subequations}
\begin{alignat}{3}
0 &= \mu \frac{\dd}{{\dd\mu }}\left[ {{m^2} - \frac{\lambda }{{2{{\left( {4\pi } \right)}^2}}}{m^2}\left( {\ln \frac{{{\mu ^2}}}{{{m^2}}} + 1} \right)} \right] & \quad\Rightarrow\quad && \mu \frac{\dd}{{\dd\mu }}{m^2} &= \frac{\lambda }{{{{\left( {4\pi } \right)}^2}}}{m^2} \,, \\
0 &= \mu \frac{\dd}{{\dd\mu }}\left[ {\lambda  - \frac{{3{\lambda ^2}}}{{2{{\left( {4\pi } \right)}^2}}}\left( {\ln \frac{{{\mu ^2}}}{{{m^2}}}} \right)} \right] & \quad\Rightarrow\quad && \mu \frac{\dd}{{\dd\mu }}\lambda  &= \frac{{3{\lambda ^2}}}{{{{\left( {4\pi } \right)}^2}}} \,.
\end{alignat}
\end{subequations}

\subsubsection{Heavy-Light Scalar Theory}\label{appsubsec:RGEphiPhi}
As a second example, we consider a theory with two real scalar fields $\phi$ and $\Phi$, with masses $m$ and $M$ respectively. This toy model was analyzed extensively using Feynman diagrams in~\cite{Cohen:2019wxr}. The Lagrangian is
\begin{equation}
{{\cal L}_{\left( {\phi ,\Phi } \right)}} = \frac{1}{2}\phi \left( { - {D^2} - {m^2}} \right)\phi  + \frac{1}{2}\Phi \left( { - {D^2} - {M^2}} \right)\Phi  - \frac{1}{{4!}}\eta {\phi ^4} - \frac{1}{4}\kappa {\Phi ^2}{\phi ^2} \,,
\end{equation}
where we are not including a $\Phi$ self-interaction for simplicity.\footnote{Of course this coupling is generated by the RGEs, and it is trivial to generalize our calculation here to incorporate this effect.} We would like to compute the RGEs for the couplings $m^2$, $M^2$, $\eta$, and $\kappa$. As before, we first compute the second variation
\begin{equation}\renewcommand\arraystretch{1.5}
{\delta ^2}{S_{\left( {\phi ,\Phi } \right)}} = \mqty( \delta \phi & \delta \Phi ) \mqty( - {D^2} - {m^2} - \frac{1}{2}\eta {\phi ^2} - \frac{1}{2}\kappa {\Phi ^2} & -\kappa \phi \Phi \\
 -\kappa \phi \Phi & -{D^2} - {M^2} - \frac{1}{2}\kappa \phi^2 ) \mqty( \delta \phi \\ \delta \Phi ) \,.
\end{equation}
This is a $2\times2$ matrix, which we can diagonalize as
\begin{align}
 - \frac{{{\delta ^2}{S_{\left( {\phi ,\Phi } \right)}}}}{{\delta {{\left( {\phi ,\Phi } \right)}^2}}} &= \mqty( {D^2} + {m^2} + \frac{1}{2}\eta {\phi ^2} + \frac{1}{2}\kappa {\Phi^2} & \kappa \phi \Phi \\
  \kappa \phi \Phi & {D^2} + {M^2} + \frac{1}{2}\kappa \phi^2 ) \notag\\
 &\to \mqty( {D^2} + {m^2} + \frac{1}{2}\eta {\phi ^2} + \frac{1}{2}\kappa {\Phi ^2} - \kappa \phi \Phi \dfrac{1}{D^2 + M^2 + \frac{1}{2}\kappa \phi^2} \kappa \phi \Phi & 0 \\
            \kappa \phi \Phi & {D^2} + {M^2} + \frac{1}{2}\kappa {\phi ^2} ) \,. \notag
\end{align}
Therefore the one-loop level 1PI effective action is given by two terms:
\begin{align}
\Gamma _{\left( {\phi ,\Phi } \right)}^{\left( 1 \right)} &= \frac{i}{2}\ln {\textstyle\det _\phi }\left[ { - \frac{{{\delta ^2}{S_{\left( {\phi ,\Phi } \right)}}}}{{\delta {{\left( {\phi ,\Phi } \right)}^2}}}} \right] \notag\\
&= \frac{i}{2}\ln {\textstyle\det _\phi }\left[ {{D^2} + {m^2} + \frac{1}{2}\eta {\phi ^2} + \frac{1}{2}\kappa {\Phi ^2} - \kappa \phi \Phi \frac{1}{{{D^2} + {M^2} + \frac{1}{2}\kappa {\phi ^2}}}\kappa \phi \Phi } \right] \nonumber \\
 &\quad + \frac{i}{2}\ln {\textstyle\det _\Phi }\left( {{D^2} + {M^2} + \frac{1}{2}\kappa {\phi ^2}} \right) \,.
\end{align}
The terms we need to keep are the kinetic terms for $\phi$ and $\Phi$, as well as the potential terms $\phi^2$, $\Phi^2$, $\phi^4$ and $\Phi^2\phi^2$. With this in mind, we evaluate the first term as
\begin{align}
&\frac{i}{2}\ln {\textstyle\det_\phi }\left[ {{D^2} + {m^2} + \frac{1}{2}\eta {\phi ^2} + \frac{1}{2}\kappa {\Phi ^2} - \kappa \phi \Phi \frac{1}{{{D^2} + {M^2} + \frac{1}{2}\kappa {\phi ^2}}}\kappa \phi \Phi } \right] \notag\\[7pt]
\supset& \frac{i}{2}\Tr_\phi \ln \left[ {1 + \frac{1}{{{D^2} + {m^2}}}\frac{1}{2}\left( {\eta {\phi ^2} + \kappa {\Phi ^2}} \right) - {\kappa ^2}\frac{1}{{{D^2} + {m^2}}}\phi \Phi \frac{1}{{{D^2} + {M^2}}}\phi \Phi } \right] \notag \\[7pt]
\supset& \frac{1}{4}i\Tr_\phi \left[ {\frac{1}{{{D^2} + {m^2}}}\left( {\eta {\phi ^2} + \kappa {\Phi ^2}} \right)} \right] - \frac{{{\kappa ^2}}}{2}i\,\Tr_\phi \left[ {\frac{1}{{{D^2} + {m^2}}}\phi \Phi \frac{1}{{{D^2} + {M^2}}}\phi \Phi } \right] \notag \\
&\hspace{50pt} - \frac{1}{{16}}i\Tr_\phi \left[ {\frac{1}{{{D^2} + {m^2}}}\left( {\eta {\phi ^2} + \kappa {\Phi ^2}} \right)\frac{1}{{{D^2} + {m^2}}}\left( {\eta {\phi ^2} + \kappa {\Phi ^2}} \right)} \right] \notag \\[7pt]
\supset& \int {\dd^4}x\, \frac{1}{{{{\left( {4\pi } \right)}^2}}} \Bigg\{ \dfrac{1}{4}{m^2}\left( {\ln \dfrac{{{\mu ^2}}}{{{m^2}}} + 1} \right)\left( {\eta {\phi ^2} + \kappa {\Phi ^2}} \right) + \dfrac{1}{{16}}\ln \dfrac{{{\mu ^2}}}{{{m^2}}}{\left( {\eta {\phi ^2} + \kappa {\Phi ^2}} \right)^2} \notag\\[-3pt]
&\hspace{50pt} + \dfrac{{{\kappa ^2}}}{2}\left[ {1 + \dfrac{1}{{{M^2} - {m^2}}}\left( {{M^2}\ln \dfrac{{{\mu ^2}}}{{{M^2}}} - {m^2}\ln \dfrac{{{\mu ^2}}}{{{m^2}}}} \right)} \right]{\Phi ^2}{\phi ^2} \Bigg\} \,,
\end{align}
where the last line was obtained by using \cref{eqn:TraceSingle,eqn:TraceMixed}. For the second term, we use \cref{eqn:UniversalDU} with $U=\frac{1}{2}\kappa\phi^2$ to obtain
\begin{align}
\frac{i}{2}\ln {\textstyle\det _\Phi }\left( {{D^2} + {M^2} + \frac{1}{2}\kappa {\phi ^2}} \right) = \int {{\dd^4}x\frac{1}{{{{\left( {4\pi } \right)}^2}}}\left[ {{M^2}\left( {\ln \frac{{{\mu ^2}}}{{{M^2}}} + 1} \right)\frac{1}{4}\kappa {\phi ^2} + \left( {\ln \frac{{{\mu ^2}}}{{{M^2}}}} \right)\frac{1}{{16}}{\kappa ^2}{\phi ^4}} \right]} \,. \notag
\end{align}
Now putting everything together, we find
\begin{align}
\Gamma_{\left( {\phi ,\Phi } \right)}^{\left( 0 \right)} + \Gamma_{\left( {\phi ,\Phi } \right)}^{\left( 1 \right)} &\supset \int \dd^4x\,\Bigg\{
\dfrac{1}{2}\phi \left( { - {D^2}} \right)\phi  + \dfrac{1}{2}\Phi \left( { - {D^2}} \right)\Phi \notag \\
 &\hspace{50pt} - \dfrac{1}{2}\left[ {{m^2} - \dfrac{\eta }{{2{{\left( {4\pi } \right)}^2}}}{m^2}\left( {\ln \dfrac{{{\mu ^2}}}{{{m^2}}} + 1} \right) - \dfrac{\kappa }{{2{{\left( {4\pi } \right)}^2}}}{M^2}\left( {\ln \dfrac{{{\mu ^2}}}{{{M^2}}} + 1} \right)} \right]{\phi ^2} \notag\\
 &\hspace{50pt} - \dfrac{1}{2}\left[ {{M^2} - \dfrac{\kappa }{{2{{\left( {4\pi } \right)}^2}}}{m^2}\left( {\ln \dfrac{{{\mu ^2}}}{{{m^2}}} + 1} \right)} \right]{\Phi ^2} \notag\\
 &\hspace{50pt} - \dfrac{1}{4} \Bigg[ \kappa - \dfrac{{2{\kappa ^2}}}{{{{\left( {4\pi } \right)}^2}}}\left( {1 + \dfrac{1}{{{M^2} - {m^2}}}\left( {{M^2}\ln \dfrac{{{\mu ^2}}}{{{M^2}}} - {m^2}\ln \dfrac{{{\mu ^2}}}{{{m^2}}}} \right)} \right) \notag\\
 &\hspace{100pt} - \dfrac{{\eta \kappa }}{{2{{\left( {4\pi } \right)}^2}}}\ln \dfrac{{{\mu ^2}}}{{{m^2}}} \Bigg] \Phi^2 \phi^2 \notag\\
 &\hspace{50pt} - \dfrac{1}{{4!}}\left[ {\eta  - \dfrac{{3{\eta ^2}}}{{2{{\left( {4\pi } \right)}^2}}}\ln \dfrac{{{\mu ^2}}}{{{m^2}}} - \dfrac{{3{\kappa ^2}}}{{2{{\left( {4\pi } \right)}^2}}}\ln \dfrac{{{\mu ^2}}}{{{M^2}}}} \right]{\phi ^4}
 \Bigg\} \,.
\end{align}
Similar to the single scalar case, the kinetic terms are already canonically normalized. We do not need any further field redefinitions and directly obtain the RGEs as
\begingroup
\allowdisplaybreaks
\begin{subequations}
\begin{align}
0 &= \mu \frac{\dd}{{\dd\mu }}\left[ {{m^2} - \frac{\eta }{{2{{\left( {4\pi } \right)}^2}}}{m^2}\left( {\ln \frac{{{\mu ^2}}}{{{m^2}}} + 1} \right) - \frac{\kappa }{{2{{\left( {4\pi } \right)}^2}}}{M^2}\left( {\ln \frac{{{\mu ^2}}}{{{M^2}}} + 1} \right)} \right]  \notag\\
&\qquad\Longrightarrow\quad \mu \frac{\dd}{{\dd\mu }}{m^2} = \frac{\eta }{{{{\left( {4\pi } \right)}^2}}}{m^2} + \frac{\kappa }{{{{\left( {4\pi } \right)}^2}}}{M^2} \,,  \\[10pt]
0 &= \mu \frac{\dd}{{\dd\mu }}\left[ {{M^2} - \frac{\kappa }{{2{{\left( {4\pi } \right)}^2}}}{m^2}\left( {\ln \frac{{{\mu ^2}}}{{{m^2}}} + 1} \right)} \right] \notag\\
&\qquad\Longrightarrow\quad \mu \frac{\dd}{{\dd\mu }}{M^2} = \frac{\kappa }{{{{\left( {4\pi } \right)}^2}}}{m^2} \,,  \\[10pt]
0 &= \mu \frac{\dd}{{\dd\mu }}\left[ {\eta  - \frac{{3{\eta ^2}}}{{2{{\left( {4\pi } \right)}^2}}}\ln \frac{{{\mu ^2}}}{{{m^2}}} - \frac{{3{\kappa ^2}}}{{2{{\left( {4\pi } \right)}^2}}}\ln \frac{{{\mu ^2}}}{{{M^2}}}} \right] \notag \\
&\qquad\Longrightarrow\quad \mu \frac{\dd}{{\dd\mu }}\eta = \frac{{3{\eta ^2}}}{{{{\left( {4\pi } \right)}^2}}} + \frac{{3{\kappa ^2}}}{{{{\left( {4\pi } \right)}^2}}} \,,  \\[10pt]
0 &= \mu \frac{\dd}{{\dd\mu }}\left\{ {\kappa  - \frac{{2{\kappa ^2}}}{{{{\left( {4\pi } \right)}^2}}}\left[ {1 + \frac{1}{{{M^2} - {m^2}}}\left( {{M^2}\ln \frac{{{\mu ^2}}}{{{M^2}}} - {m^2}\ln \frac{{{\mu ^2}}}{{{m^2}}}} \right)} \right] - \frac{{\eta \kappa }}{{2{{\left( {4\pi } \right)}^2}}}\ln \frac{{{\mu ^2}}}{{{m^2}}}} \right\} \notag\\
&\qquad\Longrightarrow\quad \mu \frac{\dd}{{\dd\mu }}\kappa  = \frac{{4{\kappa ^2}}}{{{{\left( {4\pi } \right)}^2}}} + \frac{{\eta \kappa }}{{{{\left( {4\pi } \right)}^2}}} \,.
\end{align}
\end{subequations}
\endgroup

\subsubsection{Gauge Theory}\label{appsubsec:RGEQCDQED}
Finally, we will show how this formalism can be applied to gauge theories with charged fermions. We will perform the calculation for the generalization of QCD as a non-Abelian $SU(N_c)$ gauge theory coupled to $N_f$ quarks. The Lagrangian is
\begin{equation}
{\cal L}_\text{QCD} = - \frac{1}{4} G_{\mu \nu}^a G^{\mu \nu,a} + {\cal L}_\text{gf} + {\cal L}_\text{gh} +\bar{q} i{\slashed D} q \,.
\end{equation}
The trace over gluons gives the adjoint Casimir
\begin{equation}
\tr_G\left( {{T^a}{T^b}} \right) = N_c \delta^{ab} \,.
\end{equation}
We denote the quark piece as $q$ and assume that all $N_f$ flavors transform under the same representation $Q$ of $SU(N_c)$. Accordingly, we have
\begin{equation}
\tr_q\left( {{T^a}{T^b}} \right) = N_f d_Q \delta^{ab} \,,
\end{equation}
with the factor $d_Q$ depending on the representation $Q$. For example, $d_Q=\frac{1}{2}$ if $Q$ is in the fundamental representation, and $d_Q=N_c$ if $Q$ is in the adjoint representation.

We will take variation with respect to the gauge fields using the background field method described in~\cref{appsubsec:BackgroundField}. The Lagrangian is
\begin{equation}
{\cal L}_\text{QCD} \supset \frac{1}{2}A_\mu ^a\left[ {{\eta ^{\mu \nu }}{{\left( {{D^2}} \right)}^{ab}} - 2g{f^{abc}}{G^{\mu \nu ,c}}} \right]A_\nu ^b + {{\bar c}^a}{\left( { - {D^2}} \right)^{ab}}{c^b} + \bar q\left( {i\slashed D + g\slashed A} \right)q \,.
\end{equation}
Taking the second variation, we obtain
\begin{equation}
\renewcommand\arraystretch{1.5}
{\delta ^2}{S_{{\text{QCD}}}} = \left( {\begin{array}{*{20}{c}}
{\delta A_\mu ^a}&{\delta {q^T}}&{\delta \bar q}
\end{array}} \right)
\mqty( C^{\mu\nu, ab} & \bar{\Gamma}^{\mu ,a} & -\left(\Gamma^{\mu, a}\right)^T \\
-\left(\bar{\Gamma}^{\nu, b}\right)^T & 0 & - B^T \\ \Gamma^{\nu, b} & B & 0 )
\left( {\begin{array}{*{20}{c}}
{\delta A_\nu ^b}\\
{\delta q}\\
{\delta {{\bar q}^T}}
\end{array}} \right) + 2\delta {{\bar c}^a}{\left( { - {D^2}} \right)^{ab}}\delta {c^b} \,,
\renewcommand\arraystretch{1.0}
\end{equation}
where we have defined
\begin{subequations}
\begin{align}
{C^{\mu \nu ,ab}} &= {\eta ^{\mu \nu }}{\left( {{D^2}} \right)^{ab}} - 2g{f^{abc}}{G^{\mu \nu ,c}} = {\left( {{\eta ^{\mu \nu }}{D^2} + 2{G^{D,\mu \nu }}} \right)^{ab}} \,, \\
B &= i\slashed D \quad,\quad {\Gamma ^{\mu ,a}} = g{\gamma ^\mu }{T^a}q \quad,\quad {{\bar \Gamma}^{\mu ,a}} = g\bar q{\gamma ^\mu }{T^a} \,.
\end{align}
\end{subequations}

To derive the RGE for the gauge coupling, we only need to keep the kinetic term of the gauge boson. Dropping all term with quark fields, we find
\begin{equation}
\renewcommand\arraystretch{1.5}
\frac{{{\delta ^2}{S_{{\text{QCD}}}}}}{{\delta {{\left( {A_\mu ^a,\bar q,q} \right)}^2}}} = \left( {\begin{array}{*{20}{c}}
{{C^{\mu \nu ,ab}}}&{{{\bar \Gamma }^{\mu ,a}}}&{ - {{\left( {{\Gamma ^{\mu ,a}}} \right)}^T}}\\
{ - {{\left( {{{\bar \Gamma }^{\nu ,b}}} \right)}^T}}&0&{ - {B^T}}\\
{{\Gamma ^{\nu ,b}}}&B&0
\end{array}} \right) \supset \left( {\begin{array}{*{20}{c}}
{{C^{\mu \nu ,ab}}}&0&0\\
0&0&{ - {B^T}}\\
0&B&0
\end{array}} \right) \,.
\renewcommand\arraystretch{1.0}
\end{equation}
The one-loop effective action is then
\begin{align}
\Gamma_{{\text{QCD}}}^{\left( 1 \right)} &= \frac{i}{2} \ln \text{Sdet} \left[ { - \dfrac{{{\delta ^2}{S_{{\text{QCD}}}}}}{{\delta {{\left( {A_\mu ^a,\bar q,q,\bar c,c} \right)}^2}}}} \right] \nonumber \\
 &\supset \frac{i}{2}\ln {\textstyle\det_G}\left( { - {C^{\mu \nu ,ab}}} \right) - i\ln {\textstyle\det _c}\left( {{D^2}} \right) - \frac{i}{2}\ln {\textstyle\det _q}\left( {\begin{array}{*{20}{c}}
0&{ {B^T}}\\
{ - B}&0
\end{array}} \right) \nonumber \\
 &\equiv {\Gamma_G} + {\Gamma _c} + {\Gamma _q} \,.
\end{align}

We now provide a detailed calculation for each of these three terms. For the gluon piece, we have
\begingroup
\allowdisplaybreaks
\begin{align}
{\Gamma _G} &\equiv \frac{i}{2}\ln {\textstyle\det _G}\left( { - {C^{\mu \nu ,ab}}} \right) = \frac{i}{2}\ln {\textstyle\det _G}\left( { - {\eta ^{\rho \nu }}{D^2} - 2{G^{D,\rho \nu }}} \right) \nonumber \\
 &= \frac{i}{2}\ln {\textstyle\det _G}\left( { - {D^2}{\eta ^{\rho \beta }}} \right) + \frac{i}{2}{{\Tr}_G}\ln \left( {\delta _\beta ^\nu  + \frac{1}{{{D^2}}}{\eta _{\beta \mu }}2{G^{D,\mu \nu }}} \right) \nonumber \\
 &\supset 2i\ln {\textstyle\det _G}\left( {{D^2}} \right) + \frac{i}{2}\Tr_G\left( { - \frac{1}{2}{\eta _{\nu \alpha }}\frac{1}{{{D^2}}}2{G^{D,\alpha \beta }}\frac{1}{{{D^2}}}{\eta _{\beta \mu }}2{G^{D,\mu \nu }}} \right) \nonumber \\
 &= 2i\ln {\textstyle\det _G}\left( {{D^2}} \right) + i\Tr_G\left( {\frac{1}{{{D^2}}}G_{\mu \nu }^D\frac{1}{{{D^2}}}{G^{D,\mu \nu }}} \right) \,.
\end{align}
\endgroup
The functional traces above yield scaleless loop integrals, which evaluate to zero when using dim.\ reg.. Therefore, we must isolate the UV divergences in order to compute the RGEs. To this end, we introduce a mass $m^2$ for the gluons which serves as an explicit IR regulator:
\begingroup
\allowdisplaybreaks
\begin{align}
{\Gamma _G} &\to 2i\ln {\textstyle\det _G}\left( {{D^2} + {m^2}} \right) + i{\Tr_G} \left( {\frac{1}{{{D^2} + {m^2}}}G_{\mu \nu }^D\frac{1}{{{D^2} + {m^2}}}{G^{D,\mu \nu }}} \right) \nonumber \\
 &\supset \int {{\dd^4}x\,\frac{1}{{{{\left( {4\pi } \right)}^2}}}\left[ {2\left( {\ln \frac{{{\mu ^2}}}{{{m^2}}}} \right){\tr_G}\left( {\frac{1}{{12}}G_{\mu \nu }^D{G^{D,\mu \nu }}} \right) + \left( {\ln \frac{{{\mu ^2}}}{{{m^2}}}} \right){\tr_G}\left( { - G_{\mu \nu }^D{G^{D,\mu \nu }}} \right)} \right]} \nonumber \\
 &= \int {{\dd^4}x\,\frac{1}{{{{\left( {4\pi } \right)}^2}}}\left( {\ln \frac{{{\mu ^2}}}{{{m^2}}}} \right){\tr_G}\left( { - \frac{5}{6}G_{\mu \nu }^D{G^{D,\mu \nu }}} \right)} \nonumber \\
 &= \int {{\dd^4}x\,\frac{1}{{{{\left( {4\pi } \right)}^2}}}\left( {\ln \frac{{{\mu ^2}}}{{{m^2}}}} \right)\frac{5}{6}{N_c}{g^2}G_{\mu \nu }^a{G^{a,\mu \nu }}} \,.
\end{align}
\endgroup
In deriving the second line above, we have used \cref{eqn:UniversalDU,eqn:T2}, keeping only the gluon kinetic terms. Similarly, the ghost pieces give
\begin{align}
{\Gamma _c} &\equiv  - i\ln {\textstyle\det_c}\left( {{D^2}} \right) \to  - i\ln {\textstyle\det_c}\left( {{D^2} + {m^2}} \right) \supset  - \int {{\dd^4}x\,\frac{1}{{{{\left( {4\pi } \right)}^2}}}\left( {\ln \frac{{{\mu ^2}}}{{{m^2}}}} \right)\tr_c\left( {\frac{1}{{12}}G_{\mu \nu }^D{G^{D,\mu \nu }}} \right)} \nonumber \\
 &= \int {{\dd^4}x\,\frac{1}{{{{\left( {4\pi } \right)}^2}}}\left( {\ln \frac{{{\mu ^2}}}{{{m^2}}}} \right)\frac{1}{{12}}{N_c}{g^2}G_{\mu \nu }^a{G^{a,\mu \nu }}} \,,
\end{align}
and finally the quark piece gives
\begingroup
\allowdisplaybreaks
\begin{align}
{\Gamma _q} &\equiv  - \frac{i}{2}\ln {\textstyle\det_q}\left( {\begin{array}{*{20}{c}}
0&{ {B^T}}\\
{ - B}&0
\end{array}} \right) =  - i\ln {\textstyle\det_q}\left( B \right) \to  - i\ln {\textstyle\det_q}\left( {i\slashed D - m} \right) \nonumber \\
 &=  - \frac{i}{2}\ln {\textstyle\det_q}\left( {i\slashed D - m} \right) - \frac{i}{2}\ln {\textstyle\det_q}\left( { - i\slashed D - m} \right) \nonumber \\
 &=  - \frac{i}{2}\ln {\textstyle\det_q}\left[ { - {{\left( {i\slashed D} \right)}^2} + {m^2}} \right] =  - \frac{i}{2}\ln {\textstyle\det_q}\left( {{D^2} + {m^2} - \frac{i}{2}{\sigma ^{\mu \nu }}G_{\mu \nu }^D} \right) \nonumber \\
 &\supset  - \frac{1}{2}\int {{\dd^4}x\,\frac{1}{{{{\left( {4\pi } \right)}^2}}}\left( {\ln \frac{{{\mu ^2}}}{{{m^2}}}} \right){\tr_q}\left( { - \frac{1}{8}{\sigma ^{\mu \nu }}{\sigma ^{\alpha \beta }}G_{\mu \nu }^DG_{\alpha \beta }^D + \frac{1}{{12}}G_{\mu \nu }^D{G^{D,\mu \nu }}} \right)} \nonumber \\
 &= \int {{\dd^4}x\,\frac{1}{{{{\left( {4\pi } \right)}^2}}}\left( {\ln \frac{{{\mu ^2}}}{{{m^2}}}} \right){N_f}{d_Q}\left[ { - \frac{1}{4}\left( {{\eta ^{\mu \alpha }}{\eta ^{\nu \beta }} - {\eta ^{\mu \beta }}{\eta ^{\nu \alpha }}} \right){g^2}G_{\mu \nu }^aG_{\alpha \beta }^a + \frac{1}{6}{g^2}G_{\mu \nu }^a{G^{a,\mu \nu }}} \right]} \nonumber \\
 &= \int {{\dd^4}x\,\frac{1}{{{{\left( {4\pi } \right)}^2}}}\left( {\ln \frac{{{\mu ^2}}}{{{m^2}}}} \right)\left( { - \frac{1}{3}{N_f}{d_Q}{g^2}G_{\mu \nu }^a{G^{a,\mu \nu }}} \right)} \,.
\end{align}
\endgroup
Note that in deriving the second line above, we have used the fact that one can flip the sign of the Dirac $\gamma$ matrices within the trace since only even powers will contribute. In the third line, we have applied the usual notation $\sigma^{\mu\nu} \equiv \frac{i}{2} \left[\gamma^\mu, \gamma ^\nu \right]$. To obtain the fourth line, we have used the universal formula \cref{eqn:UniversalDU}, again only keeping gauge boson kinetic term. To get the fifth line, we have used the trace property of the Dirac $\gamma$ matrices $\tr \left( \sigma^{\mu\nu} \sigma^{\alpha\beta} \right) = 4 \left( \eta^{\mu\alpha}\eta^{\nu\beta} - \eta^{\mu\beta}\eta^{\nu\alpha} \right)$.

Putting all the three pieces together, we get
\begin{align}
\Gamma_{{\text{QCD}}}^{\left( 1 \right)} &\supset {\Gamma_G} + {\Gamma_q} + {\Gamma_c} \supset \int {{d^4}x\,\frac{1}{{{{\left( {4\pi } \right)}^2}}}\left( {\ln \frac{{{\mu ^2}}}{{{m^2}}}} \right)\left( {\frac{{11}}{{12}}{N_c} - \frac{1}{3}{N_f}{d_Q}} \right){g^2}G_{\mu \nu }^a{G^{a,\mu \nu }}} \nonumber \\
 &= \int {{\dd^4}x\,\frac{1}{{{{\left( {4\pi } \right)}^2}}}\left( {\ln \frac{{{\mu ^2}}}{{{m^2}}}} \right)\left[ { - \left( {\frac{{11}}{{12}}{N_c} - \frac{1}{3}{N_f}{d_Q}} \right)G_{\mu \nu }^{D,a}{G^{D,a,\mu \nu }}} \right]} \,,
\end{align}
where we are using our notation
\begin{equation}
\left[ {{D_\mu },{D_\nu }} \right] \equiv G_{\mu \nu }^D = G_{\mu \nu }^{D,a}{T^a} \quad,\quad G_{\mu \nu }^{D,a} =  - igG_{\mu \nu }^a \,.
\end{equation}
Adding the tree-level piece
\begin{equation}
\Gamma _{{\text{QCD}}}^{\left( 0 \right)} \supset \int {{\dd^4}x\left( { - \frac{1}{4}G_{\mu \nu }^a{G^{a,\mu \nu }}} \right)}  = \int {{\dd^4}x\,\frac{1}{{4{g^2}}}\,G_{\mu \nu }^{D,a}{G^{D,a,\mu \nu }}} \,,
\end{equation}
we obtain the one-loop effective action:
\begin{equation}
{\Gamma _{{\text{QCD}}}} \supset \Gamma _{{\text{QCD}}}^{\left( 0 \right)} + \Gamma _{{\text{QCD}}}^{\left( 1 \right)} \supset \int {{\dd^4}x\left[ {\frac{1}{{4{g^2}}} - \frac{1}{{{{\left( {4\pi } \right)}^2}}}\ln \frac{{{\mu ^2}}}{{{m^2}}}\left( {\frac{{11}}{{12}}{N_c} - \frac{1}{3}{N_f}{d_Q}} \right)} \right]G_{\mu \nu }^{D,a}{G^{D,a,\mu \nu }}} \,.
\end{equation}
Following the discussion in~\cref{appsubsec:RGEgauge}, we obtain the RGE equation for the gauge coupling:
\begin{align}
0 &= \mu \frac{\dd}{{\dd\mu }}\left[ {\frac{1}{{4{g^2}}} - \frac{1}{{{{\left( {4\pi } \right)}^2}}}\ln \frac{{{\mu ^2}}}{{{m^2}}}\left( {\frac{{11}}{{12}}{N_c} - \frac{1}{3}{N_f}{d_Q}} \right)} \right] \notag\\
&\hspace{100pt} \Longrightarrow\quad \beta \left( g \right) = \mu \frac{{\dd g}}{{\dd\mu }} =  - \frac{{{g^3}}}{{{{\left( {4\pi } \right)}^2}}}\left( {\frac{{11}}{3}{N_c} - \frac{4}{3}{N_f}{d_Q}} \right) \,. \label{eqn:RGESUN}
\end{align}

Now taking the quarks to form the fundamental representations $d_Q=\frac{1}{2}$, we obtain the familiar QCD beta function
\begin{equation}
\beta_\text{QCD} \left( g \right) =  - \frac{{{g^3}}}{{{{\left( {4\pi } \right)}^2}}}\left( {\frac{{11}}{3}{N_c} - \frac{2}{3}{N_f}} \right) \,.
\label{eq:QCDbeta}
\end{equation}
In the case of QED, the gauge symmetry is Abelian. Since the structure constants vanish $f^{abc}=0$,
\begin{equation}
\tr_G \left(T^a T^b\right) = N_c \delta^{ab} = 0 \,,
\end{equation}
which corresponds to taking $N_c=0$. Hence, only the quark piece contributes, where the generator matrix is $\mathds{1}$:
\begin{equation}
\tr_q \left(T^a T^b\right) = d_Q \delta^{ab} = \mathds{1} \,.
\end{equation}
This corresponds to taking $d_Q=1$. Plugging into Eq.~\eqref{eqn:RGESUN}, we derive the QED beta function:
\begin{equation}
\beta_\text{QED} \left( g \right) = \frac{4}{3}\frac{{{g^3}}}{{{{\left( {4\pi } \right)}^2}}}{N_f} \,.
\end{equation}

\section{\boldmath Heavy-Heavy Current Matching at Finite Recoil}
\label{sec:HHMatch2}
In this appendix, we present the calculation of the heavy-heavy current matching for the case $w\ne1$. This is a generalization of the calculation performed in \cref{sec:HHMatch}. The resulting matching coefficients will now be functions of $w \equiv v_1 \cdot v_2 \ne 1$. Our results can be compared to \Ref{Neubert:1992tg} (see also the appendix of \Ref{Bernlochner:2017jka}).

The UV Lagrangian is the same as for $w=1$ case in \cref{eq:LagUVHH},
\begin{align}
  \Lag_\text{QCD} =\,& \bar{Q}_1\, \big(i\s \slashed{D} - m_1\big)\, Q_1 + \bar{Q}_2\, \big(i\s \slashed{D} - m_{2}\big)\, Q_2
  + \Big[\bar{Q}_1\, \Big(J^+_\alpha\, \gamma^\alpha + J^+_{5\alpha}\,\gamma^\alpha\,\gamma^5\Big)\, Q_2 + \text{h.c.}\Big] \notag\\[5pt]
 & - \frac{1}{4} G_{\mu\nu}^a G^{\mu\nu,a} + \Lag_\text{gf} + \Lag_\text{gh} \,.
\end{align}
However, in solving the equations of motion for the short distance modes, we now need to carefully distinguish the two velocities. The solution up to linear order in $J$ is now (\emph{cf.} \cref{eq:HSolUVHH})
\begin{subequations}
\begin{align}
 H_{v_1} &= \frac{1}{i\s v_1 \cdot D + 2\,m_1}\, i\s\slashed{D}_{\perp1}\, h_{v_1} + \frac{1}{i\s v_1 \cdot D + 2\,m_1}\, J^+
                \pqty{ 1 + \frac{1}{i\s v_2 \cdot D + 2\,m_{2}}\, i\s\slashed{D}_{\perp2} }\, h_{v_2} \,, \\[5pt]
 H_{v_2} &= \frac{1}{i\s v_2 \cdot D + 2\, m_{2}}\, i\s\slashed{D}_{\perp2}\, h_{v_2} + \frac{1}{i\s v_2 \cdot D + 2\,m_{2}} J^-
                \pqty{ 1 + \frac{1}{iv_1 \cdot D + 2m_1} i\slashed{D}_{\perp1} } h_{v_1} \,.
\end{align}
\end{subequations}
Note that the $D_\perp^\mu$ operators depend on the choice of velocity label as well. We have also introduced the shorthand
\begin{align}
J^\pm \equiv \Big(J^\pm_\alpha\,\gamma^\alpha + J^\pm_{5\alpha}\,\gamma^\alpha\,\gamma^5\Big)\, e^{\pm i\s\Delta p\cdot x} \,.
\end{align}
with $\Delta p^\mu \equiv m_1v_1^\mu - m_{2}v_{2}^\mu$. Taking the second variation of the action proceeds as before, yielding the same matrix structure
\begin{small}
\begin{align}
  \delta^2 \SHQETnl =
    \pmqty{ \var{A_\mu^a} & \var{h_{v_1}^T} & \var{\bar{h}_{v_1}} & \var{h_{v_2}^{T}} & \var{\bar{h}_{v_2}} }
    \pmqty{ C^{\mu\nu,ab} & \bar{\Gamma}_1^{\mu,a} & -\Big(\Gamma_1^{\mu,a}\Big)^T
            & \bar{\Gamma}_2^{\mu,a} & -\Big(\Gamma_2^{\mu,a}\Big)^T \\[5pt]
            -\Big(\bar{\Gamma}_1^{\nu,b}\Big)^T & 0 & -B_1^T & 0 & -S_2^T \\[8pt]
            \Gamma_1^{\nu,b} & B_1 & 0 & S_1 & 0 \\[5pt]
            -\Big(\bar{\Gamma}_2^{\nu,b}\Big)^T & 0 & -S_1^T & 0 & -B_2^T \\[8pt]
            \Gamma_2^{\nu,b} & S_2 & 0 & B_2& 0}
    \pmqty{ \var{A_\nu^b} \\[10pt] \var{h_{v_1}} \\[10pt] \var{\bar{h}^T_{v_1}} \\[10pt] \var{h_{v_2}} \\[9pt] \var{\bar{h}^{T}_{v_2}} } \,.\notag\\[3pt]
\end{align}
\end{small}\noindent
However, the matrix elements are now generalized to (dropping terms that do not contribute to the matching)
\begingroup
\allowdisplaybreaks
\begin{subequations}
\begin{align}
 \hspace{-7pt} C^{\mu\nu,ab}\! &= \eta^{\mu\nu} \big(D^2\big)^{ab} - 2\,\Big(U_1^{\mu\nu,ab} + 1 \leftrightarrow 2\Big) \,, \\[5pt]
 \hspace{-7pt}  U_{1,2}^{\mu\nu,ab}\! &= -g_s^2\, \bar{h}_{v_{1,2}}\, T^a\s T^b\, \frac{\gamma_{\perp1,2}^\mu}{i\s v_{1,2} \cdot D + 2 m_{1,2}}
                     \,J^\pm\,
                     \frac{\gamma_{\perp2,1}^\nu}{i\s v_{2,1} \cdot D + 2m_{2,1}}\, h_{v_{2,1}} \,, \\[5pt]
 \hspace{-7pt}  B_{1,2} &= i\s v_{1,2} \cdot D + i\s \slashed{D}_{\perp1,2}\, \frac{1}{i\s v_{1,2} \cdot D + 2m_{1,2}} i\s \slashed{D}_{\perp1,2} \,, \\[5pt]
 \hspace{-7pt}  \Gamma_{1,2}^{\mu,a} &= g_s\, T^a\, \Bigg\{ \bqty{ v_{1,2}^\mu + i\s \slashed{D}_{\perp1,2}
                                       \frac{\gamma_{\perp1,2}^\mu}{i\s v_{1,2} \cdot D + 2m_{1,2}} }\, h_{v_{1,2}} \notag\\[-3pt]
    &\hspace{50pt}+ \bqty{ 1 + i\slashed{D}_{\perp1,2} \frac{1}{iv_{1,2} \cdot D + 2m_{1,2}} }
        \,J^\pm\,
        \frac{\gamma_{\perp2,1}^\mu}{i\s v_{2,1} \cdot D + 2m_{2,1}}\, h_{v_{2,1}} \Bigg\} \,, \\[5pt]
 \hspace{-7pt}  \bar{\Gamma}_{1,2}^{\mu,a} &= g_s\, \Bigg\{ \bar{h}_{v_{1,2}}\, \bqty{ v_{1,2}^\mu
    + \frac{\gamma_{\perp1,2}^\mu}{i\s v_{1,2} \cdot D + 2m_{1,2}} i\slashed{D}_{\perp1,2} }\notag\\[-3pt]
    &\hspace{50pt}
    + \bar{h}_{v_{2,1}} \frac{\gamma_{\perp2,1}^\mu}{iv_{2,1} \cdot D + 2m_{2,1}}
                   \,J^\mp\,
                                      \bqty{ 1 + \frac{1}{i\s v_{1,2} \cdot D + 2m_{1,2}} i\s \slashed{D}_{\perp1,2} } \Bigg\}\s T^a \,, \\[5pt]
 \hspace{-7pt}  S_{1,2} &= \bqty{ 1 + i\s \slashed{D}_{\perp1,2} \frac{1}{i\s v_{1,2} \cdot D + 2m_{1,2}} }
           \,J^\pm\,
           \bqty{ 1 + \frac{1}{i\s v_{2,1} \cdot D + 2m_{2,1}} i\s \slashed{D}_{\perp2,1} } \,.
\end{align}
\end{subequations}
\endgroup

The relevant piece of the one-loop effective action is still abstractly given by the first line in \cref{eq:HHSDetForm}, since it derives from a matrix with the same form. However, now one must be careful to track the non-trivial interplay between the two velocities. Plugging in the concrete expressions of the matrix elements yields
\begin{align}
   \MoveEqLeft S^{(1)}_\text{HQET}
     \supset
      i\Tr\bqty{ \frac{1}{(i\s D)^2}\, \delta^{ab}\, \eta_{\mu\nu}\, \Big(U_1^{\mu\nu,ab}
                         + \bar{\Gamma}_1^{\mu,a}\, B_1^{-1}\, \Gamma_1^{\nu,b}
                         - \bar{\Gamma}_1^{\mu,a}\, B_1^{-1}\, S_1\, B_2^{-1}\, \Gamma_2^{\nu,b}\Big) } + 1 \leftrightarrow 2 \notag\\[8pt]
     &\supset -i\s g_s^2\, \frac{4}{3} \Tr\left\{w\, \frac{1}{(i\s D)^2}\, \bar{h}_{v_1}\, \frac{2m_1 + 2i\s v_1\cdot D}{(i\s D)^2 + 2m_1\, i\s v_1 \cdot D}\,J^+\, \frac{2m_{2} + 2i\s v_2\cdot D}{(i\s D)^2 + 2m_{2}\, i\s v_2 \cdot D}\, h_{v_2} \right.\notag \\[5pt]
 &\hspace{30pt}+ \frac{1}{(i\s D)^2}\, \bar{h}_{v_1}\, \frac{2m_1 + 2i\s v_1\cdot D}{(i\s D)^2 + 2m_1\, i\s v_1 \cdot D}\,J^+\, \frac{i\s \slashed{D}_{\perp2} - i\s v_2\cdot D}{(i\s D)^2 + 2m_{2}\, i\s v_2 \cdot D}\,\slashed{v}_1\, h_{v_2} \notag \\[5pt]
 &\hspace{30pt}+ \frac{1}{(i\s D)^2}\, \bar{h}_{v_1}\, \slashed{v}_2\, \frac{i\s \slashed{D}_{\perp1} - i\s v_1\cdot D}{(i\s D)^2 + 2m_1\, i\s v_1 \cdot D}\,J^+\, \frac{2m_{2} + 2i\s v_2\cdot D}{(i\s D)^2 + 2m_{2}\, i\s v_2 \cdot D}\, h_{v_2} \notag \\[5pt]
&\hspace{30pt} \left.+ \frac{1}{(i\s D)^2}\, \bar{h}_{v_1}\, \gamma^\mu\, \frac{i\s \slashed{D}_{\perp1} - i\s v_1 \cdot D}{(i\s D)^2 + 2m_1\, i\s v_1 \cdot D} \,J^+\, \frac{i\s \slashed{D}_{\perp2} - i\s v_2 \cdot D}{(i\s D)^2 + 2m_{2}\, i\s v_2 \cdot D}\, \gamma_\mu\, h_{v_2}
\right \} \notag\\[5pt]
&\hspace{13pt}+ \left( 1 \leftrightarrow 2 ,\, J^+ \leftrightarrow J^-\right) \,.
\end{align}
This expression reduces to the second line in \cref{eq:HHSDetForm} under the special case $v_1=v_2$, as it must. Next, we apply the CDE prescription discussed in~\cref{appsubsubsec:naiveCDE} to evaluate these functional traces (the analog of \cref{eq:HHIntForm}), which eventually simplifies to
\begin{align}
 S^{(1)}_\text{HQET}&\supset \int \dd[d]{x} \bqty{ \bar{h}_{v_1} \, \big(J^+_\alpha\, I_\text{HH}^\alpha + J^+_{5\alpha}\, I_{\text{HH},5}^\alpha\big)\,e^{i\s\Delta p\cdot x}\, h_{v_2} + \text{h.c.} } \,,
 \label{eqn:SHHw}
\end{align}
where the loop integrals are given by
\begin{subequations}
\begin{align}
    I_\text{HH}^\alpha &\equiv
      -i\s g_s^2\, \mu^{2\eps}\, \frac{4}{3} \int \ddp{p}\,\frac{1}{p^2 \big(p^2 + 2m_1\, v_1 \cdot p\big) \big(p^2 + 2m_{2}\, v_2 \cdot p\big)}\notag\\[2pt]
        &\hspace{107pt}\times\bigg\{w\, \big(2m_1 + 2v_1 \cdot p\big)\, \gamma^\alpha\, \big(2m_{2} + 2v_2 \cdot p\big) \notag\\[-3pt]
              &\hspace{130pt} + \big(2m_1 + 2v_1 \cdot p\big)\, \gamma^\alpha\, \big(\slashed{p}_{\perp2} - v_2 \cdot p\big)\, \slashed{v}_1 \notag\\[3pt]
              &\hspace{130pt} + \slashed{v}_2\, \big(\slashed{p}_{\perp1} - v_1 \cdot p\big)\, \gamma^\alpha\, \big(2m_{2} + 2v_2 \cdot p\big) \notag\\[-3pt]
              &\hspace{130pt} + \gamma^\mu \big(\slashed{p}_{\perp1} - v_1 \cdot p\big)\, \gamma^\alpha \big(\slashed{p}_{\perp2} -  v_2 \cdot p\big)\, \gamma_\mu \bigg\} \,,
 \\[15pt]
    I_{\text{HH},5}^\alpha &\equiv I_\text{HH}^\alpha\big|_{\gamma^\alpha \rightarrow\, \gamma^\alpha\,\gamma^5} \,.
\end{align}
\end{subequations}
Evaluating these integrals is straightforward, although significantly more tedious than in the equal velocity case. For example, evaluating $I_\text{HH}^\alpha$ with dim.\ reg.\ in the \MSbar scheme, we obtain
\begin{equation}
 I^\alpha = \tilde{C}_1\left(z, w \right)\,\gamma^\alpha - C_2\left(z, w\right)\,v_1^\alpha - C_3\left(z, w\right)\, v_2^\alpha \,,
\end{equation}
where $z\equiv\frac{m_2}{m_1}$, and the three coefficients are given by the following Feynman parameter integrals
\begin{subequations}
\begin{align}
 \tilde{C}_1 \left(z, w\right) &= \frac{2\s\alpha_s}{3\s\pi} \int_0^1 \dd x \int_0^{1-x} \dd y\, \left(\frac{4\s\pi\,\mu^2}{m_1^2}\,\frac{1}{x^2 + y^2\,z^2 + 2\,x\,y\,z\,w}\right)^\eps\,\Gamma(\eps)\, \Bigg[ 1 - 3\,\eps + 2\,\eps^2 \notag\\[-3pt]
 &\hspace{60pt} + \eps\,\frac{x + y\,z^2}{x^2 + y^2\,z^2 + 2\,x\,y\,z\,w} - \eps\,\frac{2\,w - \left(1+2\,w\right)\,\left(x+y\right)}{x^2 + y^2\,z^2 + 2\,x\,y\,z\,w}\,z \Bigg] \,, \\[10pt]
 C_2 \left(z, w\right) &= -\frac{2\s\alpha_s}{3\s\pi}\, 2 \int_0^1 \dd x \int_0^{1-x} \dd y\, \frac{x^2 - y\,\left(1-x\right)\,z}{x^2 + y^2\,z^2 + 2\,x\,y\,z\,w} \,, \\[10pt]
 C_3 \left(z, w\right) &= -\frac{2\s\alpha_s}{3\s\pi}\, 2 \int_0^1 \dd x \int_0^{1-x} \dd y\, \frac{y^2\,z^2 - x\,\left(1-y\right)\,z}{x^2 + y^2\,z^2 + 2\,x\,y\,z\,w}  = C_2\left(z^{-1}, w\right) \,.
\end{align}
\end{subequations}
Note that for the integrands of $C_2\left(z, w\right)$ and $C_3\left(z, w\right)$, we have taken the $\eps\to0$ limit and subtracted the $\frac{1}{\eps}-\gamma_\text{E}+\ln 4\s\pi$ counter terms. We are allowed to do so before performing the Feynman parameter integral when the $\eps\to0$ limit is finite. Further evaluation of these integrals is again straightforward, but will yield lengthy expressions that are less enlightening. As an explicit cross check, we can evaluate $C_3\left(z, w\right)$:
\begin{align}
C_3\left(z, w\right) &= \frac{2\s\alpha_s}{3\s\pi}\, \Bigg\{ \frac{z}{\left(1-2\,w\,z+z^2\right)^2}\, \Bigg[ 2\,(w-1)\,z\,(1+z)\,\ln z \notag\\
                                &\hspace{75pt} -\pmqty{ (w+1)- 2\,w\,(2\,w+1)\,z \\ - \left(1- 5\,w-2\,w^2\right) z^2 - 2\,z^3 }\,\frac{1}{\sqrt{w^2-1}}\,\ln\left(w+\sqrt{w^2-1}\,\right) \Bigg] \notag\\
                     &\hspace{50pt} -\frac{z}{1-2\,w\,z+z^2}\,\left(\ln z -1 + z\right) \Bigg\} \,.
\end{align}
This agrees with Eq.~(19) and Eq.~(A1) in~\Ref{Neubert:1992tg}. As noted above, the function $C_2\left(z, w\right)$ and $C_3\left(z, w\right)$ are related to each other by $m_1\leftrightarrow m_{2}$, as they must. The same relation holds for the results obtained in~\Ref{Neubert:1992tg}, as explicitly stated in Eq.~(A3) therein. So this verifies our derivation of $C_2\left(z, w\right)$ as well. We note that our coefficient $\tilde{C}_1 \left(z, w\right)$ defined above is slightly different from the $C_1$ defined in~\Ref{Neubert:1992tg}, because in our approach, the residue difference contribution needs to be added on top of the result obtained in \cref{eqn:SHHw}, \emph{cf.} \cref{eqn:hhresult}. We leave the evaluation of $\tilde{C}_1 \left(z, w\right)$ and $I_{\text{HH},5}^\alpha$ as well as the comparison of them with~\Ref{Neubert:1992tg} for future work, as we believe the evidence that functional methods can be applied to HQET is sufficient.

\addcontentsline{toc}{section}{\protect\numberline{}References}%
\bibliographystyle{jhep}
\bibliography{CDE_HQET}
\end{document}